\documentclass[useAMS,usenatbib]{mn2e}
\usepackage{graphicx}
\usepackage{caption}
\usepackage[fleqn]{amsmath}
\usepackage{lipsum}
\DeclareCaptionFormat{empty}{#1}

\title[Proper motions and kinematics of selected bulge globular clusters]
{Proper motions and kinematics of selected bulge globular clusters}

\author[Rossi et al.]{ L. J. Rossi$^{1}$, S. Ortolani$^{2,3}$
\footnote{Observations collected at the European Southern Observatory at La Silla, Chile;
Proposals 087.D-0218(A), 089.D-0194(A), 091.D-0711(A), PI: S. Ortolani},
 B. Barbuy$^{4}$,  E. Bica$^{5}$ \& A. Bonfanti$^{2}$\\
$^{1}$ Centre for Astrophysics and Supercomputing, Swinburne University of Technology, Hawthorn, Victoria 3122, Australia\\
$^{2}$ Universit\`a di Padova, Dipartimento di Fisica e Astronomia,  I-35122 Padova, Italy\\
$^{3}$ INAF-Osservatorio Astronomico di Padova, Vicolo dell'Osservatorio 5,
I-35122 Padova, Italy\\
$^{5}$ Universidade de S\~ao Paulo, IAG, Rua do Mat\~ao 1226,
 Cidade Universit\'aria, S\~ao Paulo 05508-900, Brazil\\
$^{4}$ Universidade Federal do Rio Grande do Sul, Departamento de Astronomia, CP 15051, Porto Alegre 91501-970, Brazil}

\begin{document}

\date{  }

\pagerange{\pageref{firstpage}--\pageref{lastpage}} \pubyear{2002}

\maketitle

\label{firstpage}

\begin{abstract}
We computed proper motions of a selected sample of globular clusters projected on the central bulge,
 employing CCD images gathered along the last 25 years
 at the ESO-NTT, ESO-Danish and HST telescopes. We presented a method to derive their proper motions,
 and a set of coordinate transformations to obtain 3D Galactic velocity vectors of the clusters.
We analysed 10 globular clusters, namely Terzan 1, Terzan 2, Terzan 4, Terzan 9, NGC 6522, NGC 6558, NGC 6540, AL~3, ESO456--SC38 and Palomar 6. For comparison purposes we also studied the outer bulge cluster NGC 6652. 
We discuss the general properties of the proper-motion-cleaned Colour-Magnitude Diagrams,
derived for the first time for most of them. A general conclusion is that the inner bulge globular clusters
 have clearly
 lower transverse motions (and spatial velocities) than halo clusters, and appear to be trapped 
in the bulge bar.
\end{abstract}


\section{Introduction}

Globular clusters (GCs) in the Galactic bulge  preserve in their spatial distribution and 
orbital evolution essential information to probe the early the early formation
 stages of the Galaxy central parts.
The combination of dynamical properties of globular clusters, 
with their ages and chemical composition, provides a new tool to investigate the bulge stellar populations,
and to build a consistent scenario of the Galactic bulge formation.

 

Stars and globular clusters in the Galactic halo present very elliptical orbits, with low angular momentum,
while disk objects show circular orbits with high angular momentum, and small
vertical velocity. The bulge instead shows an intermediate angular momentum,
with higher perpendicular velocities than disc stars, but not going as deep into the halo
as genuine halo stars.
For the field bulge stars, kinematics reveals two different behaviours
(Babusiaux et al. 2010): (1) a metal poor component, enriched in [Mg/Fe],
consistent with an isotropic rotating population belonging to an old spheroid, and
(2) a metal rich one with a vertex deviation consistent with that expected from a population
with orbits supporting a bar.
The spatial separation of these two components is not well determined.
More recently, Babusiaux et al. (2014) and Zoccali et al. (2014)
carried out large surveys on stars in the Galactic bar, inner disk
and central bulge, by measuring radial velocities and metallicities from
CaT lines, and traced the bar with more precision.
Of interest to us is the conclusion that there is a concentration of metal-poor stars,
that could have formed before the bar instability, and
 could have been trapped by the bar when it formed,
staying confined in the innermost regions of the boxy bulge (Babusiaux et al.
2014; Di Matteo et al. 2014).
Their conclusion is that the exact distribution of metal-poor stars, and their
connection with the formation history of the thick disk, inner halo and the bar 
 need further investigations.

Most of previous efforts in bulge cluster proper motion measurements were carried out using HST data 
as those e.g. by Zoccali et al. (2001) for NGC 6553, 
 Feltzing \& Johnson (2002) for NGC 6528,
Bellini et al. (2013) for NGC 6338 and NGC 6441, 
 and Kuijken \& Rich (2002) for field stars. Ground-based large telescopes equipped with
Multi-Conjugate Adaptive Optics also allowed proper-motion-cleaning to be performed, such
as those
 carried out by Ferraro et al. (2009) for Terzan 5, and Ortolani et al. (2011) for HP~1.

Calculations of Galactic orbits, based on proper motions and radial velocities,
for some inner Galaxy clusters were carried out by Dinescu et al. (2003). 

In previous works we studied the inner bulge globular clusters, 
by means of Colour-Magnitude Diagrams (CMDs) and spectroscopy 
(e.g. Ortolani et al. 1995; Barbuy et al. 1998, 2009).
 We now have an unprecedented archive of CCD images taken since the 90s with 
  European Southern Observatory (ESO) NTT and Danish telescopes, and 
the Hubble Space Telescope (HST).
Through the combination of this archive data with more recent CCD images, this
data set allows cluster proper motions to be derived, and therefore
proper-motion-cleaned CMDs and kinematics (e.g. Ortolani et al. 2011).

We focus on  moderately metal-poor GCs
([Fe/H]$\sim -1$) with blue horizontal branch, 
 projected at low galactic latitude,
because they might be the oldest population in the Galaxy
(e.g. Barbuy et al. 2009).
 They might belong to the same generation of stars 
as the central RR Lyrae (D\'ek\'any et al. 2013), the latter identified
as an old and spheroidal component of the bulge.
We also study a few more metal-rich GCs.

Our main goal is to derive accurate ($\sim$0.5 mas/yr)
 absolute proper motions.
Kinematic properties of the sample clusters
 should allow bulge members and halo intruders to be distinguished. 
The proper-motion results, coupled with radial velocities,
 allow 3D-orbit determinations in the Galactic potential
 (e.g. Ortolani et al. 2011).
The orbits will be used to constrain different stellar-population
 components.
In a forthcoming paper the orbits will be presented in detail. 


In this work we derive the  initial state vector for a total of ten GCs together with 
a recompilation of their metallicities. Sect. 2 describes the selection of the sample. In Sect. 3
 we report the observations.
In Sect. 4  proper motions are derived. In Sect. 5 we present an analysis of errors,
 the results are presented in Sect. 6 and conclusions are drawn in Sect. 7.

\begin{table*}
\label{log}
\small
 \centering
 \begin{minipage}{140mm}
 \caption{Log of observations of first and second epochs.}
  \begin{tabular}{llllllll}
  \hline
Cluster & Telescope & Instrument & Date &Filter  & Exposure & Seeing &  \\
\hline
Terzan 1 1st    & Danish & 0.47''/pix  & 06/1990 &I & 180s & 1.1 &  \\
   & Danish & 0.47''/pix  & 06/1990 &Gunn z & 120s & 1.1 &  \\
Terzan 1 2nd & NTT     & EFOSC2 0.24''/pix & 05/2012 &I & 10s & 1.1 &  \\
 & NTT     & EFOSC2 0.24''/pix & 05/2012 &Gunn z & 180s & 1.1 &  \\
\hline
Terzan 2 1st & NTT     & SUSI 0.13''/pix & 05/1994 &V & 30,720s & 0.8, 0.9 &  \\
             & NTT     & SUSI 0.13''/pix & 05/1994 &I & 60,420s & 0.8, 0.8 &  \\
Terzan 2 2nd & NTT     & EFOSC2 0.24''/pix & 06/2011 &V & 60,300,20s & 1.1,1.1,1.1 &  \\
 & NTT     & EFOSC2 0.24''/pix & 06/2011 &I & 40,180,10s & 1.1,1.1,1.1 &  \\
\hline
Terzan 4 1st & NTT & SUSI 0.13''/pix & 05/1994 &V & 60,600s & 0.55, 0.55 &  \\
 & NTT & SUSI 0.13''/pix & 05/1994 &I & 60,300s & 0.4,0.4 &  \\ 
 & NTT & SUSI 0.13''/pix & 05/1994 &Gunn z & 120s & 0.5 &  \\
Terzan 4 2nd  & NTT & EFOSC2 0.24''/pix & 05/2012 &I & 30s & 0.6 &  \\
 & NTT & EFOSC2 0.24''/pix & 05/2012 &V & 30s & 0.6 &  \\
\hline
Terzan 9 1st & Danish & DFOSC 0.39''/pix & 07/1998 &V & 60,900,60s & 1.3,1.3,1.3 &  \\
& Danish & EFOSC2 0.39''/pix & 07/1998 &I & 60,90s & 1.1,1.1 &  \\
Terzan 9 2nd  & NTT & DFOSC 0.24''/pix & 05/2012 &V & 30,90s & 0.5,0.7 &  \\
 & NTT & EFOSC2 0.24''/pix & 05/2012 &I & 10,20s & 0.55,0.6 &  \\
\hline
NGC 6522 1st & HST & WFPC2 & 09/1995 & F439W & 50,160,160s & ---& \\
 & HST & WFPC2 & 09/1995 & F555W & 10,50s & ---& \\
NGC 6522 1st & Danish & 0.47''/pix & 06/1992 &V & 60,480s & 1.2,1.3 & \\
 & Danish & 0.47''/pix & 06/1992 &I & 20,300s & 1.2,1.2 & \\
NGC 6522 2nd & NTT & EFOSC2 0.24''/pix & 05/2012 & V & 30,60s & 0.55,0.75 & \\
& NTT & EFOSC2 0.24''/pix & 05/2012 & I & 15,30s & 0.5,0.7 & \\
\hline
NGC 6558 1st  & NTT & EMMI 0.35''/pix & 06/1993 & I & 7,120s &1.2,1.2 & \\
 & NTT & EMMI 0.35''/pix & 06/1993 & V & 10,180s &1.2,1.2 & \\
NGC 6558 2nd  & NTT & EFOSC2 0.24'/pix & 05/2012 & V & 90,300s & 0.6,0.6 & \\
 & NTT & EFOSC2 0.24'/pix & 05/2012 & I & 60,180,20s & 0.5,0.55,0.55-0.6 & \\
\hline
NGC 6540 1st & Danish  & 0.47''/pix & 06/1990 & V & 60s & 1.3 & \\
& Danish  & 0.47''/pix & 06/1990 & I & 15s & 1.0 & \\
NGC 6540 2nd & NTT & EFOSC2 0.24''/pix & 05/2012 & V & 30s & 0.9 &  \\
& NTT & EFOSC2 0.24''/pix & 05/2012 & I & 10s & 0.9 &  \\
\hline
AL 3  1st & Danish & DFOSC 0.39''/pix  & 03/2000 &V & 60,180s & 1.2,1.2 & \\
& Danish & DFOSC 0.39''/pix  & 03/2000 &I & 10,40s & 1.2,1.2 & \\
AL 3 2nd & NTT & EFOSC2 0.24''/pix & 05/2012 & V & 30,60,240,60s & 0.6 for all &  \\
 & NTT & EFOSC2 0.24''/pix & 05/2012 & I & 10,30,180s & 0.6 &  \\
\hline
ESO456--SC38 1st & NTT & SUSI 0.13''/pix &  05/1994 & I & 60s & 0.9 & \\
& NTT & SUSI 0.13''/pix &  05/1994 & V & 180s & 0.8 & \\
& NTT & SUSI 0.13''/pix &  05/1994 & Gunn z & 60s & 0.8 & \\
ESO456--SC38 2nd & NTT &EFOSC2 0.24''/pix & 05/2012 &V & 900,60s &1.0,1.0 & \\  
& NTT &EFOSC2 0.24''/pix & 05/2012 &I & 30s &0.8 & \\ 
\hline
Palomar 6  1st & NTT & EMMI 0.35''/pix & 06/1993 &V & 900s & 1.4 & \\
 & NTT & EMMI 0.35''/pix & 06/1993 &I & 300s & 1.4 & \\
Palomar 6 2nd & NTT & EFOSC2/HR 0.12''/pix& 05/2013 & Gunn z & 300s & 0.9 &  \\
 & NTT & EFOSC2/HR 0.12''/pix& 05/2013 & I & 300s & 1.1 &  \\
\hline
\end{tabular}
\end{minipage}
\end{table*}

\begin{table*}
\label{distance}
 \centering
 \begin{minipage}{140mm}
\caption{Distances from the literature (all values in kpc).}
  \begin{tabular}{lllllllllll}
  \hline
Cluster & Bica+06 & Valenti+07 & Harris96 & Barbuy+98 & other &   Reference  \\
  \hline
Terzan 1 & 6.2 & 6.6 &6.7      &4.90    & 5.2  &  Ortolani et al. (1999a)  &   \\
Terzan 2 & 8.7 & 7.4 &7.5      & 6.64   & 7.7  & Ortolani et al. (1997)  &   \\
Terzan 4 & 9.1 & 6.7 &7.2      & 7.28   & 8.0  & Ortolani et al. (2007)  &  \\
Terzan 9 & 7.7 & 5.6 &7.1      &---     & 4.9  &  Ortolani et al. (1999b)  &   \\
NGC 6522 & 7.8 & 7.4 &7.7      &6.05    & 6.2  &  Barbuy et al. (1994)  &   \\
NGC 6558 & 7.4 & --- &7.4      &--      & 6.3  &  Rich et al. (1998)  &  \\
NGC 6540  & 3.7 & 5.2 & 5.3      &3.02    & 3.5 &  Bica et al. (1994)  &   \\
AL 3 & --- & --- & 6.5         &---     & 6.0  &  Ortolani et al. (2006)  &   \\
ESO456-SC38 & 6.7 & 7.0 & 6.3      &3.02    & 5.5  &  Ortolani et al. (1997)  &  \\
Palomar 6 & 7.3 & --- & 5.8     &--      & 8.9 &  Ortolani et al. (1995) &  \\
NGC 6652  & 9.6 & ... & 10.0    & ---    & 9.3 &  Ortolani et al. (1994) &  \\
\hline          
\end{tabular}
\end{minipage}
\end{table*}

%
                                        
\section{Sample selection}       
       
\subsection{Metallicity and spatial distribution}

\begin{figure}
\centering
\includegraphics[scale=0.45]{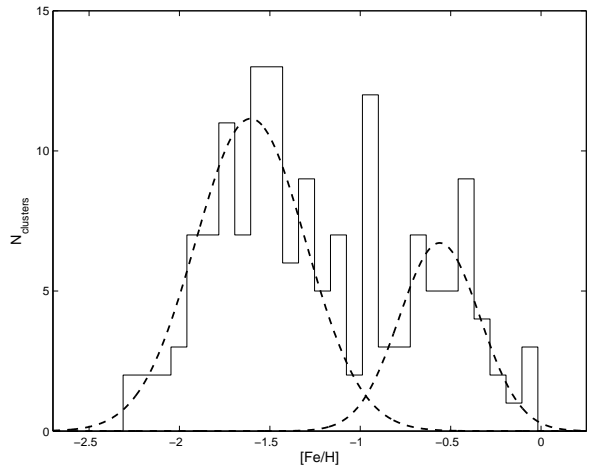}
\caption{Metallicity distribution of the Milky Way globular clusters, updated with recent literature.
Dashed lines are gaussian fits for the metal-rich and metal-poor populations.}
\label{fig:FE_H_dist}
\end{figure}

The metallicity distribution of the Milky Way globular clusters is shown in Figure \ref{fig:FE_H_dist},
where metallicities are adopted from Bica et al. (2006), in some cases
updated with more recent spectroscopic analyses (Barbuy et al. 2009, 2014).

Figure \ref{fig:FE_H_dist} shows a Gaussian distribution centred around $[\mbox{Fe}/\mbox{H}]\simeq-1.6$, representing the metal--poor population, and a Gaussian distribution around the value $[\mbox{Fe}/\mbox{H}]\simeq-0.55$, corresponding to the metal--rich population (e.g. Morgan 1959; Kinman 1959; Zinn 1985; C\^ot\'e 1999).
A moderately metal--poor sub--population characterized by $[\mbox{Fe}/\mbox{H}]\simeq -1.0$ is evident (see also C\^ot\'e 1999).

With this criterion in mind we subdivided the globular clusters into three sub--samples:
\begin{itemize}
\item metal--poor GCs (95 objects): $[\mbox{Fe}/\mbox{H}] < -1.1$
\item metal--rich GCs (42 objects): $[\mbox{Fe}/\mbox{H}] > -0.9$
\item moderately metal--poor GCs (13 objects): $-1.1 \leq [\mbox{Fe}/\mbox{H}]-\leq -0.9$.     
\end{itemize}

In this work we prioritize clusters with
a blue horizontal branch together with a metallicity of [Fe/H]$\sim -1.0$,
including NGC 6522, NGC 6558, AL~3, ESO456-SC38, NGC 6540.
We also study two mainly red horizontal branch and moderately metal-poor
clusters, Terzan 1 and Palomar 6, two metal-poor globular  clusters projected on
 the inner bulge (Terzan 4, Terzan 9), and one metal-rich  (Terzan 2).
For comparison purposes we also included the outer bulge  NGC 6652,
with proper motion values taken from Sohn et al. (2014).

The log of observations is presented in Table \ref{log}.
In Table 2 we report the cluster distances from the literature. This Table shows a good
agreement among authors, which gives confidence on the distance values.

\subsection{The distance of the Galactic centre}

The distance of the Sun from the Galactic centre ($R_\odot$) is a crucial parameter that we
use to establish a cluster initial state vector.

One of the most common methods adopted to determine the value of this parameter is the determination of the mean distance of Galactic globular clusters (Bica et al. 2006), or the distribution of RR Lyrae stars (D\'ek\'any et al. 2013). Another method is based on the distribution of field clump stars (Bobylev et al. 2014; Nishiyama et al. 2006; Cao et al. 2013).  
Finally, Gillessen et al. (2009), and Do et al. (2013) reported recent values from the orbit of
 stars around SGR A*, and derived a geometrical distance.
A recent review of these methods and results was presented by Malkin (2013). The distances
of the Sun to the Galactic center essentially vary from 7.5 kpc to 8.5 kpc.
In the present paper we considered both 7.5 kpc, for the reason that this distance is consistent with
the distribution of bulge globular clusters, and  8.3 kpc, as reported by
Brunthaler et al. (2011), Sch\"onrich (2012), Reid et al. (2014) and Hou \& Han (2014), among others.
In Table \ref{tab:all_clusters} we give the proper motion values, and Galactic distance components 
and velocity vectors,  computed for both
 $R_{\odot}=7.5 \; \mbox{kpc}$, and  $R_{\odot}=8.3 \; \mbox{kpc}$ (see Sect. 4).     

\subsection{Distribution of clusters in the Galaxy}

\begin{figure}
\centering
\includegraphics[scale=0.55]{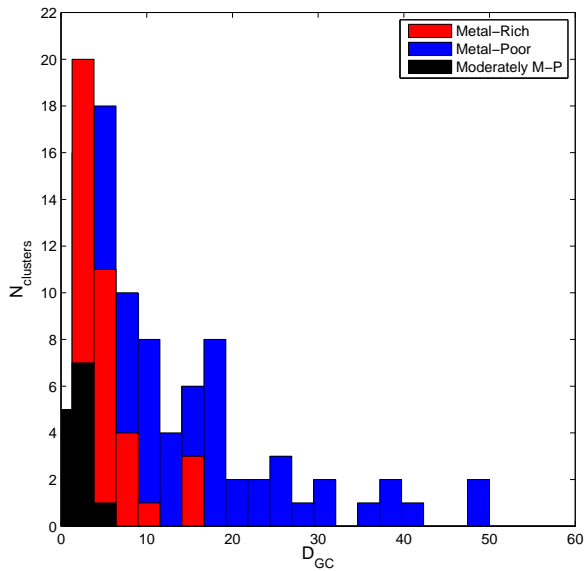}
\caption{Galactocentric distance (in kpc) distribution
 of the three sub--samples of globular clusters.}
\label{fig:distance}
\end{figure}

\begin{figure}
\includegraphics[scale=0.58]{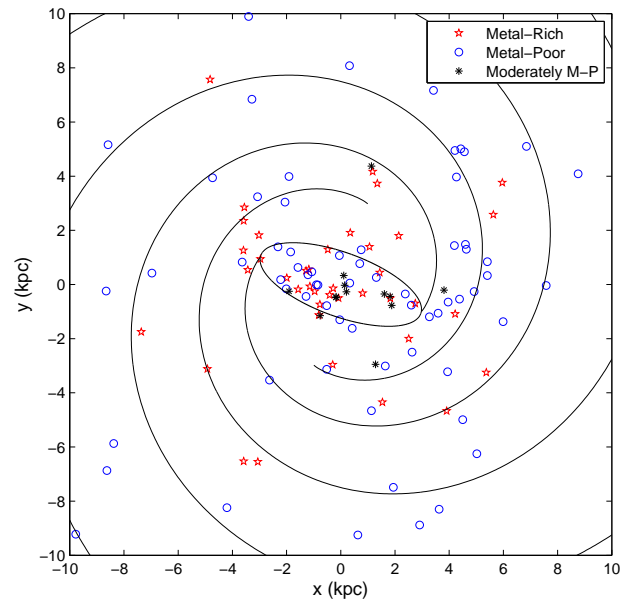}
\caption{Spatial distribution of globular clusters projected on the Galactic plane
on the x--y plane, viewed from the North Galactic Pole.}
\label{fig:plane_distr}
\end{figure}

\begin{figure}
\includegraphics[scale=0.58]{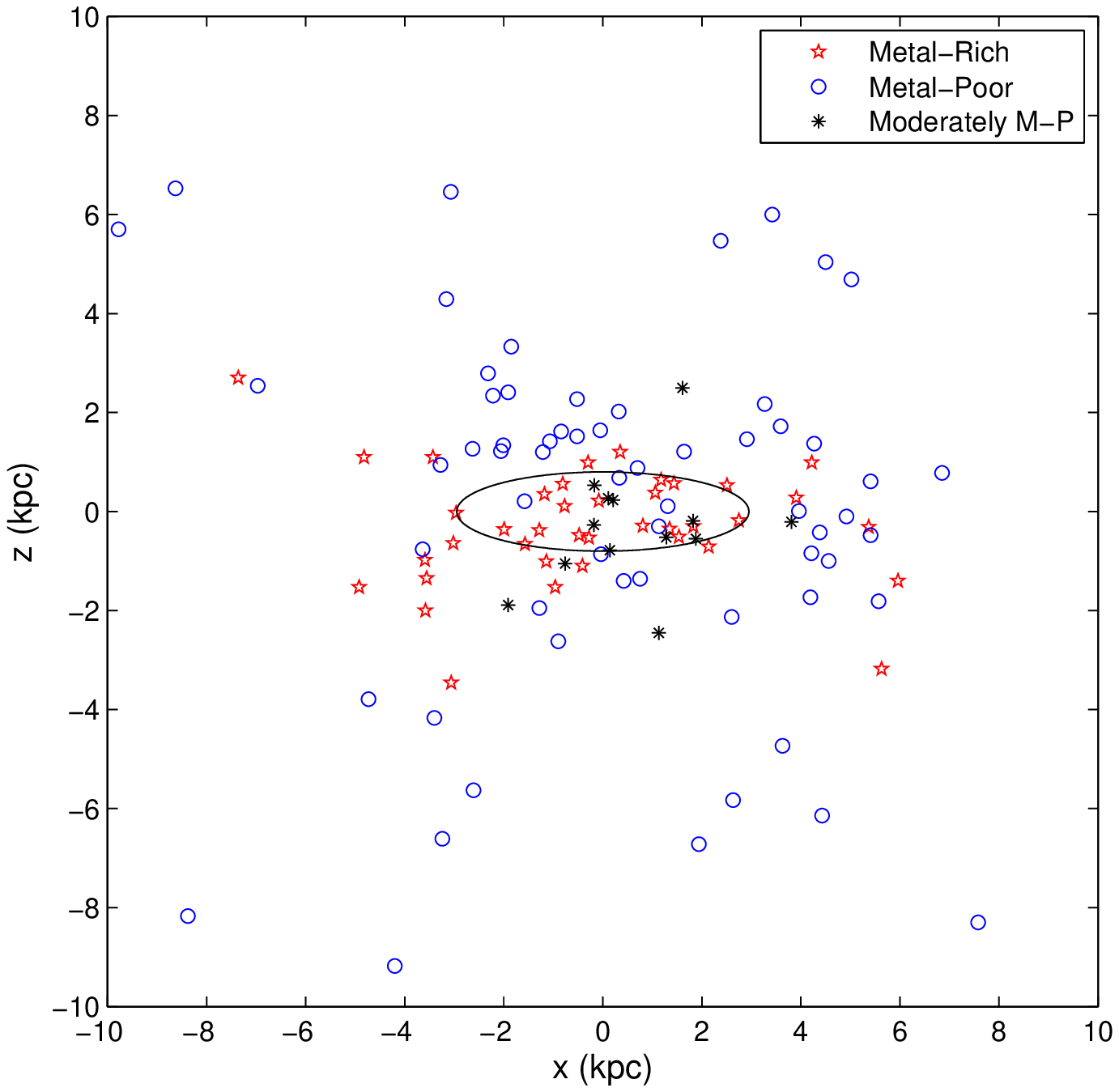}
\caption{Spatial distribution of globular clusters projected on the x--z plane
 for a selection of $|l|\leq 18^{\circ}$ and $|b|\leq 7.5^{circ}$.}
\label{fig:xz_distr}
\end{figure}

Figure \ref{fig:distance} shows the Galactocentric-distance distribution of the three
metallicity sub--samples of GCs. Figures \ref{fig:plane_distr} and \ref{fig:xz_distr} show the Galactic distribution of the three sub--samples. The coordinates of the clusters are expressed in the inertial right--handed Galactocentric frame of reference, in which the $x$--axis points towards the Sun. The model of the Galactic bar and of the spiral pattern have been chosen according to Pichardo et al. (2004) and Pichardo et al. (2003), respectively.
  In Figs. \ref{fig:distance},  \ref{fig:plane_distr} and
 \ref{fig:xz_distr} we used a Galactocentric distance of $R_\odot$ = 7.5 kpc.

We identified the following features of the three sub--samples. Metal poor clusters are distributed in a nearly spherical volume with a radius of about 30 kpc, consistently with a halo population. 
The spatial distribution of metal--rich clusters is more centrally concentrated than the metal-poor ones, as expected. 
The moderately metal--poor sub--sample is confined within the inner 5 kpc of the Milky Way.

\begin{figure}
\hskip -0mm
\includegraphics[scale=0.45]{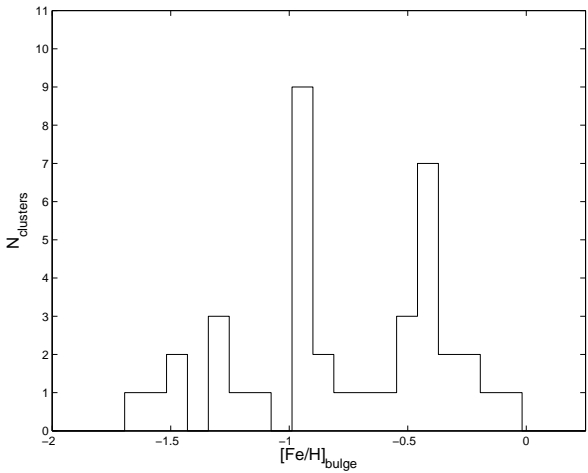}
\caption{Metallicity distribution of the GCs projected on the Galactic bulge.}
\label{fig:bulge_metallicity}
\end{figure}

In order to identify the clusters located in the Galactic bulge, we selected a region on the celestial
sphere included between $|l|\leq 18^{\circ}$ and $|b|\leq 7.5^{\circ}$. 
The metallicity distribution of the 39 clusters located in this region of the sky is shown in Figure \ref{fig:bulge_metallicity}.
 The moderately metal-poor clusters show a peak comparable to that
of the metal-rich clusters.

\begin{figure}[h!]
\hskip -10mm
\begin{tabular}{l}
\includegraphics[scale=0.5]{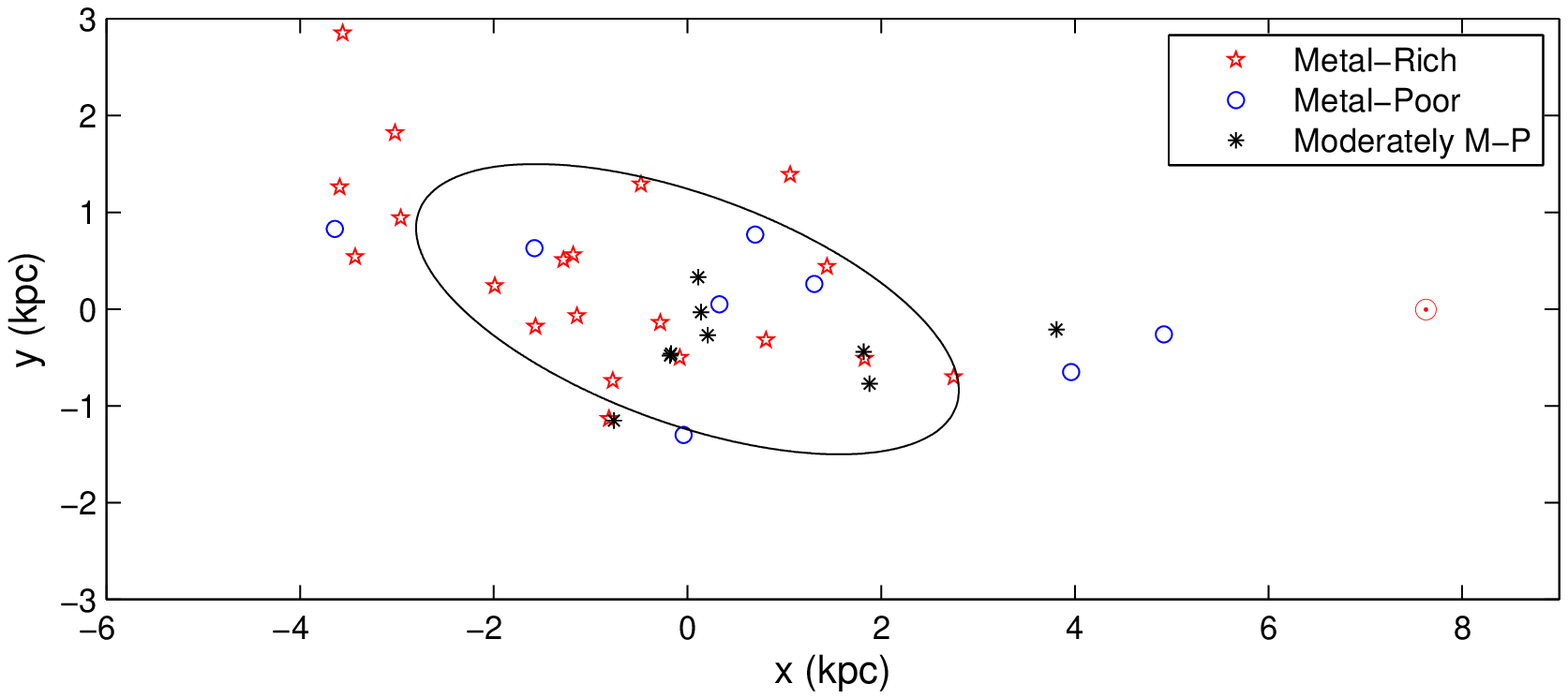}\\
\includegraphics[scale=0.5]{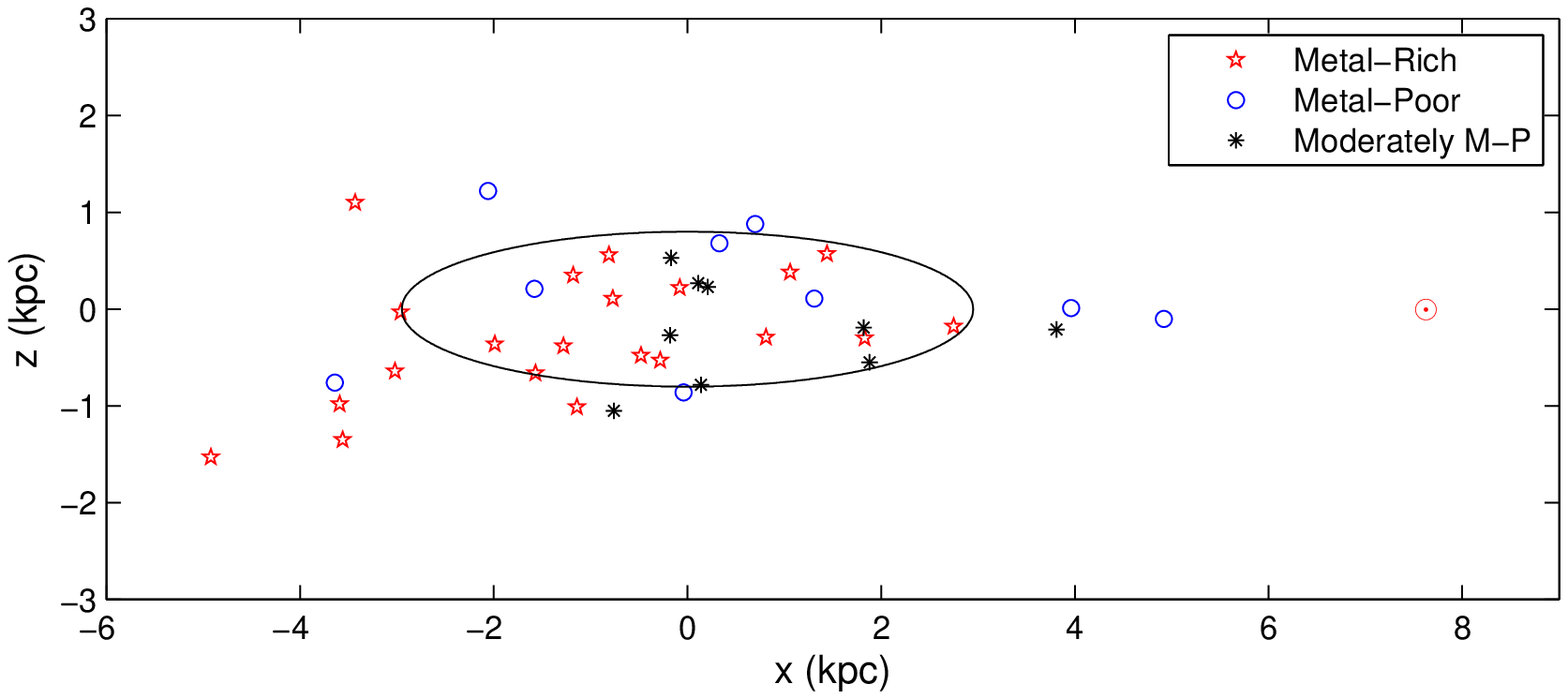}
\end{tabular}
\caption{Spatial distribution of Galactic bulge globular clusters overlapped on the central bar model.}
\label{fig:bulge_distr}
\end{figure}

In Figure \ref{fig:bulge_distr} we show the projection of their coordinates on the Galactic plane and on 
the x--z plane, where a distance to the Galactic center of
$R_\odot$ = 7.5 kpc was adopted. They appear centrally concentrated, and possibly trapped within the bar.

\section{Observations}

The first epoch archival data are based on the ESO NTT telescope,
obtained in 1993 and 1994, the ESO Danish 1.5m telescope in 1990, 1998 and 2000,
 and the Hubble Space Telescope (HST). The latter  
observations were acquired in 1995, under the proposal GO 9065, PI: S. Djorgowski,
and the  photometry is available as reported in  
 Piotto et al. (2002)\footnote{http://www.astro.unipd.it/globulars/}.

The NTT 1993 data used the EMMI spectrograph/focal reducer equipped with the CCD Loral ESO \#34,
of 2048$\times$2048, giving a $0\farcs35$/pixel scale. The NTT 1994 images employed the
 SUSI camera equipped with the CCD ESO \#25, with  $0\farcs13$/pixel scale.
The Danish 1.5m 1990 images employed the
direct focal camera with CCD ESO \# 5, of 512$\times$320  30 $\mu$m pixels,
 of scale  $0\farcs47$/pixel. The Danish 1.5m 1998 and 2000 data
come from the DFOSC spectrograph/focal reducer equipped with the Loral/Lesser 
C1W7 2052$\times$2052 15 $\mu$m pixels, with scale  $0\farcs39$/pixel.

The  second-epoch observations were obtained with the ESO NTT+EFOSC2 instrument,
in V, I and Gunn z. The CCD ESO \# 40 UV-flooded, of 2048$\times$2048,
with 15 $\mu$m pixels was used, giving a projected scale of  $0\farcs12$/pixel.
The total field has $4\farcm1\times 4\farcm1$. In the 2011 and 2012 observations
the CCD was read in two by two pixel binned mode, which gives the
best duty cycle observing time.  In 2013 the 
full resolution was used, because of its higher astrometric accuracy
and higher dynamical range. 

The log of observations for the two epochs (Table \ref{log}) summarize the
essential information on the data.


\section[Proper motions]{Proper motion derivation}

We describe below the method applied to derive the proper motions of the sample.
 The basic idea is to transform the coordinates of cluster and field stars
 in the two different epochs, into a common frame of reference (main--frame) and compute the
 proper motion in terms of coordinate changes in the main--frame.
The configuration of stars in the two epochs in principle 
should be different because of the different proper motion of single stars. 
Therefore a suitable choice of the stars to be used to determine the
 coordinate transformation is required.
 We assumed that cluster members have similar proper motion and, because of the gravitational bound,
 they maintain a similar configuration in the two epochs.
 We used the cluster members to determine the coordinate transformation in the main--frame and
 subsequently we applied the same transformation to all stars, both field and cluster.
 In this work we chose second epoch images as main--frame.
Subtracting the value of the coordinates of stars in the second epoch image
 from those in the first, we expect to find (if the cluster has a significant proper motion)
 two distinct aggregations of stars in the
 ($\Delta x, \Delta y$) plane: one centred around ($\Delta x, \Delta y$)=($0,0$)
 (the cluster) and the other around another point different from the origin (the field). 

\subsection{The technique for computing proper motions}

 The raw data have been preprocessed, using MIDAS commands, trimming the
images and subtracting the bias. Many tests have been
carried out on the flatfields obtained during the same observing runs,
both inside the dome and on the sky twilight, and the latter proved to be 
better distributed. An average of 10 flats have been used for the flatfield
correction.
We used the code  \texttt{DAOPHOT} (Stetson 1987) for the PSF photometry choosing
a uniform number of bright and isolated stars for the average reference
PSF used for fitting. A two degree polynomial was used across the field
accounting for the PSF variations in the second epoch NTT+EFOSC2 data.  Some
crowded fields needed a local subtraction of
 faint, close-by sources to the reference stars used to
model the PSF. The fitting was performed using the
ALLSTAR routine which generates the relative photometry, the refined
positions of the centroids and other parameters useful to clean the
generated list from spurious objects or bad measurements. This routine
creates the  \texttt{.als} tables that we used as starting data for the proper
motions.

We used single-image data after several tests on the selected
best images, as compared with the average of multiple observations. 
This is due to  the peculiarity of the targets, that
are very crowded and contaminated. The quality of the results is more dependent
on the resolution (seeing, focus, tracking),
 and the capability to isolate the cluster CMD sequences, rather than
on the number of used frames. In addition, in most cases,
the errors are dominated by the first epoch, where only a single 
colour image was available.

 The stars in common in the two images have been identified and
 sorted out using the codes \texttt{DAOMATCH} and \texttt{DAOMASTER}.
 In the following analysis we use the quantities
\begin{itemize}
\item $(x_1, y_1)$: coordinates of the star (in pixels) in the first image
\item $(x_2,y_2)$: coordinates of the star (in pixels) in the second image
\item $(x_1^\mathrm{t}, y_1^\mathrm{t})$, $(x_2^\mathrm{t}, y_2^\mathrm{t})$: coordinates of the star (in pixels)
 in the main--frame, obtained applying the transformation.
\item $(\Delta{}x,\Delta{}y)=(x_2^\mathrm{t}-x_1^\mathrm{t},y_2^\mathrm{t}-y_1^\mathrm{t})$: 
difference in the position of the star in the main--frame in the two epochs. 
\end{itemize}

\subsubsection{Field distortion and coordinate transformation}

Before proceeding further with the description of the present technique to compute proper motions, 
we propose an analysis of the effects of the field distortion. In fact, the distortion of star positions
 is a systematic effect that has to be treated carefully. 
A systematic displacement of the coordinates might result in a misinterpretation
 of the proper motions. A first solution to minimize this effect is to select stars in the central 
regions of the CCD, where the distortion is expected to be lower than in the peripheral regions.
 As discussed below, the field distortion is the reason why a
 conformal transformation including shift, rotation and scale is not enough to transform the
 coordinates in the two epochs into the main--frame.

\begin{figure}
\centering
\includegraphics[scale=0.5]{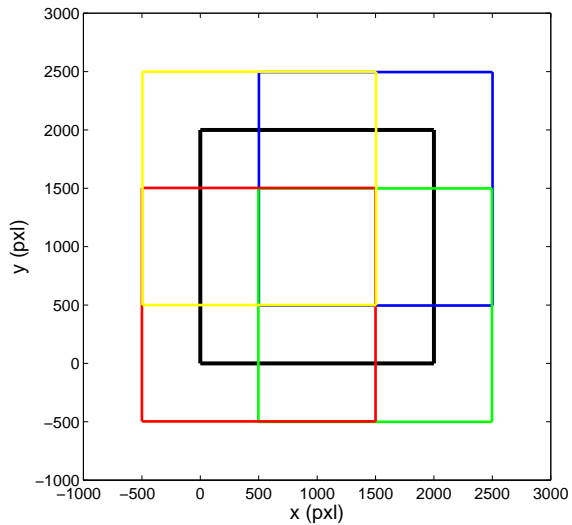}
\caption{Dither pattern of four of the eight images used to evaluate the field--distortion.}
\label{fig:dither_pattern}
\end{figure}

We evaluated the field--distortion affecting the second--epoch images from the NTT.
(for the first epoch we do not have distortion maps available).
 In particular, we reduced a set of images of a field
 located in the Baade's Window with 
uniformly-distributed stars including a central frame and
 eight shifted images of the same field.
 The shift of the eight images is 60 arcsec, 
corresponding to approximately 500 pixels,
 both in right ascension and/or in declination, depending on the image. 
Figure \ref{fig:dither_pattern} shows the dithering pattern for the four images shifted 
both in right ascension and declination. For a better visualization of the dithering pattern
 we did not show the configuration of the images shifted only in right ascension or declination.
We identified stars in common in all nine frames (419 stars) and computed the distortion 
vector for each star. We selected as main--frame of reference the system of coordinates of the central image.
 According to Anderson and King (2003), in order to transform the coordinates of the shifted images into
 the main--frame we applied a conformal transformation
 on the coordinates of the stars in the shifted images.
\begin{equation}
\begin{bmatrix}
x_i^\mathrm{t} \\ 
y_i^\mathrm{t}
\end{bmatrix}
 = \lambda
 \begin{bmatrix}
 \cos \theta & \sin \theta \\ -\sin \theta & \cos \theta
 \end{bmatrix} 
 \begin{bmatrix}
 x_i \\ y_i
 \end{bmatrix}
  + 
   \begin{bmatrix}
 x_0 \\ y_0
 \end{bmatrix} 
\end{equation}
 where ($x_i,y_i$) are the coordinates of the stars in each shifted frame,
 ($x_i^\mathrm{t},y_i^\mathrm{t}$) are the 
transformed coordinates, $\lambda$ is the scale parameter, $\theta$ is the rotation parameter 
and $(x_0,y_0)$ are the shift parameters. 

\begin{figure}
\hskip -10mm
\begin{tabular}{c}
\includegraphics[scale=0.5]{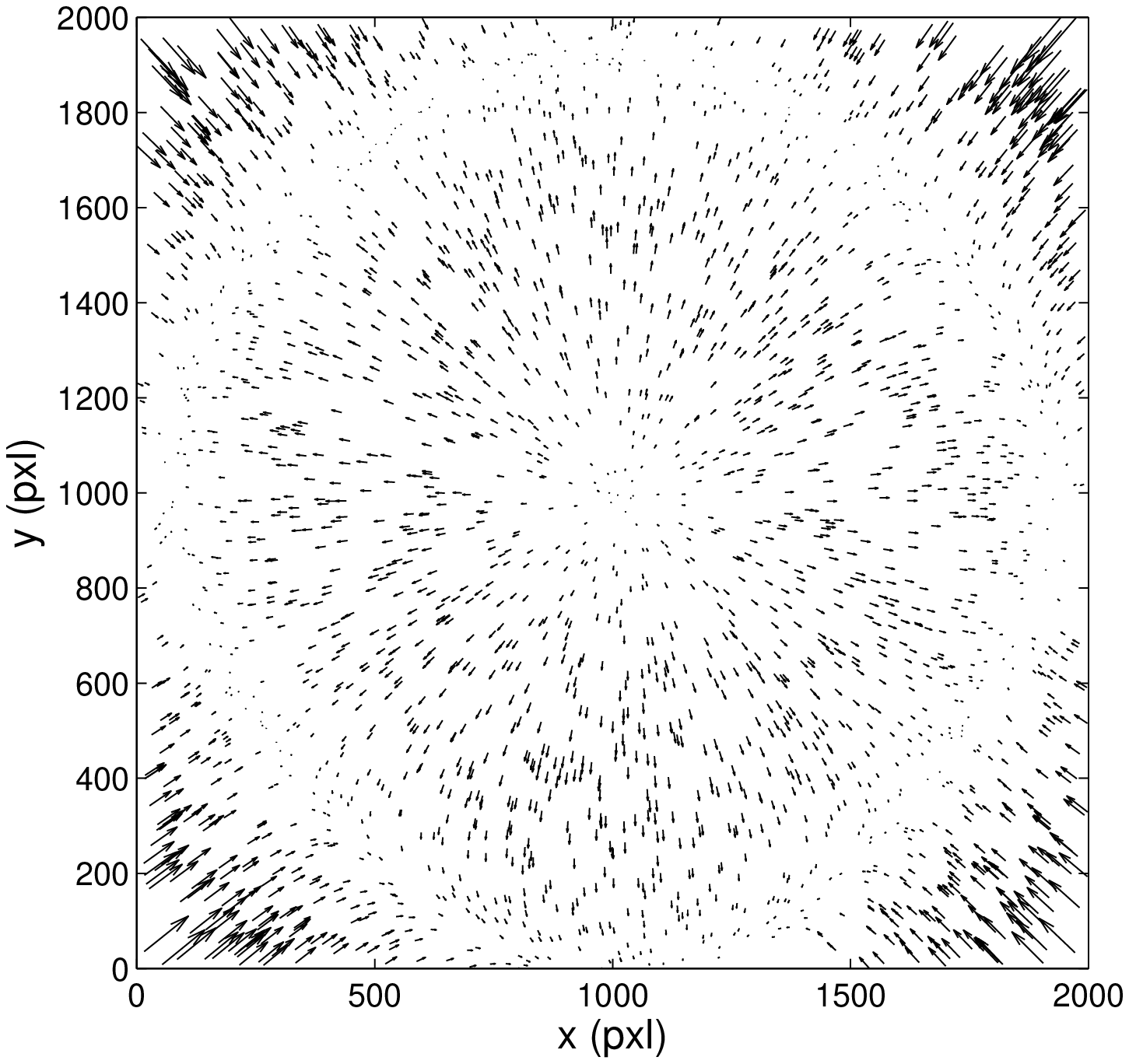}\\
\includegraphics[scale=0.5]{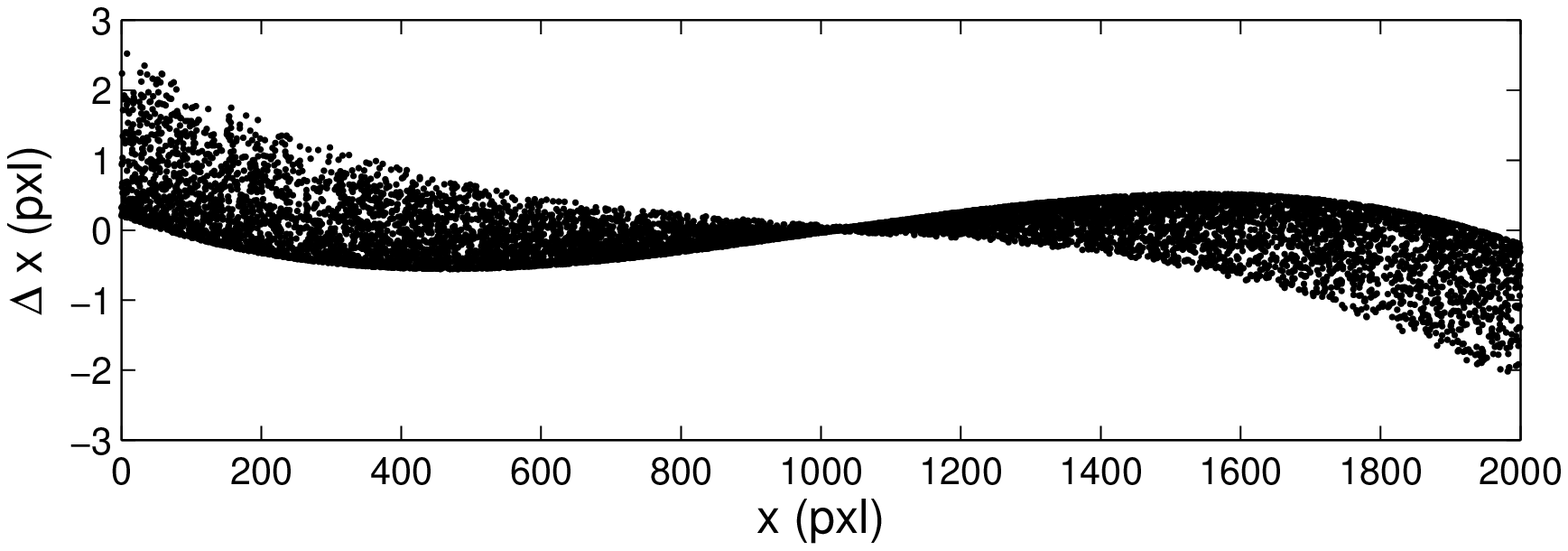}\\
\includegraphics[scale=0.5]{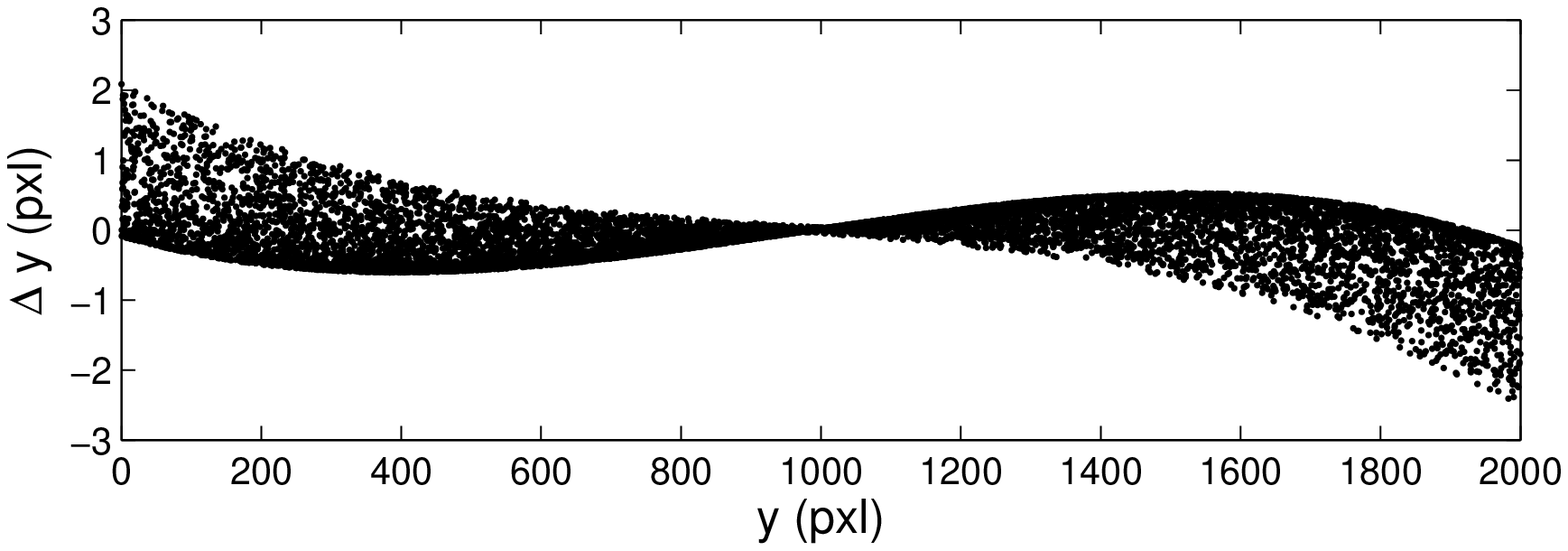}
\end{tabular}
\caption{Top panel:Model of the field distortion for NTT. For a better visualization,
 the distortion vectors are magnified by a factor 40. Central and middle panels: 
field distortion as function of the ($x,y$) coordinates.}
\label{fig:dist_map}
\end{figure}

We computed the mean position of each star in the central frame of reference as follows:
\begin{equation}
\begin{cases}
\bar{x}=\dfrac{x_{\mathrm{c}}+\sum_{i=1}^8x^\mathrm{t}_i}{9}\\
\bar{y}=\dfrac{y_{\mathrm{c}}+\sum_{i=1}^8y^\mathrm{t}_i}{9}
\end{cases}
\end{equation}
where ($x_{\mathrm{c}},y_{\mathrm{c}}$) are the coordinates of each star in the central image.
 Our main assumption is that the mean position computed represents a good approximation of the
 distortion--free coordinates of the stars, justified by the fact that we obtained an isotropic
 dithering pattern. The field distortion has been evaluated by subtracting the coordinates in each 
shifted image to the mean coordinates. We then assumed that the distortion map can be well
 represented by a  third-order-polynomial transformation of the form
\begin{equation}
\begin{cases}
\Delta x (x,y) =& a_{00} + a_{10}x + a_{01}y + a_{20}x^2 + a_{11}xy + a_{02}y^2\\
                & a_{30}x^3 + a_{21}x^2y + a_{12}xy^2 + a_{03}y^3\\
\Delta y (x,y) =& b_{00} + b_{10}x + b_{01}y + b_{20}x^2 + b_{11}xy + b_{02}y^2\\
                & b_{30}x^3 + b_{21}x^2y + b_{12}xy^2 + b_{03}y^3\\
\end{cases}
\end{equation}
We derived the ($a_{ij},b_{ij}$) coefficients by fitting the distortion model to the computed
 distortion vectors.
The top panel of Figure \ref{fig:dist_map} shows the distortion model for NTT derived from this 
analysis, with distortion vectors  magnified by a factor 40. The central
 and the bottom panels show the values of our model of ($\Delta x,\Delta y$) as a function
 of the ($x,y$) coordinates on the CCD. 
 The field--distortion is lower in the CCD centre than in the peripheral regions,
 where the distortion of a star coordinates can be up to about 2 pixels. 

\begin{figure}
\hskip -10mm
\begin{tabular}{c}
\includegraphics[scale=0.5]{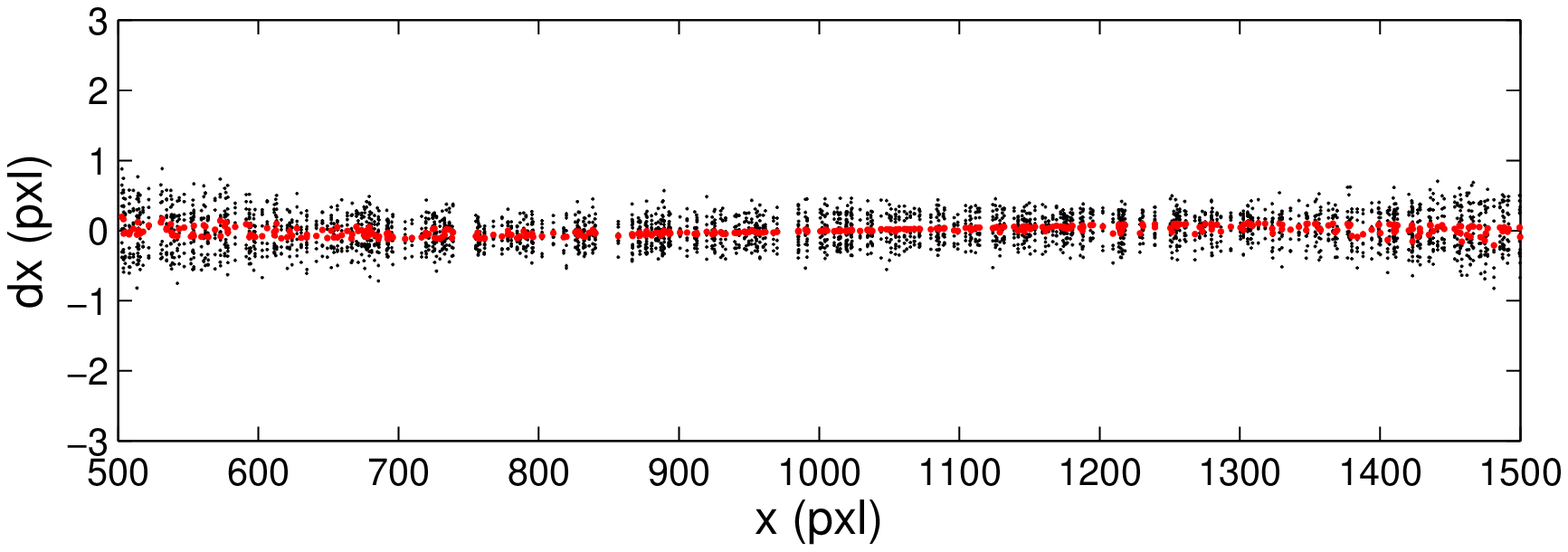} \\
\includegraphics[scale=0.5]{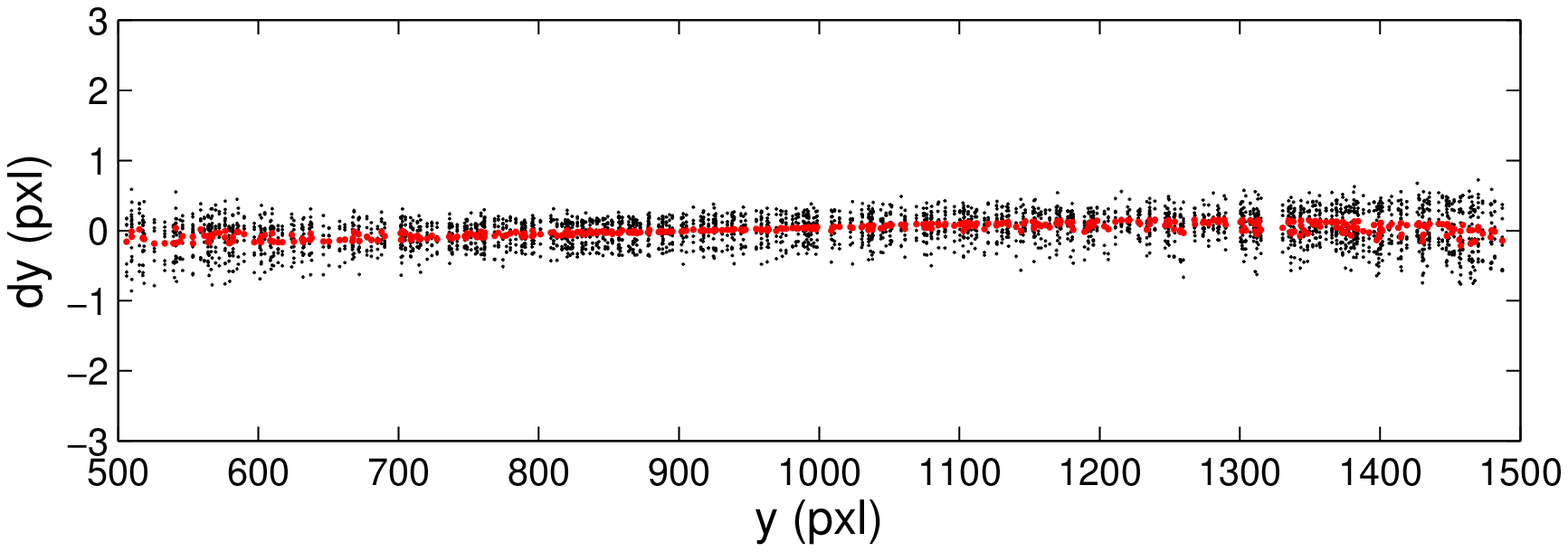}
\end{tabular}
\caption{Residuals of the position of the stars after correction for field--distortion
 as function of the mean coordinates. The red dots represent the average of the displacement.}
\label{fig:dx_dy_residuals}
\end{figure}

\begin{figure}
\hskip -10mm
\begin{tabular}{c}
\includegraphics[scale=0.5]{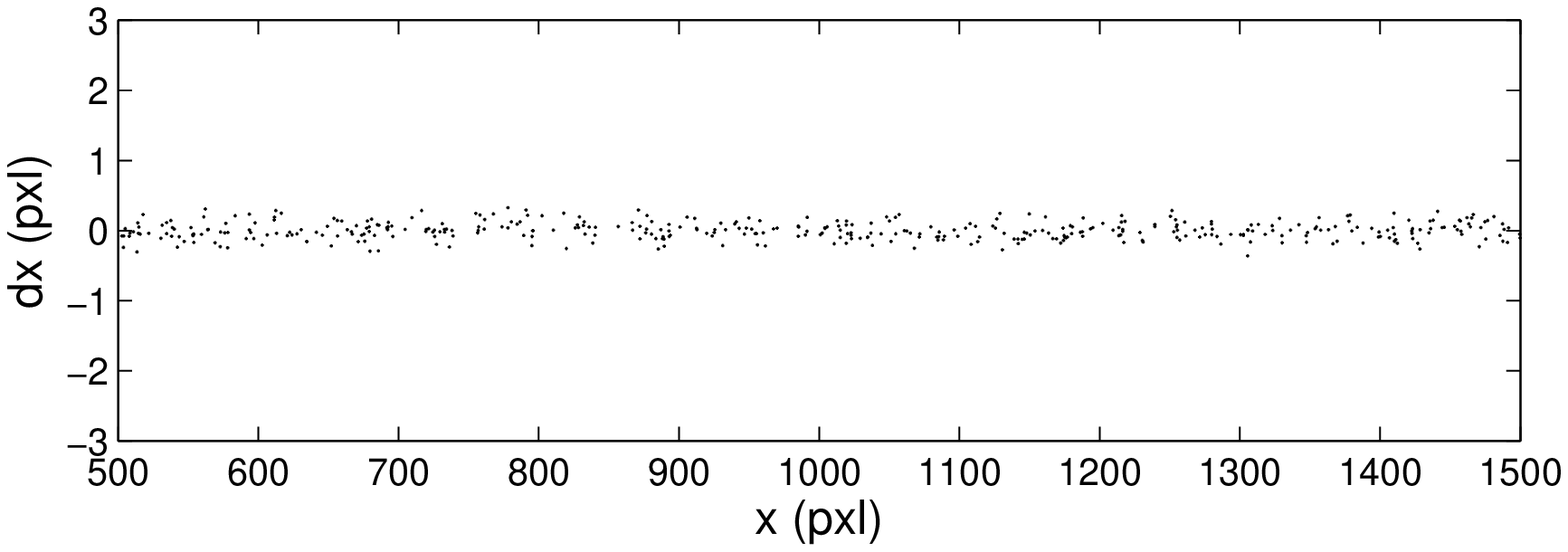} \\
\includegraphics[scale=0.5]{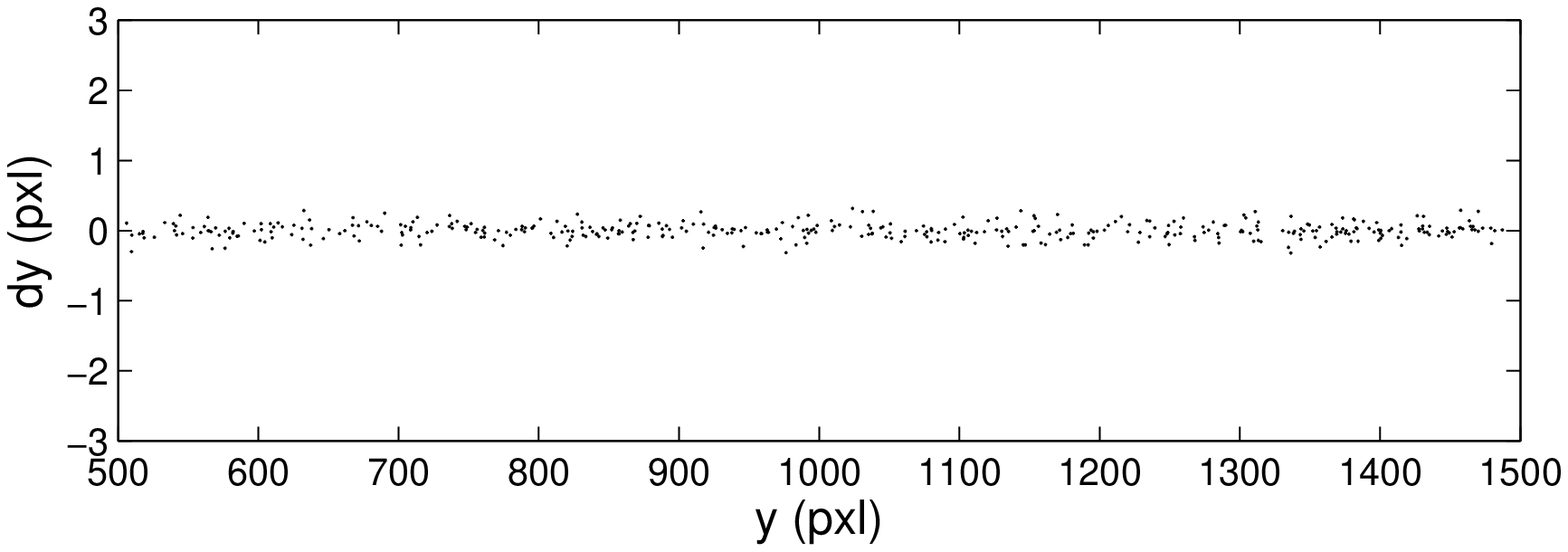}
\end{tabular}
\caption{Residuals of the position of the stars in the central frame after correcting
 for field--distortion with a second order polynomial transformation as function of the mean coordinates.}
\label{fig:dx_dy_residuals2}
\end{figure}

In order to test the reliability of the field--distortion model, we applied the correction to the
 stellar positions of the 9 exposures of the Baade's Window. We recalculated the mean position of the stars 
in the main--frame using conformal transformation and evaluated the residuals ($\mathrm{d}x,\mathrm{d}y$)
 between the main--frame average position and those of the single exposures transformed on the 
main--frame as a function of the ($x,y$) main--frame positions (see Figure \ref{fig:dx_dy_residuals}).
Note that the displacement is constant to a good approximation 
along the whole CCD and we conclude that the  third-order-polynomial transformation produces a 
good distortion model.
 Given that the clusters in our images are located approximately at the 
 centre of the detector,
thus with minimal distortions,
 in order to minimize the effects of the distortion we obtained the proper motions  by selecting
 field stars included within a circular area between 500 and 1500 pixels both in $x$ and $y$,
 centred on the cluster. We test if in this region of the detector the field-- distortion
 can be well-represented by a polynomial transformation of second order. 
The most important reason for such a order decrease is
 that, for some sample clusters, the number of cluster members available to calculate
 coordinate transformation between the two epochs is low. Decreasing the order
  will result into a more robust evaluation of the transformation coefficients.
 Following this approach, we considered only the central frame and evaluated the distortion model  by applying
 the same procedure described above, decreasing the order of the transformation to the second power.
 Figure \ref{fig:dx_dy_residuals2} shows the analogue results of Figure \ref{fig:dx_dy_residuals} 
for the central regions of the CCD after correcting the coordinates with a second order transformation. 
Also in this case the displacement is constant for every value of ($x,y$) and we can conclude that the distortion map in the inner regions is well represented by a second order polynomial model.

\begin{figure}
\centering
\begin{tabular}{c}
\includegraphics[scale=0.6]{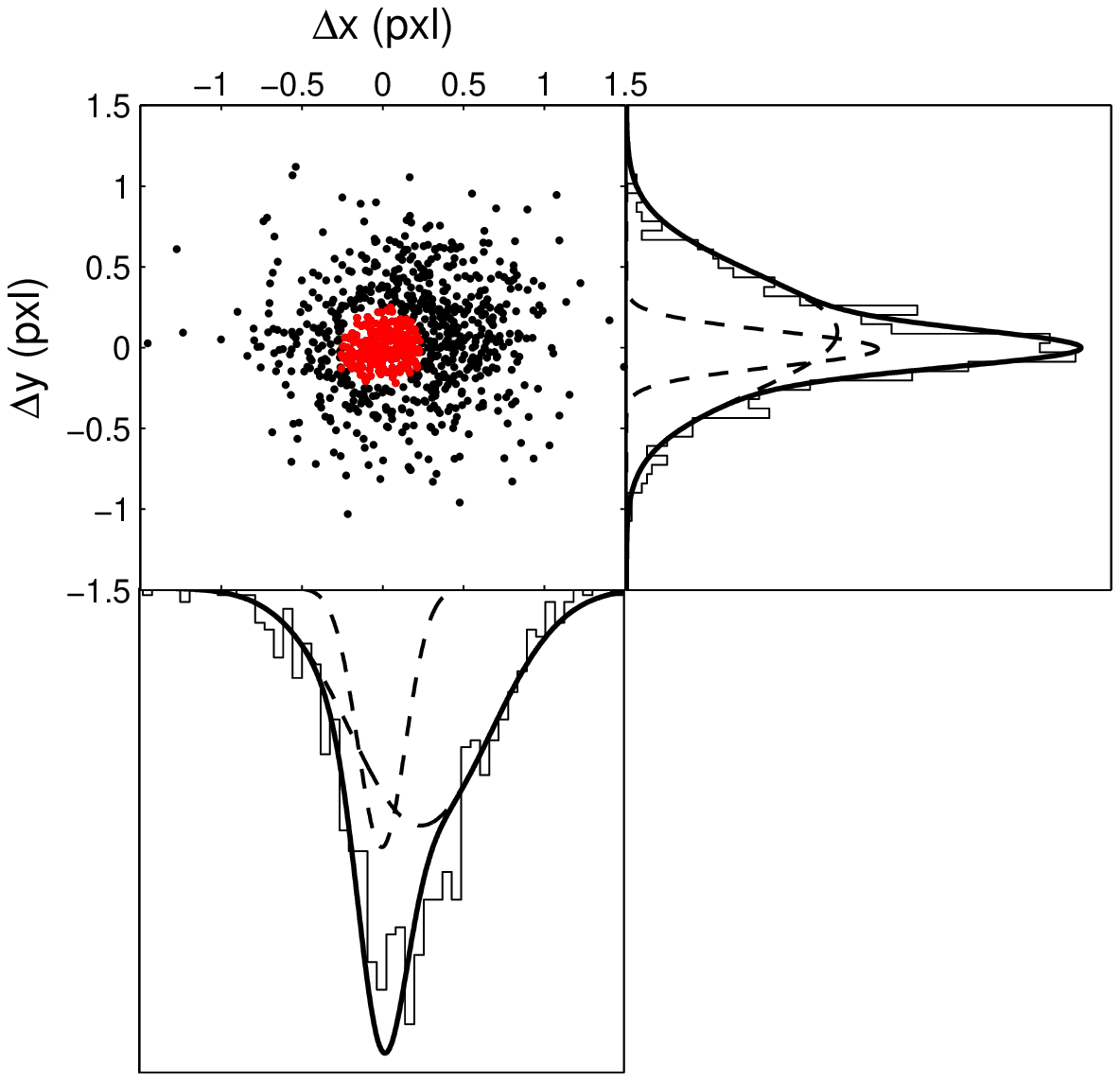} \\
\includegraphics[scale=0.6]{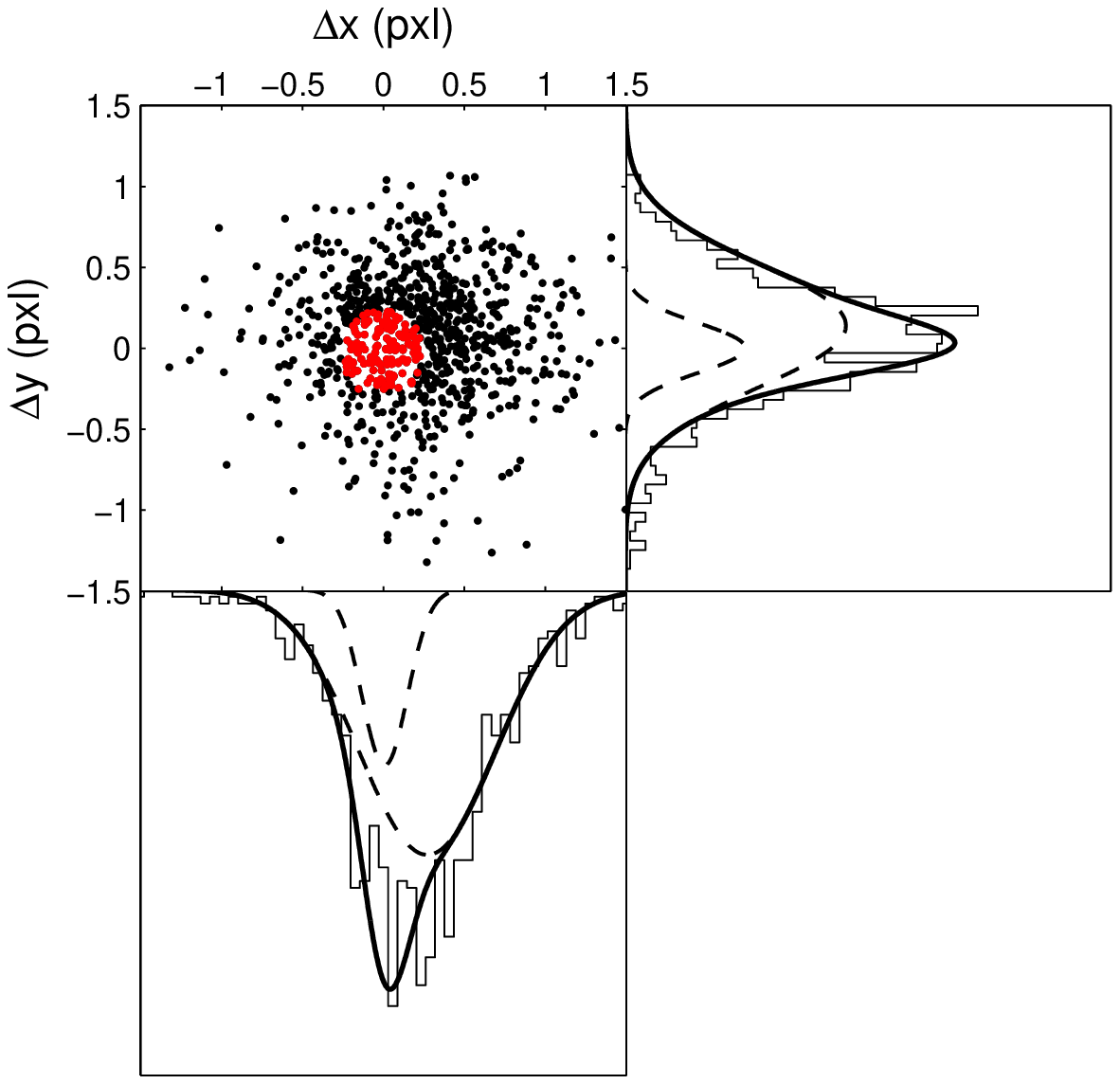}
\end{tabular}
\caption{Displacement (in pixels) of NGC 6522 derived from \textit{test 1} (upper panel) and \textit{test 2} (bottom panel). The black dots are the field stars, while the red dots  are the cluster members.}
\label{fig:compare_methods}
\end{figure}

Having derived a field--distortion model for all second--epoch exposures from NTT+EFOSC2,
 we test if a quadratic transformation is sufficient to transform the coordinates of the first--epoch images
 in the frame of reference of the second--epoch for all other instruments used in the present work.
 Explicitly:
\begin{equation}
\begin{cases}
x_1^\mathrm{t} = a_1\,{x_1}^2+a_2\,{x_1}{y_1}+a_3\,{y_1}^2+a_4\,x_1+a_5\,y_1+a_6\\
y_1^\mathrm{t} = b_1\,{x_1}^2+b_2\,{x_1}{y_1}+b_3\,{y_1}^2+b_4\,x_1+b_5\,y_1+b_6
\end{cases}
\label{eq:quadratic}
\end{equation}
\begin{equation}
\begin{cases}
x_2^\mathrm{t}=x_2+\Delta x_\mathrm{fd} (x_2,y_2)\\
y_2^\mathrm{t}=y_2 + \Delta y_\mathrm{fd} (x_2,y_2)
\end{cases}
\end{equation}
where $(x_1,y_1)$ are the raw positions of the stars in the first--epoch exposure, ($x_1^\mathrm{t},y_1^\mathrm{t}$)
 are the coordinates of the first--epoch exposure transformed in the field--distortion--corrected second--epoch frame of reference,
 $(x_2,y_2)$ are the raw positions of the stars in the second epoch--epoch exposure and $(\Delta x_\mathrm{fd} ,\Delta y_\mathrm{fd} )$
 are the corrections for field--distortion of the second--epoch exposures. 
We performed the following tests: 
\begin{itemize}
\item \textit{Test 1}: Apply our model of field--distortion correction to the second--epoch exposure.
 Use quadratic transformations to bring the first--epoch exposure raw positions on the second--epoch reference
 frame and measure the displacement between cluster members and field stars.\\
\item \textit{Test 2}: Apply the field--distortion to both the first--epoch and second--epoch exposures.
 Use a conformal transformation to bring the first--epoch coordinates in the frame of reference of the
 second--epoch image and measure the displacement between cluster members and field stars.
\end{itemize}
If the two methods give consistent results we should be confident about the correctness of our
 approach. In the sample studied the only case with known field--distortion for the first 
epoch is NGC 6522 from HST. In particular, we adopted the geometric--distortion solution proposed by
 Anderson \& King (2003) for WFPC2. Figure \ref{fig:compare_methods} shows the results of this analysis.
 The displacement obtained from \textit{test 1} is ($\Delta x, \Delta y$) = ($0.24\pm 0.04,0.10 \pm 0.03 $) pxl,
 while \textit{test 2} produced the vales   ($\Delta x, \Delta y$) = ($0.26\pm 0.05, 0.14 \pm 0.04$) pxl.
 In this case we evaluated the displacement
 by fitting a double Gaussian to the data points. 
Considering the errors associated to our measurements, we see  that the results of the two methods
 are consistent and a quadratic transformation is sufficient.

More specifically, we computed the transformation of coordinates in an iterative process.
 We identified in the  two-epoch images a circular area centred on the cluster. As discussed above,
 the radius of such a circular area has been chosen in such a way as to minimize the effects of the
 field distortion. In order to estimate, {to a} first approximation, the value of the coefficients of the 
coordinate transformation, we had to identify some stars likely belonging to the cluster. We based our
 choice on some features of the CMD. In particular, we assumed that most of 
the blue horizontal branch (HB) stars are cluster members.
 In the cases in which the HB was not well defined or absent, we selected the bright red giant branch (RGB) stars
 as cluster members. After the first fit based on the HB/RGB members, we employed an iterative process. We
 applied the transformation to the whole sample of stars and plotted their position in the ($\Delta x, \Delta y$) plane.
 The cluster members were selected as the stars included within a circular area centred on the origin of the coordinates.
 The radius of the selection was chosen making sure that the selected stars 
 were mostly located around the cluster evolutionary sequences of the CMD.
We then computed again the coefficients of the transformation using the newly identified cluster members. 
We repeated this procedure iteratively until the value of the coefficients of the coordinate transformation
 converged (we verified that usually 4 or 5 iterations were enough).  This solution leads to two results: the
 determination of the proper motion of the cluster and the decontamination of the field stars from the
cluster CMD.

\subsubsection{Gaussian interpolation}
Once  the distribution of stars in the ($\Delta x,\Delta y$) diagram
is known, 
we determined the value of proper motion (in pixels) by means of an interpolation with a double Gaussian of the form

\begin{align}
f(\Delta{}x)&=A\mbox{exp}\left[ -\left( \dfrac{(\Delta{}x-\Delta{}x_\mathrm{cl})^2}{2\sigma_\mathrm{cl}^2}\right)\right]+
 \nonumber\\ &+B\mbox{exp}\left[-\left( \dfrac{(\Delta{}x-\Delta{}x_\mathrm{field})^2}{2\sigma_\mathrm{field}^2} \right)\right]
\end{align}

where $(A,B)$, $(x_\mathrm{cl},x_\mathrm{field})$ and $(\sigma_\mathrm{cl},\sigma_\mathrm{field}$ are the amplitude, 
the center and the dispersion of the Gaussian describing the cluster stars and the field stars distribution on the
 $\Delta{}x$ axis respectively. The same was done for the distribution of stars along the $\Delta{}y$ axis). 

Finally the differences
\begin{equation}
\begin{cases}
\Delta{}X=\Delta{}x_\mathrm{cl}-\Delta{}x_\mathrm{field}\\
\Delta{}Y=\Delta{}y_\mathrm{cl}-\Delta{}y_\mathrm{field}
\end{cases}
\end{equation}
give the cluster proper motion.

\subsection{From proper motion to the state vector}
We present a method to compute the state vector into the inertial Galactocentric
 frame of reference from proper motions, radial velocities and distances.

\subsubsection{Absolute proper motion vector}
Since we measure the proper motion of the cluster with respect
to the bulge field stars (relative proper motion), 
in order to obtain the absolute proper motion of the cluster members we have
 to correct for the internal proper motion of the field stars.
Ortolani et al. (2011) derived for the cluster HP~1 an internal proper motion of the bulge field stars equal to 
$v_T=1.03 \; \mbox{km}\mbox{s}^{-1}$. 
Zoccali et al. (2001) assumed to measure the background field stars, having
those in the near side moving on average in the opposite direction, with respect to those in the far side, because of
 bulge rotation, making that their overall proper motion  is null. 
Terndrup et al. (1998) derived an internal proper motion ($\mu_\mathrm{l},\mu_\mathrm{b}$)=(0.2,0.2) mas/yr. 
We concluded that the correction for the internal proper motion is small if compared to the uncertainties 
associated to the proper motion of the cluster, so we will ignore it (see
 Ortolani et al. 2011).

\begin{figure}
\centering
\includegraphics[scale=0.4]{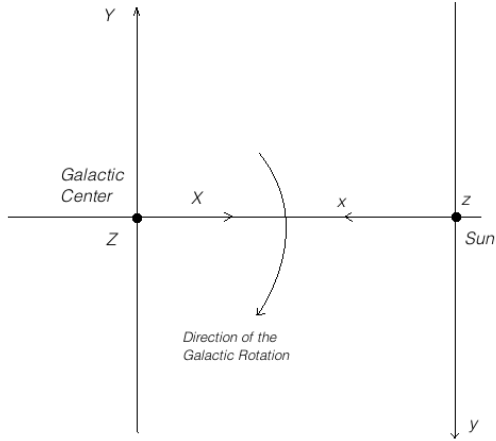}
\caption{Inertial Galactocentric frame of reference ($X,Y,Z$) and non--inertial heliocentric
 frame of reference ($x,y,z$) configuration at the present time.}
\label{fig:SoRef}
\end{figure}

\subsubsection{Initial state vector}

In order to predict the orbit of the cluster, we need the initial state vector, i.e. the initial conditions necessary 
for solving the system of differential equations represented by the equations of motion. 
The first step is to obtain the value of the velocity components of the cluster in the heliocentric frame of reference ($U,V,W$). 
We used a right-handed coordinate system, so that $U$, $V$, and $W$ are positive in the direction of the Galactic center, Galactic rotation and North Galactic Pole (NGP) respectively. 
We refer to Johnson \& Soderblom (1987) for a more complete explanation of the procedure to obtain the heliocentric velocity vector starting from distance, proper motion and radial velocity.
We adopted the value of the velocity in the local standard of rest at the Sun's Galactocentric distance equal to $v_\mathrm{LSR}=239.7 \; \mbox{km}\mbox{s}^{-1}$, according to Model III of Irrgang et al. (2013).
 The velocity components of the Sun with respect to the local standard of rest are (Sch\"onrich et al. 2010)
\begin{eqnarray}
U_{\odot} &=& 11.1 \pm 1 \; \mbox{km} \; \mbox{s}^{-1}\nonumber\\
V_{\odot} &=& 12.24 \pm 0.2 \; \mbox{km} \; \mbox{s}^{-1}\nonumber\\
W_{\odot} &=& 7.25 \pm 0.5 \; \mbox{km} \; \mbox{s}^{-1} \nonumber\\
|{v}_{\odot}| &=&18.04 \; \mbox{km} \; \mbox{s}^{-1}\nonumber
\end{eqnarray}
Thus, the Sun moves towards the Galactic center and up towards the NGP away from the plane.
 It also moves around the Galactic center faster than it would if it were on a circular orbit.
The proper motion is then measured in the heliocentric frame of reference, which is not inertial
 (see Figure \ref{fig:SoRef}). According to Terndrup et al. (1998), the perpendicular components 
of the velocity vector ($y$ and $z$ direction) expressed into the inertial heliocentric frame of reference is given by
\begin{equation}
v_{\mathrm{cl},\perp} = d_\odot \boldsymbol\mu_\mathrm{obs}+v_{\odot,\perp}\left(1-\dfrac{d_\odot}{d_\mathrm{f}}\right)
\label{eq:terndrup}
\end{equation} 
where $d_\odot$ is the heliocentric distance of the cluster, $\boldsymbol\mu_\mathrm{obs}$ is the observed proper motion
 in Galactic coordinates, $d_\mathrm{f}$ is the heliocentric distance of the field stars and v$_{\odot,\perp}$ are
 the perpendicular components of the velocity vector of the Sun with respect to the local standard of rest.
 Considering that the clusters we are studying are projected on the Galactic bulge, we assumed that 
\begin{equation}
d_\mathrm{f} = R_\odot
\end{equation}
where $R_\odot$ is the Galactocentric distance of the Sun. Applying equation \ref{eq:terndrup}, we can write
\begin{equation}
\begin{cases}
\dot{X}&=-\left(U + U_{\odot}\right)\\
		\\
\dot{Y}&=-\left[V+ (v_\mathrm{LSR}+V_{\odot})\left(1-\dfrac{d_\odot}{R_\odot}\right)\right]\\
\\
\dot{Z}&=\left[W+W_{\odot}\left(1-\dfrac{d_\odot}{R_\odot}\right)\right]\;\;\; .
\end{cases}
\end{equation}
The position vector of the cluster in the Galactocentric frame ($X,Y,Z$) is related to the measured position
 in the heliocentric frame of reference ($x,y,z$) by
\begin{equation}
\begin{cases}
X&=R_\odot-x\\
Y&=-y\\
Z&=z \;\;\; .
\end{cases}
\end{equation}

\section{Analysis of errors}

In this section we describe the errors affecting our measurements.

\subsection{Astrometric Errors}
Astrometric errors are the combination of different random and systematic errors. We identified several sources of uncertainty, 
namely the \textit{random centering errors}, the \textit{chromatic errors} (Ortolani et al. 2011) and the \textit{field distortion errors}.
 We used images of the cluster NGC 6558 as a test case, but the results are valid in a more general context.

\subsubsection*{Centering Error}
\begin{figure}
\includegraphics[scale=0.5]{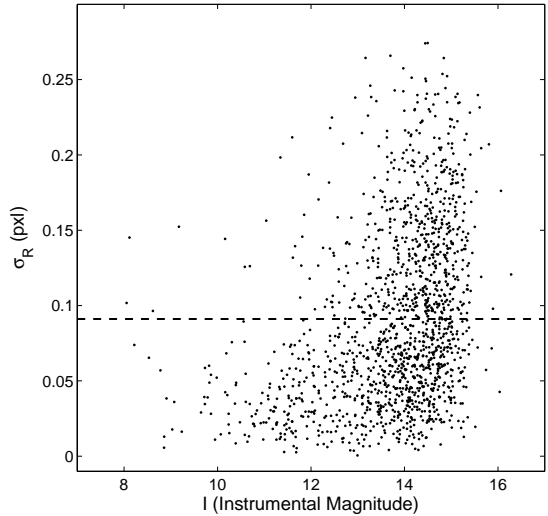}\\
\caption{Centering error vs. I band instrumental magnitude for the 1993 images (NTT+EMMI). The dashed line shows the average astrometric error.}
\label{fig:astrometric_93}
\end{figure}

\begin{figure}
\includegraphics[scale=0.5]{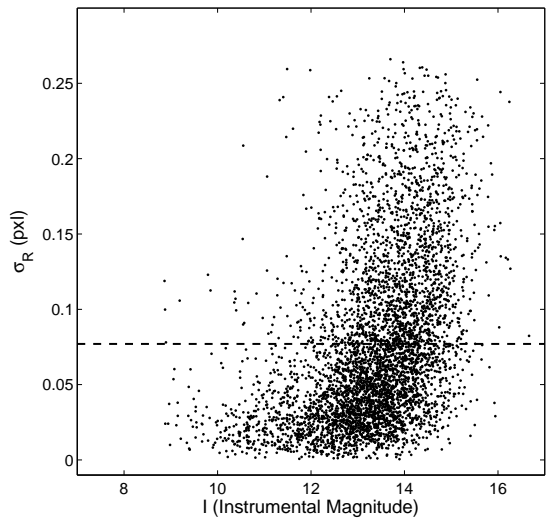}
\caption{Centering error vs. I band instrumental
 magnitude for the 2012 images (NTT+EFOSC).
 The dashed line shows the average astrometric error. }
\label{fig:astrometric_12}
\end{figure}

We considered four images of NGC 6558 from NTT+EMMI (1993):
\begin{itemize}
\item V band, 10 s
\item V band, 3 min
\item I band, 7 s
\item I band, 5 min
\end{itemize}
We selected the same region of the CCD used for the determination of the proper motion 
in each image and matched the
 \texttt{.als} tables with the \texttt{DAOMATCH} and \texttt{DAOMASTER} routines.
 We considered the I band images to evaluate the centering error, because they are more populated 
  than the images in the V band, leading to a better statistics. Furthermore, 
for most of the studied clusters,
 we used I band images for the derivation of the proper motion.  
We designated the coordinates of the single star in the first 
image $(x_1,y_1)_i$
 and in the second image $(x_2,y_2)_i$, with $i=1$ to N$_{stars}$ 
(hereafter $i$ will be always intended in this range). The mean coordinates of the stars are
\begin{equation}
\begin{cases}
\bar{x}_i=\dfrac{(x_1+x_2)_i}{2}\\
\bar{y}_i=\dfrac{(y_1+y_2)_i}{2}
\end{cases}
\end{equation}
The standard deviation associated to this mean value is
\begin{equation}
\begin{cases}
{\sigma_x}_i=\left[\dfrac{(x_1-\bar{x})_i^2+(x_2-\bar{x})_i^2}{2} \right]^{1/2}\\
{\sigma_y}_i=\left[\dfrac{(y_1-\bar{y})_i^2+(y_2-\bar{y})_i^2}{2} \right]^{1/2}
\end{cases}
\end{equation}
In Figure \ref{fig:astrometric_93} we show the error on the
radial  position of stars as a function of the instrumental I magnitude
\begin{equation}
{\sigma_\mathrm{R}}_i=\sqrt{{\sigma_x}_i^2+{\sigma_y}_i^2}
\end{equation}
We applied the same procedure for the I band images from NTT+EFOSC  (2012) and the results are shown in Figure \ref{fig:astrometric_12}.

\begin{figure}
\includegraphics[scale=0.5]{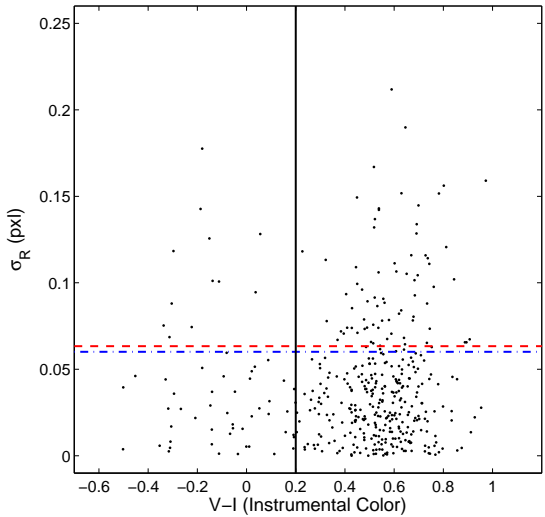}
\caption{Centering error vs. instrumental ($V-I$) colour. 
The red line shows the average centering error for the ``red'' group, while the blue line shows the average centering error
 for the ``blue'' group; these groups are separated by black vertical line.}
\label{fig:chromatic}
\end{figure}

The order of magnitude of the average value of the astrometric error is about 0.1 pixel.
 Taking into account the number of about 1000 stars used for the derivation of the proper
 motions of the clusters, the order of magnitude of the centering error is
as small as $10^{-3}$ pixel.

\subsubsection*{Chromatic Error}

Chromatic errors are due to differential chromatic refraction (DCR), both atmospheric and instrumental,
 on stars with different colours at different air masses. The refraction dependence on
 wavelength (inverse quadratic to the first approximation) makes this effect more pronounced 
in the ($V-I$) colour than in the near infrared. In order to study the displacement 
of the stars as a function of their colours, we subdivided the sample of stars into
  ``red'' and  ``blue'' groups, evaluating the average astrometric error for both.
 We show the results for NGC 6558 (NTT+EMMI 1993) in Figure \ref{fig:chromatic}. 
Note that the number of stars is lower than that of
 the previous figures as a consequence of
 the match of the I and V \texttt{.als} tables.\\ 
We associate to the chromatic error the quantity
\begin{equation}
\sigma_\mathrm{Chr}=|\bar{\sigma}_\mathrm{C,r}-\bar{\sigma}_\mathrm{C,b}|
\end{equation}
We expect to obtain a similar result for the 2012 images. It is important to notice that the chromatic effect is small if compared with the centring error. 
In fact, the order of magnitude is $10^{-4}$ pixels, and then negligible.

\begin{figure}
\hskip -8mm
\begin{tabular}{c}
\includegraphics[scale=0.5]{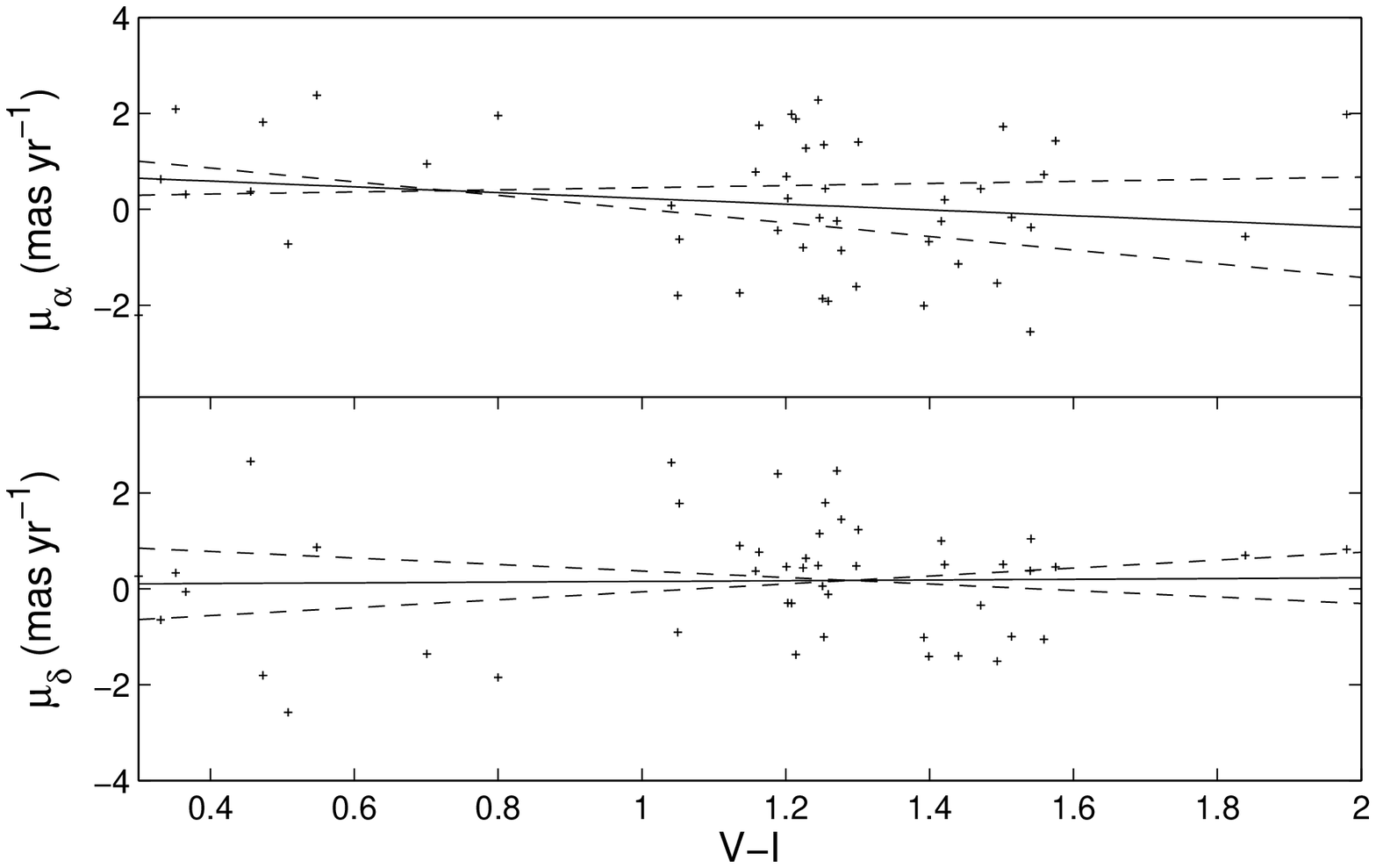}\\
\includegraphics[scale=0.5]{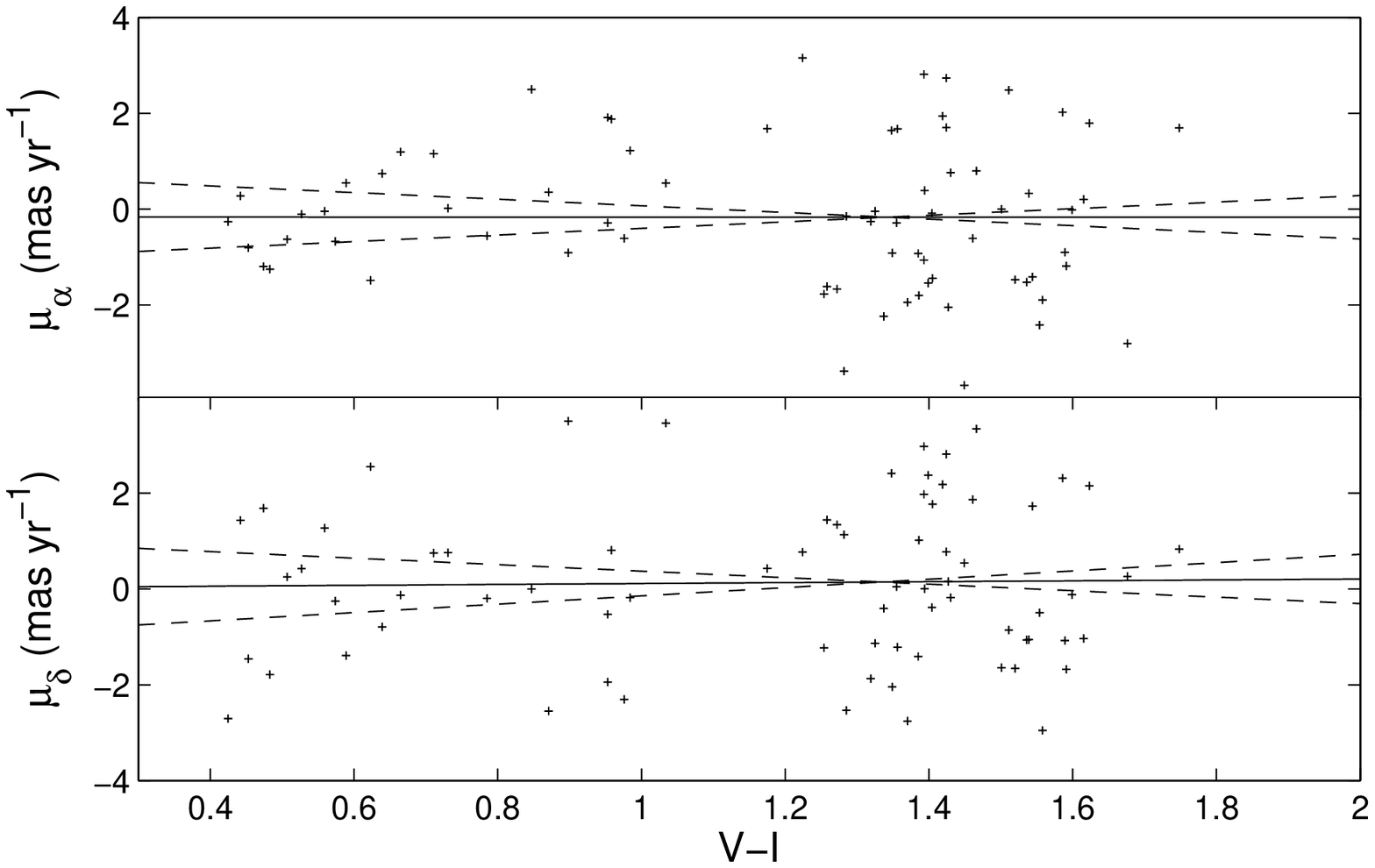}
\end{tabular}
\caption{Components of the proper motion as function of the star colours for NGC 6540 (two two panels) and for AL 3 (bottom two panels).}
\label{fig:DCR}
\end{figure}

We performed a further test in order to investigate the effects of DCR
  on our measured proper motions.
 We selected two clusters of our sample with detected blue HB stars, namely NGC 6540 (Figure \ref{fig:NGC6540pm}) 
and AL 3 (Figure \ref{fig:AL3pm}).
 In order to maximize the colour extension of the selection, we selected identified cluster members in a $V$ range that 
contains both HB  and RGB stars ($15.5< V < 16.5$ for NGC 6540 and $14.5< V < 17$ for AL 3). 
We then analysed the value of the proper motion
 as a function of the star colours. Figure \ref{fig:DCR} shows the results of this analysis. We fitted a regression line (continuous line) to the data points applying a linear least square interpolation and plotted the regression lines (dashed lines) associated to the 95\% confidence bound on the coefficient values of the fit model. The right ascension component of NGC 6540 shows the highest dependence of the proper motion on the stars' colour. The declination components of NGC 6540 and both the right ascension and declination components of AL 3 show a very low dependence on the colour. However, if we consider the uncertainties associated to the regression model, we note that the all the fitted lines are consistent with no slope within the errors.
We can conclude that we did not find any significant dependence of the
 proper motion on the star colours and hence that we can confidently exclude a strong DCR effect on our data.
The same conclusion was reached in Ortolani et al. (2011).

\subsection{Gaussian Interpolation Error}

The error associated to the double Gaussian interpolation is 
\begin{equation}
\begin{cases}
\sigma_{\Delta X}=\sqrt{\sigma_{\Delta x_\mathrm{cl}}^2+
\sigma_{\Delta x_\mathrm{field}}^2}\\
\\
\sigma_{\Delta Y}=\sqrt{\sigma_{\Delta y_\mathrm{cl}}^2+
\sigma_{\Delta y_\mathrm{field}}^2}
\end{cases}
\end{equation}
where $\Delta x_\mathrm{cl}$, $\Delta x_\mathrm{field}$, $\Delta y_\mathrm{cl}$, $\Delta y_\mathrm{field}$ are the uncertainties in the position of the cluster and field stars centroids in the ($\Delta X, \Delta Y$) diagram for the $x$ and $y$ coordinates respectively. 
The uncertainties associated to the astrometric and numerical errors are small if compared with the uncertainties associated to the Gaussian interpolation, then we can neglect them and assume that the main source of uncertainty derives from this fitting procedure. This error is estimated directly from the routine that computes the coefficients of the Gaussian curve and coincides with the difference between the best fitting value and the extremes of the 95\% confidence bounds. 

Following Johnston et al. (1987), we have all the ingredients to estimate the uncertainties in the velocity components.
By applying the error-propagation to the estimate of the velocity components we found
\begin{equation}
\begin{bmatrix}
\sigma_U^2 \\ \\\sigma_V^2 \\ \\\sigma_W^2
\end{bmatrix}
=C
\begin{bmatrix}
\sigma_{\rho}^2 \\\\ (k/\pi)^2[\sigma_{\mu_{\alpha}}^2+(\mu_{\alpha}\sigma_{\pi}/\pi)^2] \\\\ (k/\pi)^2[\sigma_{\mu_{\delta}}^2+(\mu_{\delta}\sigma_{\pi}/\pi)^2]
\end{bmatrix}+\dfrac{2\mu_{\alpha} \mu_{\delta}k^2\sigma_{\pi}^2}{\pi^4}
\begin{bmatrix}
b_{12}\cdot b_{13} \\\\ b_{22}\cdot b_{23} \\\\ b_{32}\cdot b_{33}
\end{bmatrix}
\end{equation}
The elements of the matrix $C$ are the squares of the individual elements of $B$, i.e. $c_{ij}=b_{ij}^2$ for all $i$ and $j$ (see Johnston et al. 1987).

The final step to obtain the uncertainties associated to the state vector is to propagate the errors in position and velocity in the inertial Galactocentric frame of reference. This can be easily done computing the partial derivatives of the transformations of coordinates described in Section 4.

\section[Selected clusters]{Results for individual globular clusters}

We derived the proper motion of ten bulge clusters. Table \ref{tab:all_clusters} gives
the proper motion values, velocity  components of the Galactocentric
velocity vectors, and other relevant information. The input distances and coordinates for the sample clusters are from Bica et al. (2006). The data
show that the proper motions and the velocity components of our targets
are, on average, lower than those of halo clusters (e.g. Dinescu et al. 1999),
with the exception of the two most metal-poor clusters (Terzan 4, Terzan 9).
For the more reddened clusters Terzan 1, Terzan 4 and Palomar 6, we employed the
z-band (instrumental), which provided  a better definition to the CMDs. 
The studied clusters are discussed individually as follows.

\subsection*{Terzan 1}

Terzan 1 is a moderately metal--poor ([Fe/H]$\sim-1.0$) globular cluster with a dominant red HB.
 There is evidence that it is a second parameter globular cluster showing a 
post--core collapse structure  (Ortolani et al. 1999a). 
We used as first epoch CCD camera on the Danish telescope observations taken in June 1990,
and as second epoch NTT+EFOSC2 data observed in May 2012, with a baseline of 22 years.

\begin{figure*}
\centering
\begin{tabular}{c c c}
\includegraphics[scale=0.35]{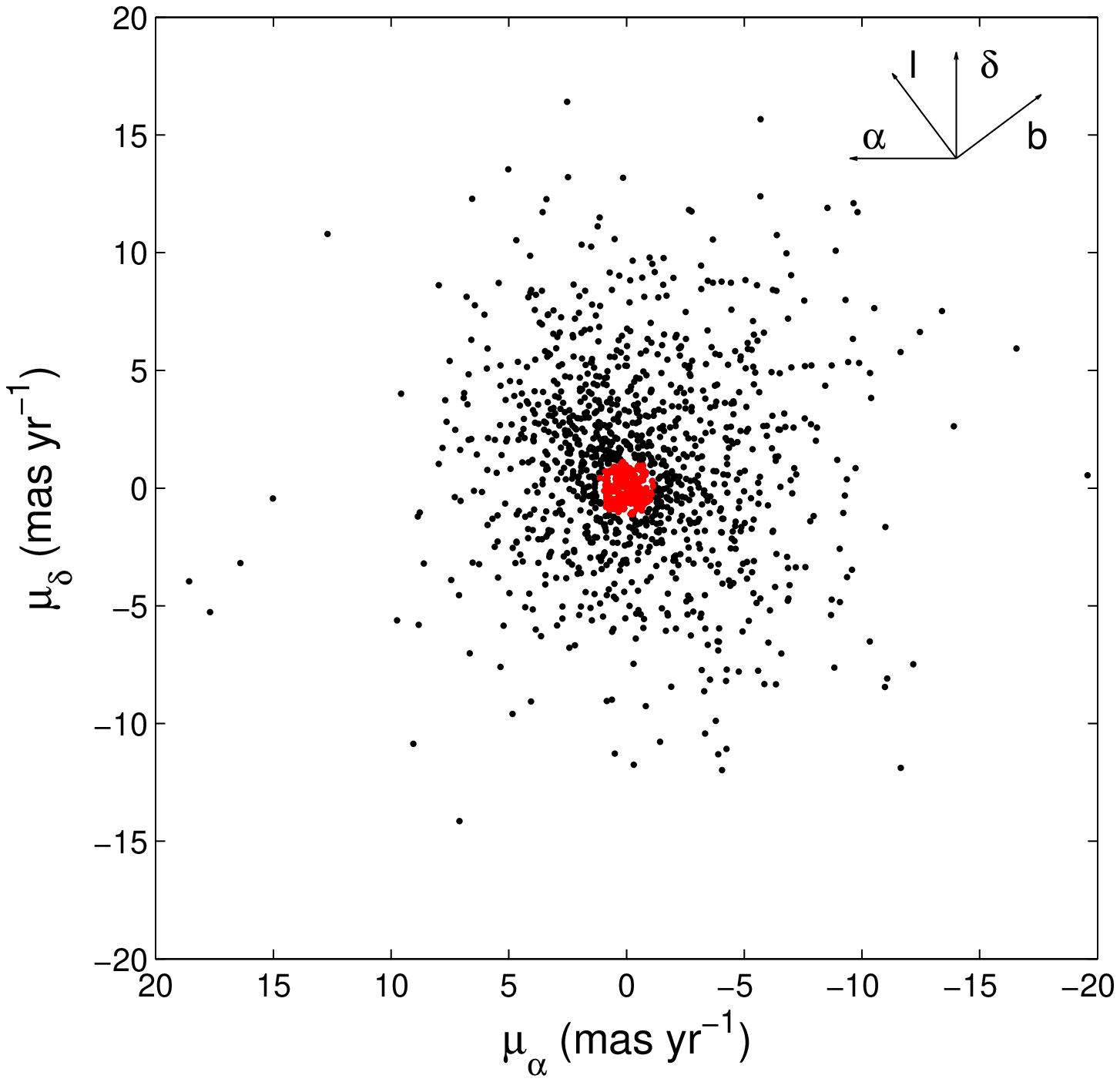} 
\includegraphics[scale=0.35]{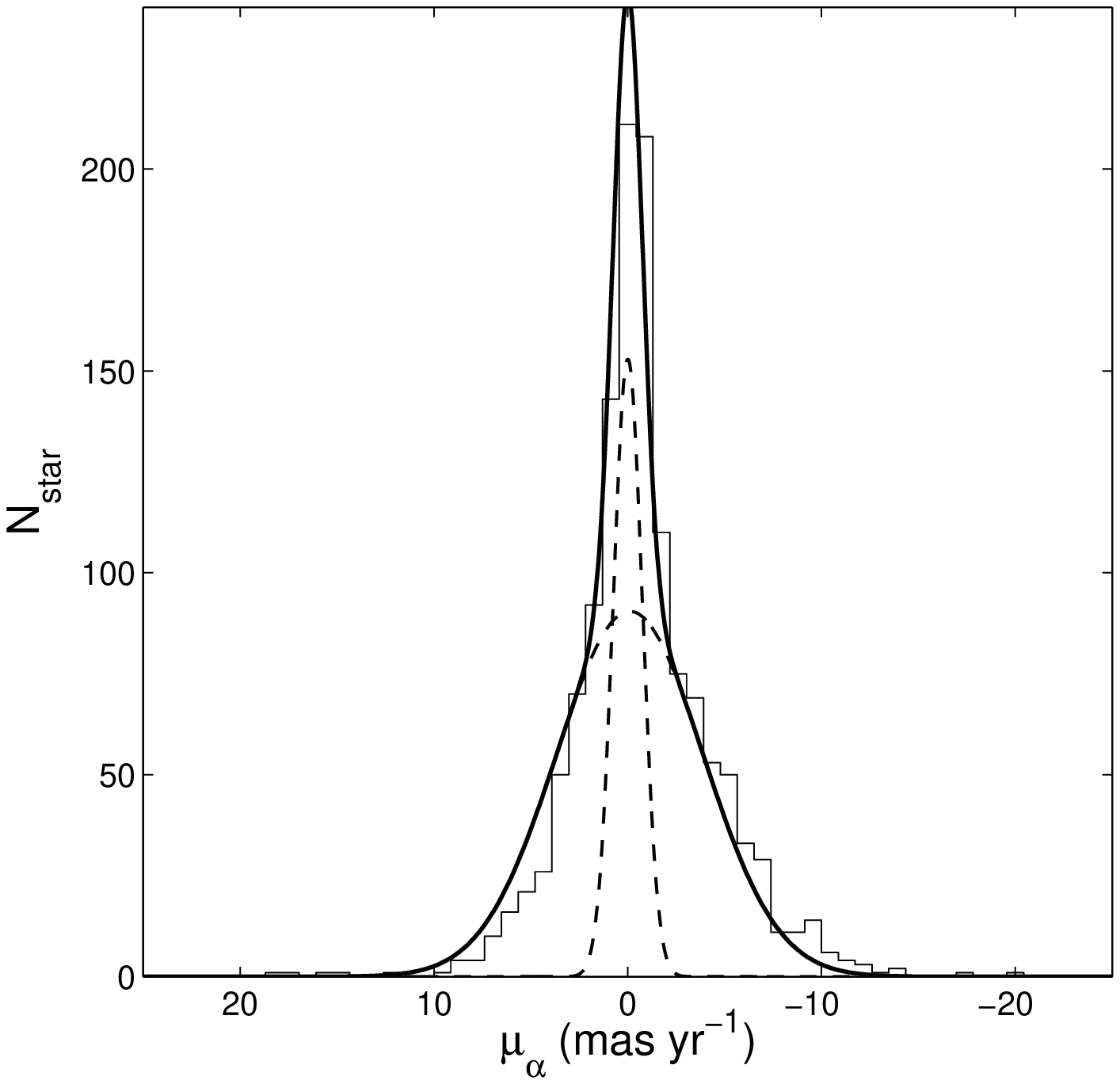} 
\includegraphics[scale=0.35]{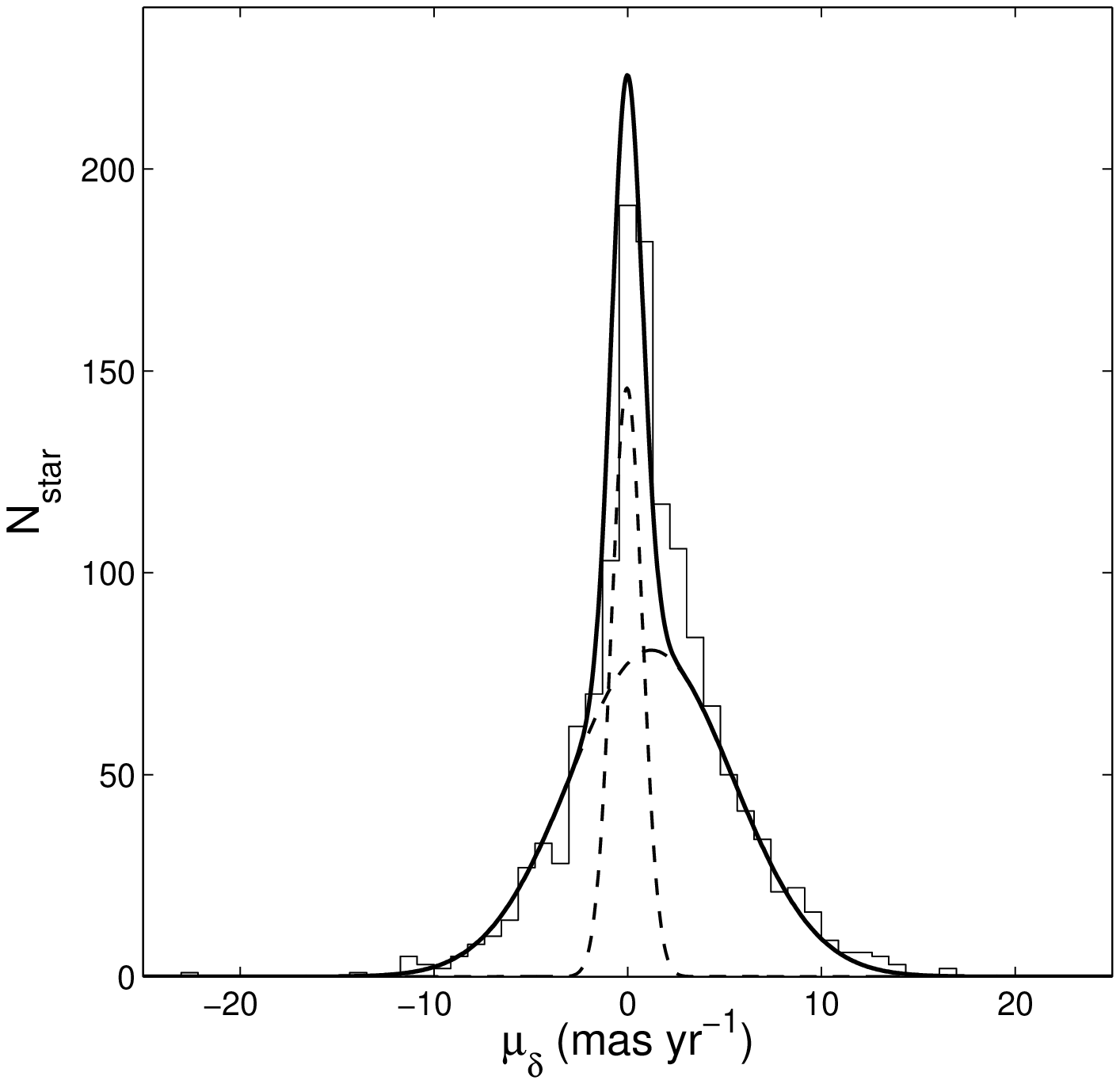}\\
\includegraphics[scale=0.35]{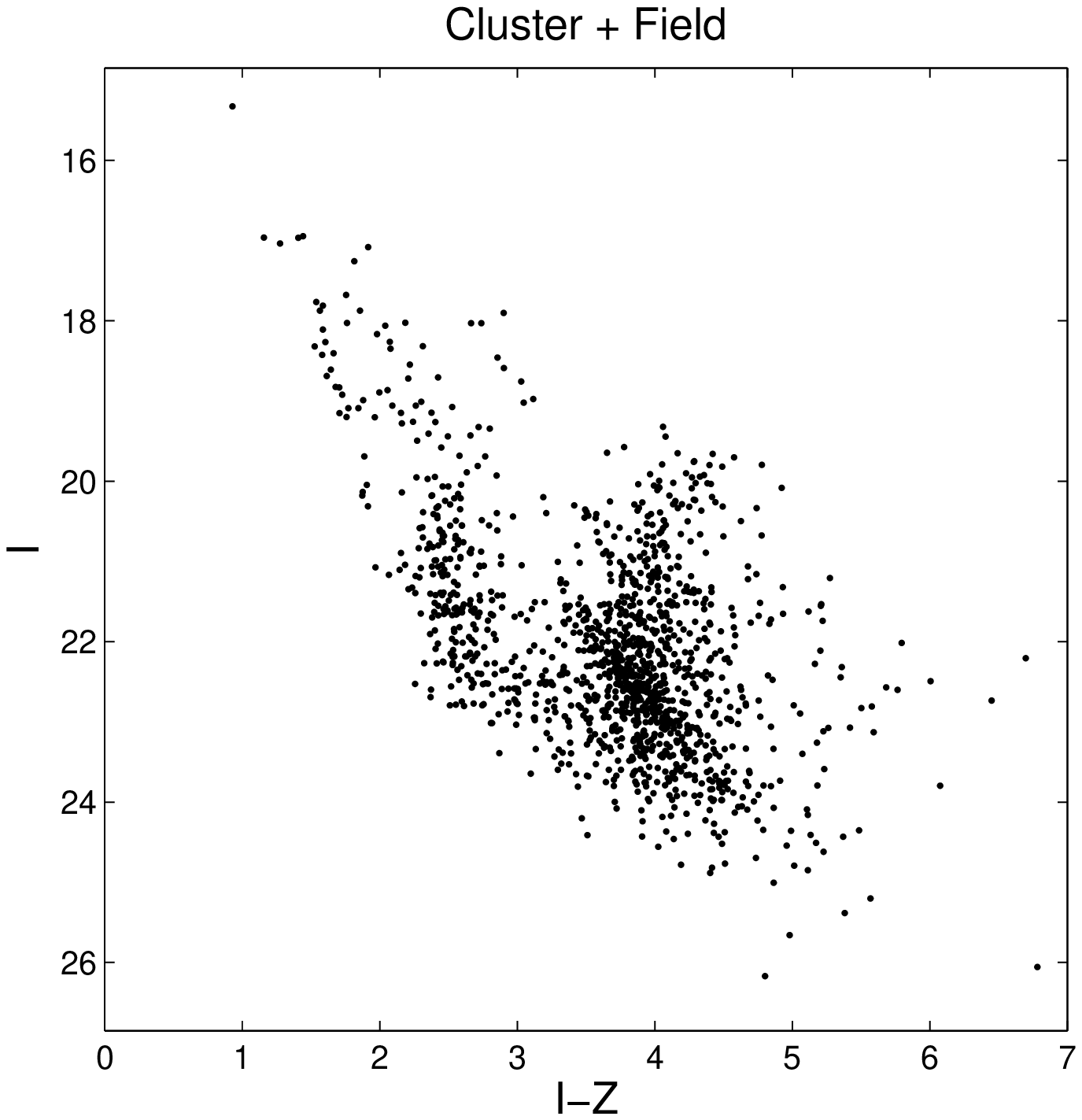} 
\includegraphics[scale=0.35]{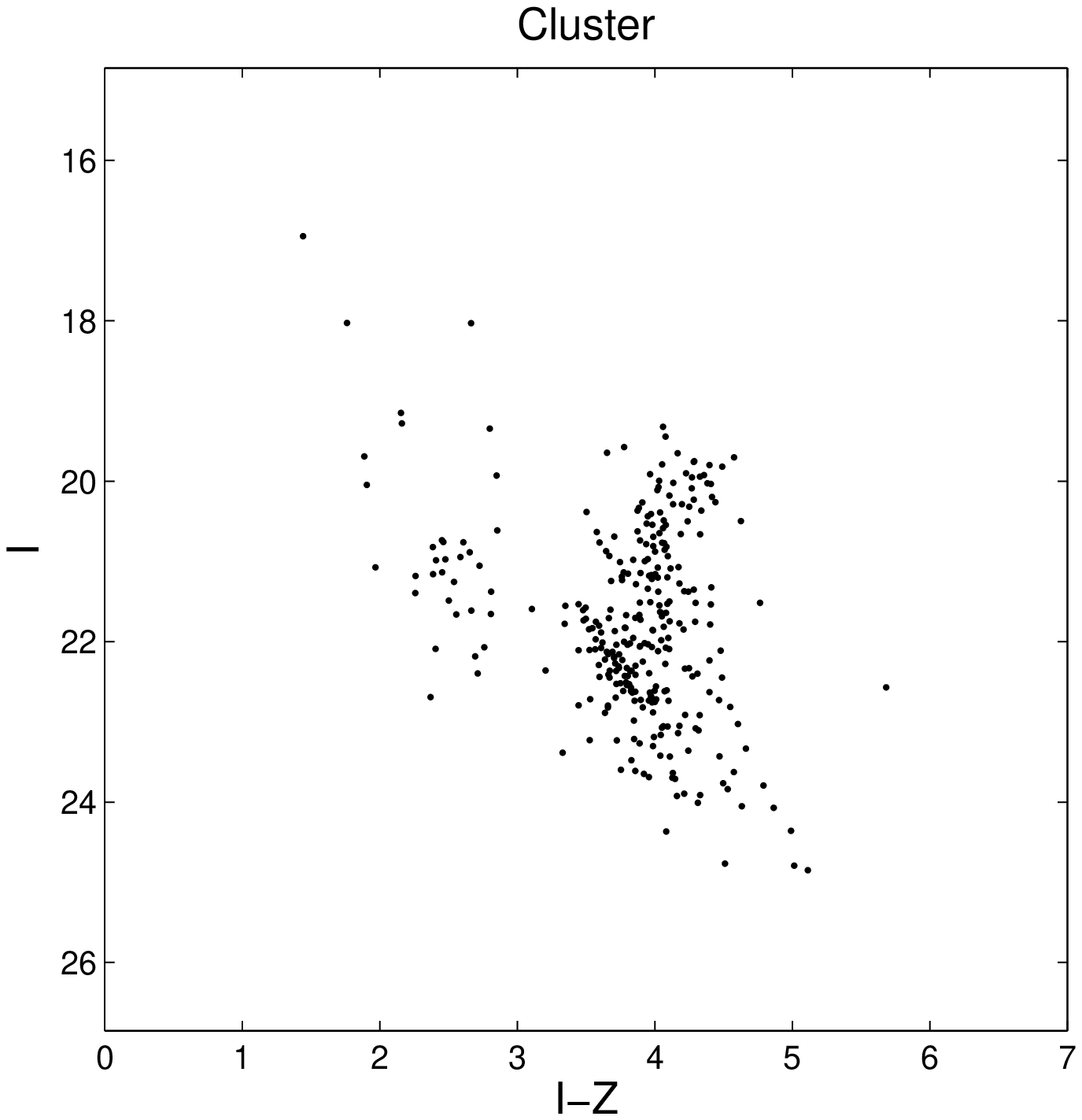} 
\includegraphics[scale=0.35]{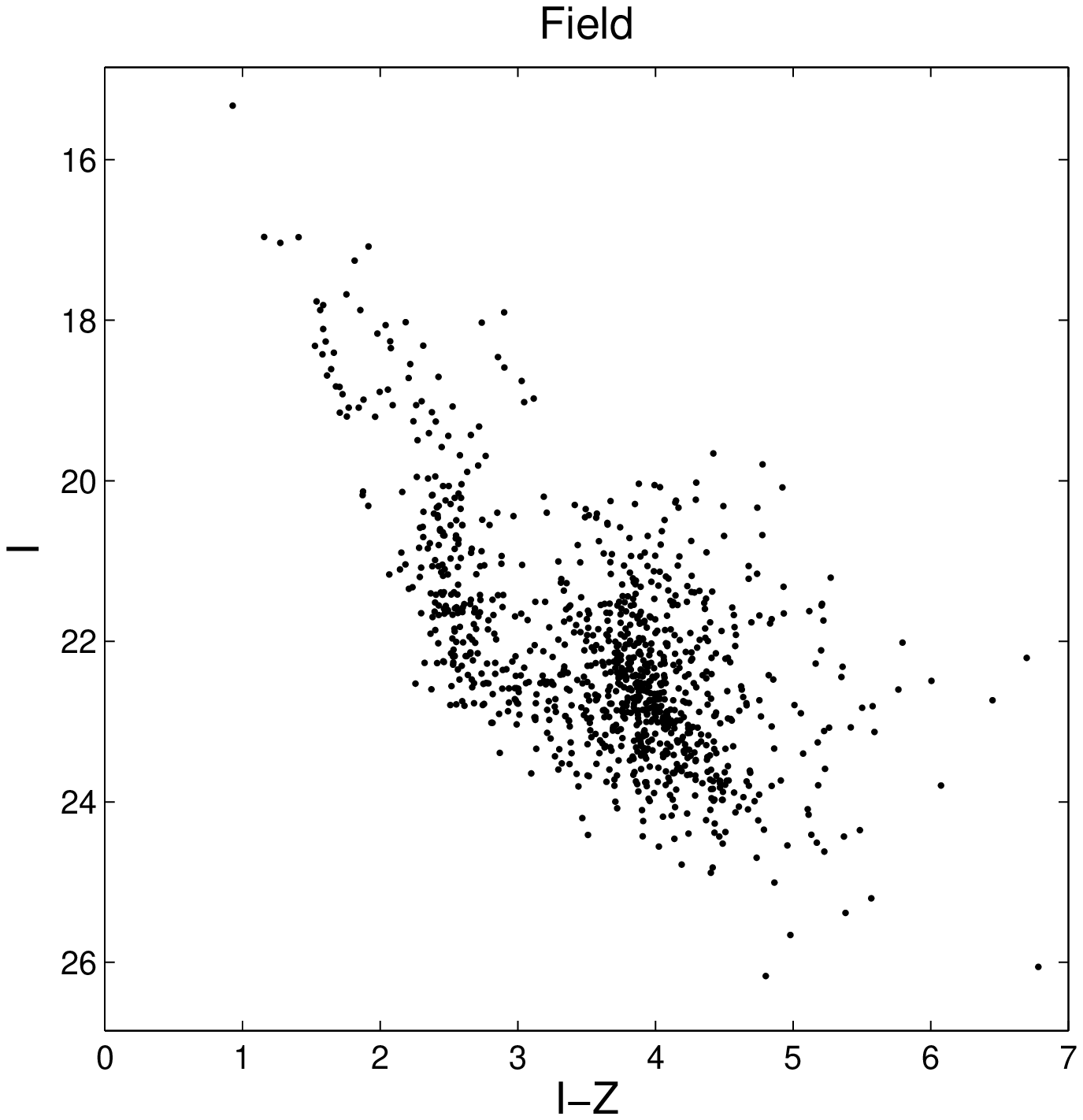}
\end{tabular}
\caption{Proper motion and CMD decontamination of Terzan 1.}
\label{fig:Terzan1pm}
\end{figure*}

In Figure \ref{fig:Terzan1pm} are shown the derived proper motions, 
the double--Gaussian interpolation of the displacements and the CMD cleaning of the cluster.
The present analysis showed a small proper motion both in right ascension and declination, that could be an indication of the very elongated orbit of the cluster. This picture would be consistent with the relatively high radial velocity of Terzan 1 ($v_\mathrm{r}=114 \pm 14  \; \mathrm{km}\; \mathrm{s}^{-1}$, Idiart et al. 2002). The values are reported in Table 3.

\subsection*{Terzan 2}

\begin{figure*}
\centering
\begin{tabular}{c c c}
\includegraphics[scale=0.35]{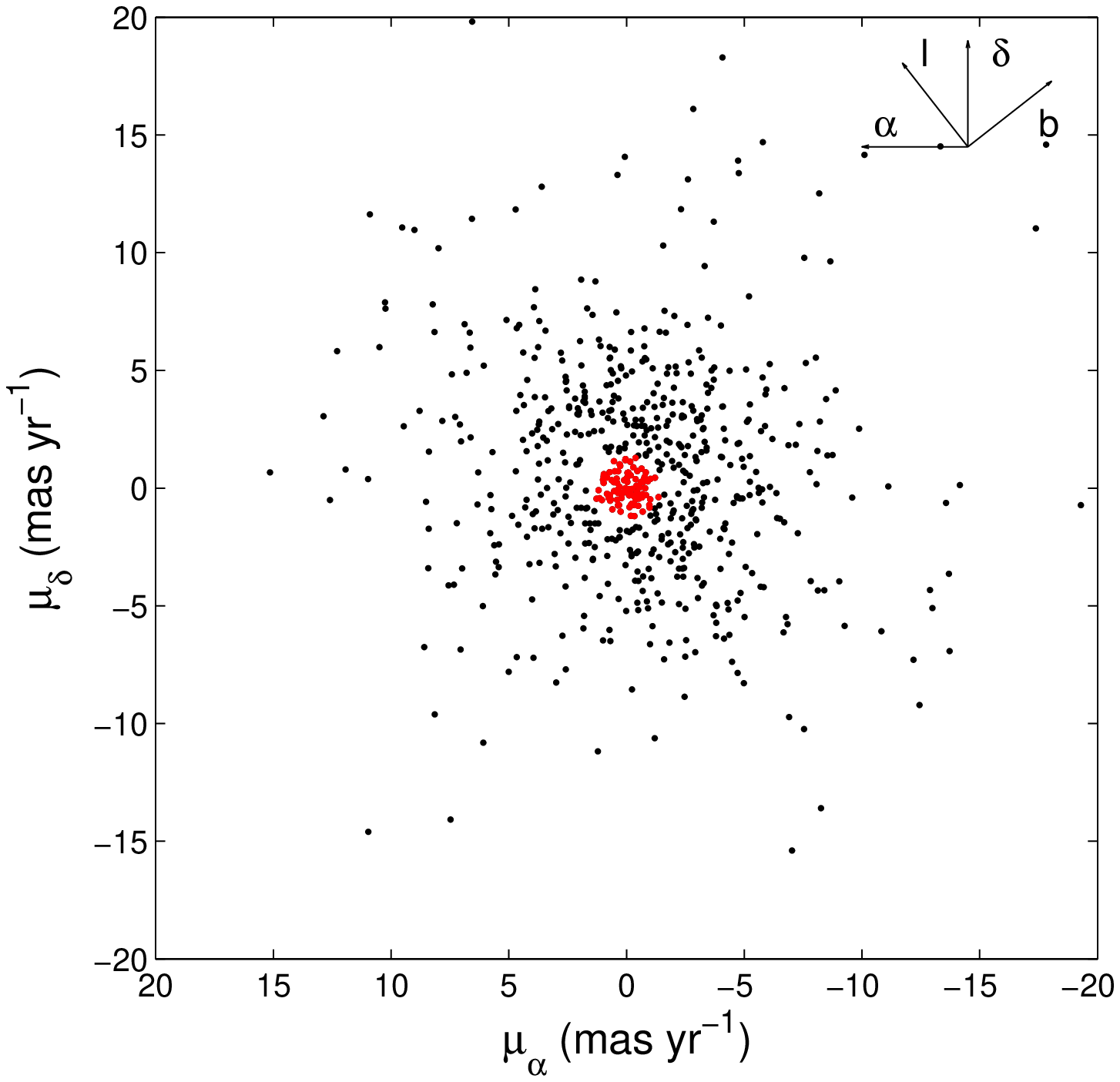} 
\includegraphics[scale=0.35]{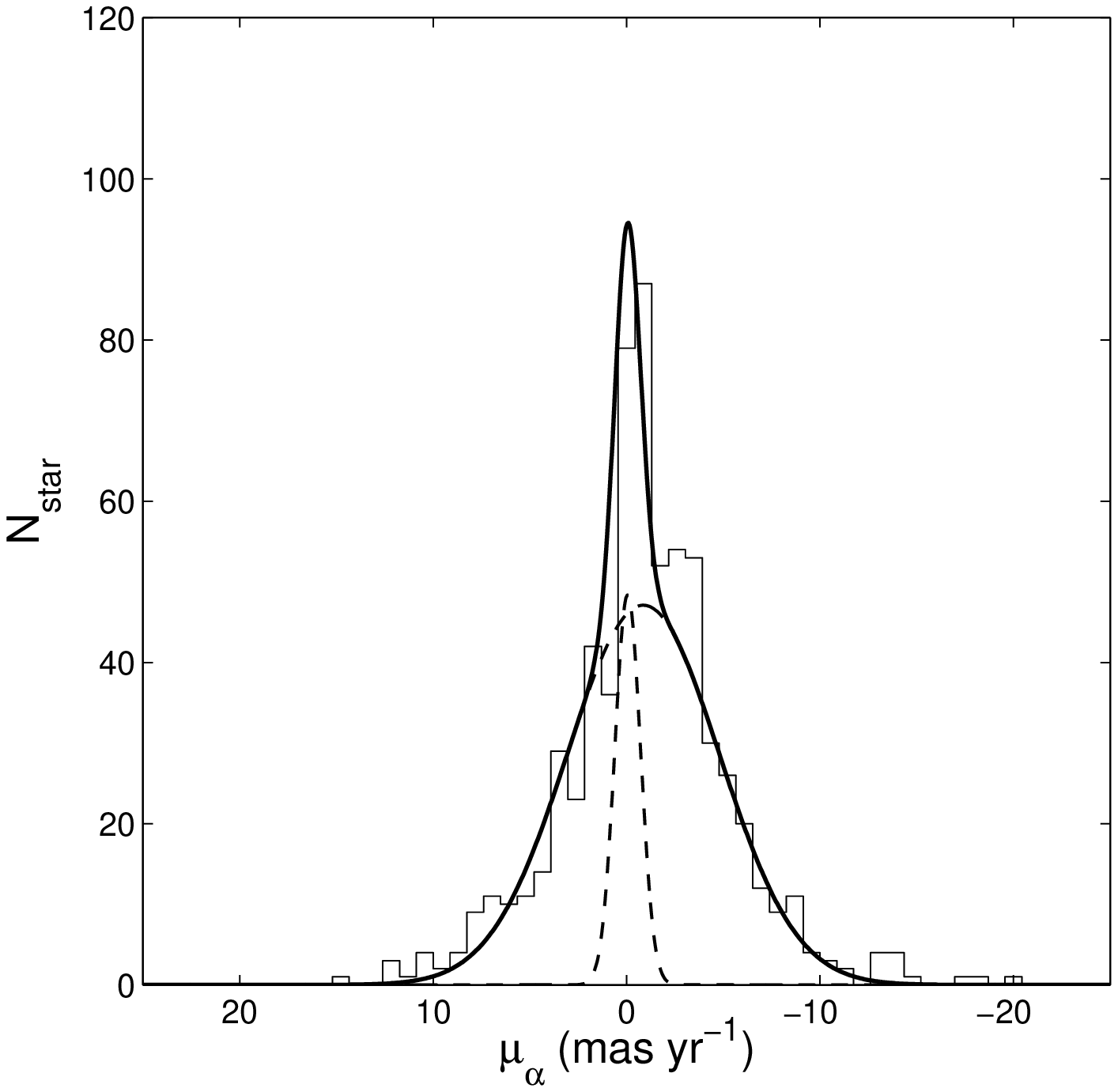} 
\includegraphics[scale=0.35]{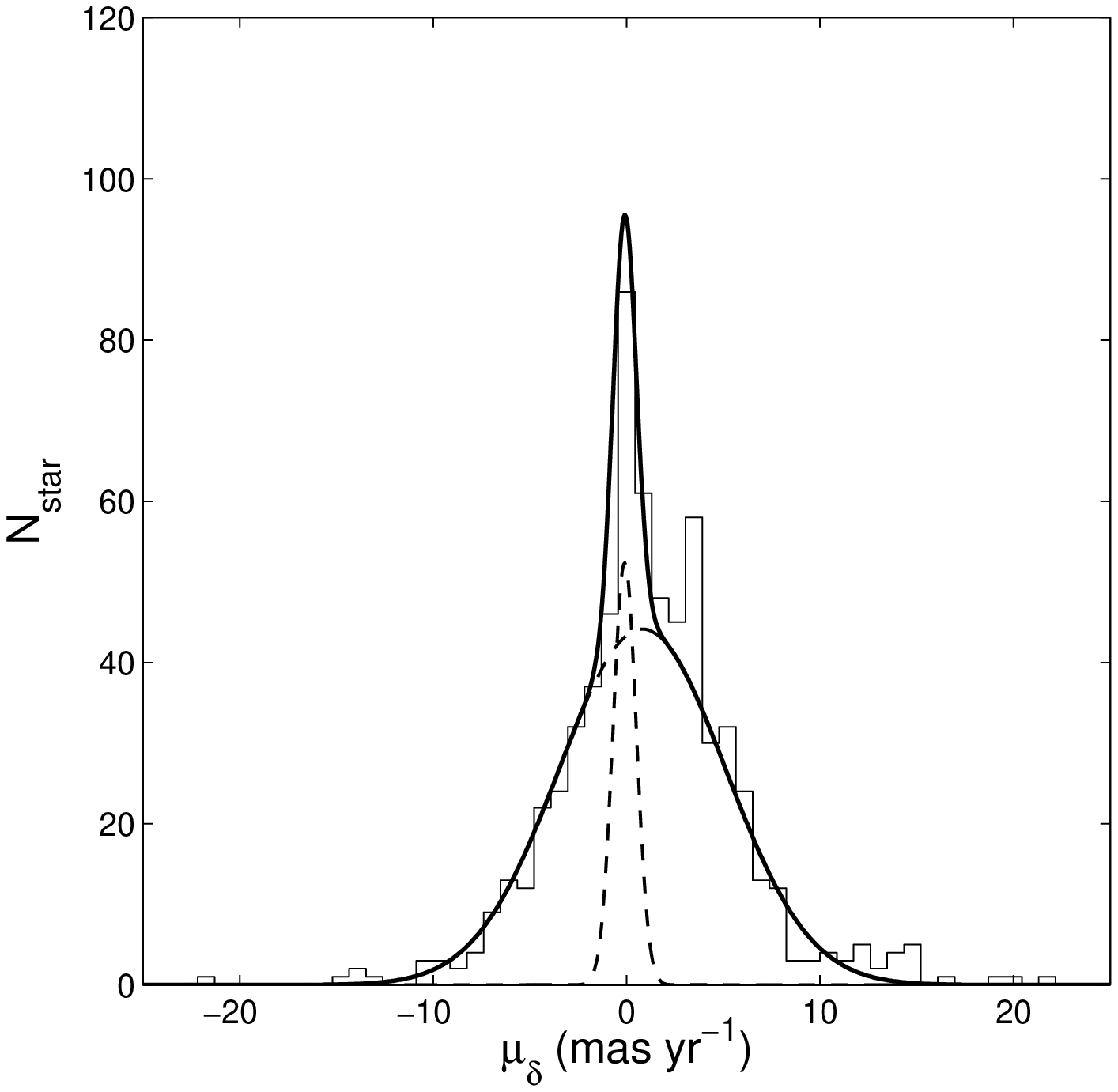}\\
\includegraphics[scale=0.35]{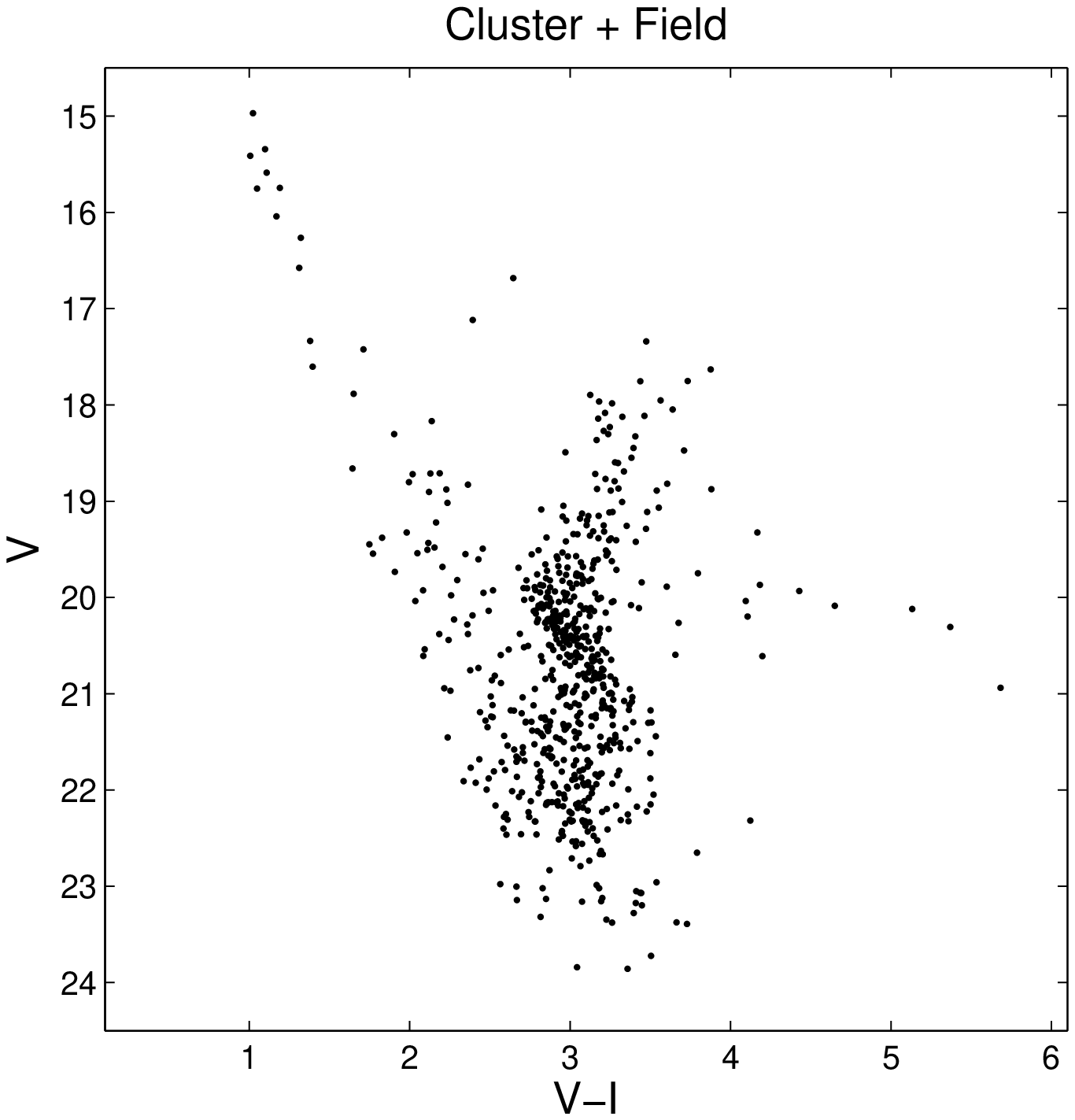} 
\includegraphics[scale=0.35]{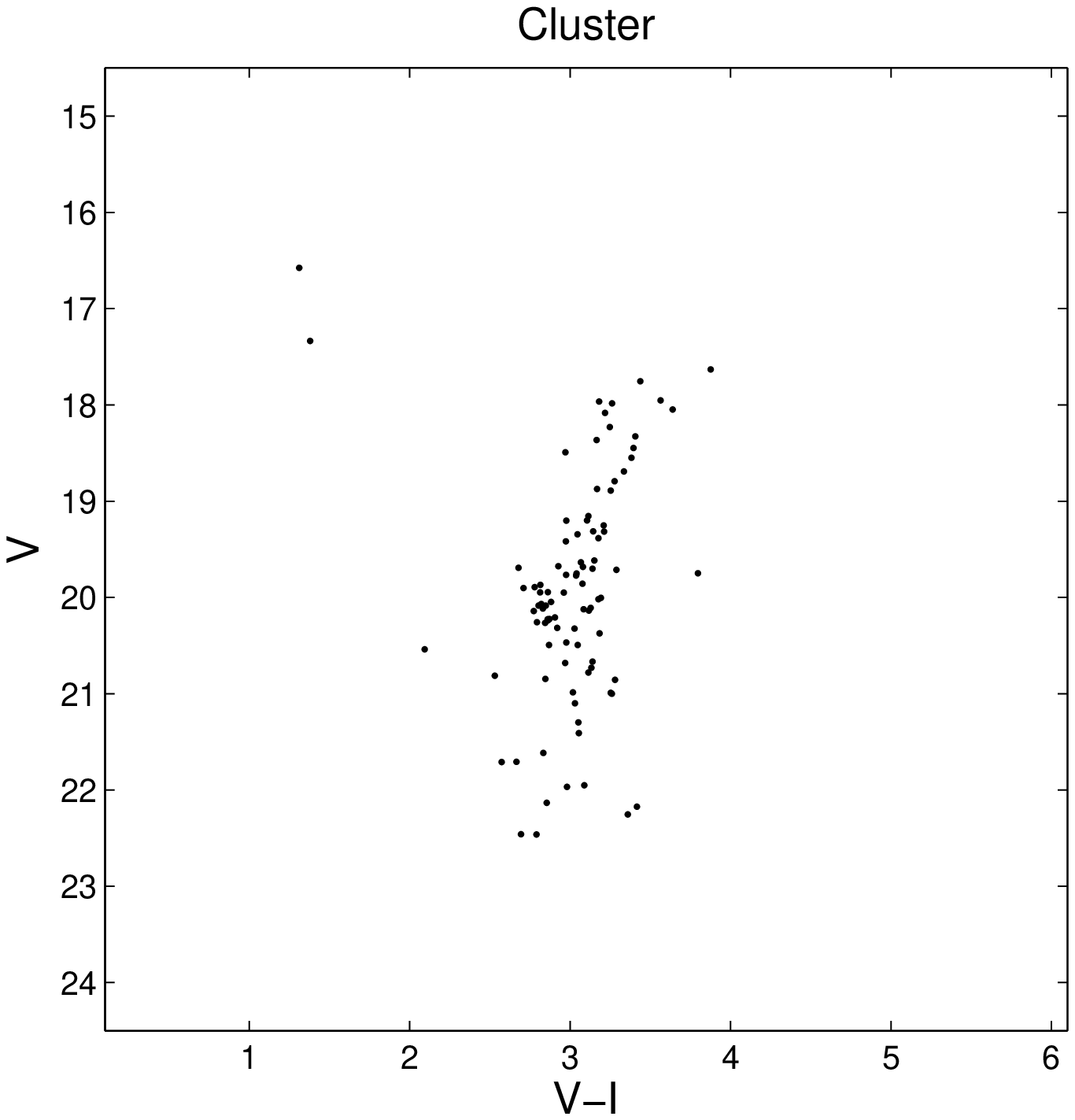} 
\includegraphics[scale=0.35]{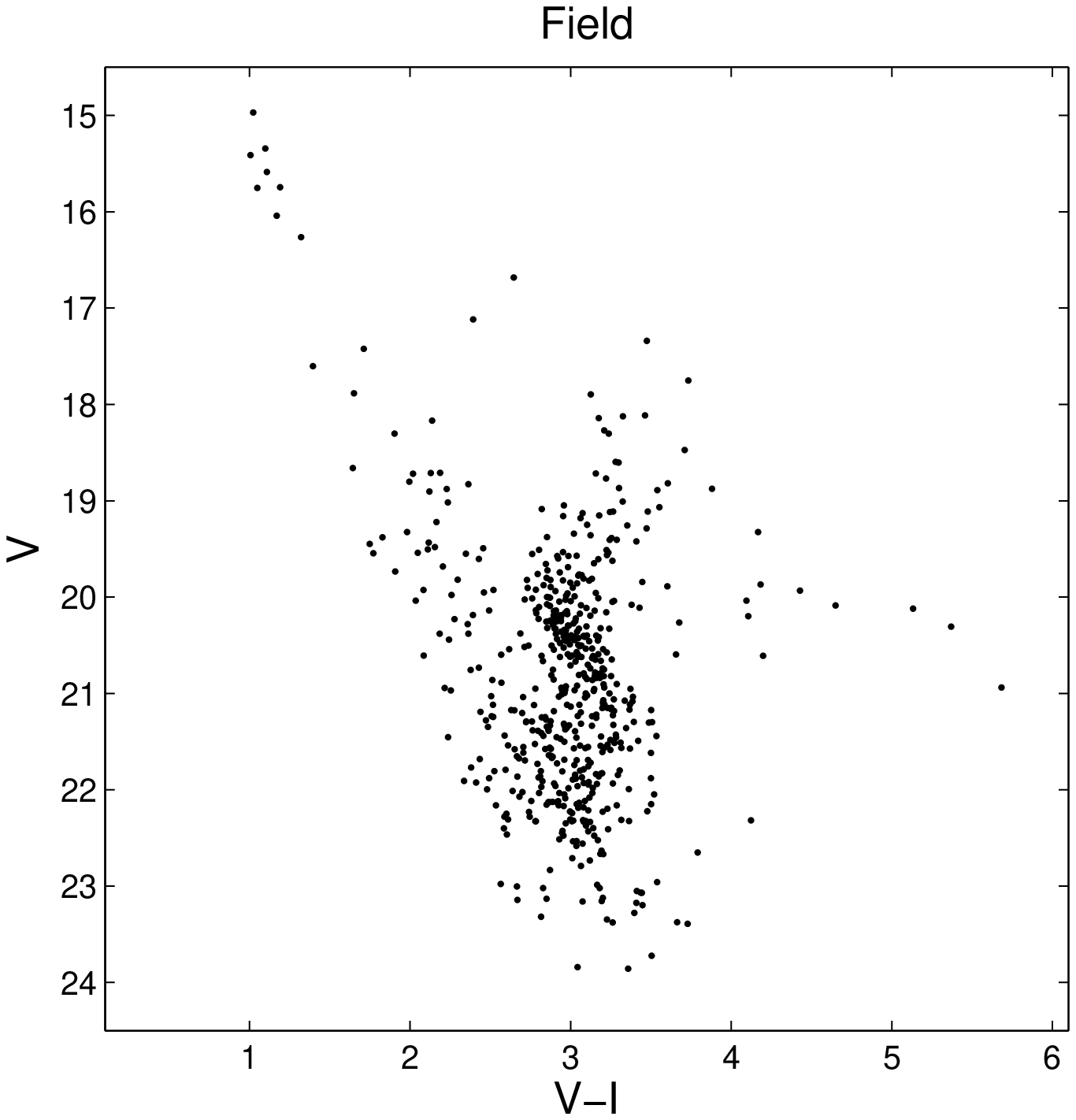}
\end{tabular}
\caption{Proper motion and CMD decontamination of Terzan 2.}
\label{fig:Terzan2pm}
\end{figure*}

Terzan 2 is a metal-rich red horizontal branch cluster, 
located very close to the Galactic center (Ortolani et al. 1997a; Valenti et al. 2007).
With a baseline of 18 years, a very good seeing and a good sampling in the first epoch observations,
taken at NTT+SUSI, in May 1994, the 
 proper-motion-cleaning is very satisfactory. The second epoch was taken with
NTT+EFOSC2 in June 2011.
  The cluster CMD given in Fig. \ref{fig:Terzan2pm}
is not richly populated, but the elimination of field sequences was effective.
Despite its location being among the closest objects to the Galactic center, it shows a very low proper motion, 
below 1 mas/yr, together with a radial velocity of 109 km/s.

\subsection*{Terzan 4}

Terzan 4 is among the most metal-poor clusters in the inner bulge 
(Ortolani et al. 1997b; Origlia et al. 2004).
First and second epoch observations were obtained respectively, at the
NTT+SUSI in May 1994, and NTT+EFOSC2 in May 2012. Both these sets of images
have been obtained under exceptionally good seeing.
The proper-motion-cleaned CMD with a baseline  of 18 years provided a very well defined RGB.
The definition of the CMD is comparable to that of NICMOS/HST one (Ortolani et al. 2007),
showing the possibility of a few blue HB stars.
The proper motion values, as well as the Galactic velocities ( lower panel of Table \ref{tab:all_clusters}) 
are somewhat higher than those of the sample average, possibly characterizing a halo-originated
 cluster (given also its much lower metallicity). The orbit prediction will bring more constraints.

\subsection*{Terzan 9}

Ortolani et al. (1999b) estimated a metallicity of [Fe/H]$\sim-2.0$,
 whereas Valenti et al. (2007) indicates [Fe/H]$\sim-1.2$. In the latter case,
 it could be added to the group of moderate metallicy and BHB. Harris (1996, 2010 version)
 reports a radial velocity of 59 km/s. 
First and second epoch observations were obtained respectively, with the DFOSC@Danish
in July 1998, and  and NTT+EFOSC2 in May 2012. For this cluster the time baseline is of 14 years.
It has a relatively high proper motion
(Table 3). Only a detailed orbital study can clarify if it is a halo or a bulge cluster.

\subsection*{NGC 6522}

NGC 6522 is a moderately metal--poor globular cluster projected on the Baade's Window.  
We derived the proper motion from two sets of observations: the first epoch images have been
 obtained with the Danish Telescope in June 1992, and HST/WFPC2 in 
September 1995,
while the second epoch images come from NTT (May 2012). 
Both diagrams are well-cleaned, with HST first epoch including fainter stars.
The dispersion of the proper motion plots is tighter for the Danish data, possibly due
to the inclusion of fainter stars in the HST comparison.
The abundance analysis and  radial velocity of 8 stars was presented in Barbuy et al. (2009, 2014).
 Its proper motion  has  been investigated also by Terndrup et al. (1998), who derived values with respect
 to a star at rest in the nuclear bulge equal to
 $(\mu_l,\;\mu_b)=(1.4\pm0.2,\;-6.2\pm0.2) \;\mbox{mas}\;\mbox{yr}^{-1}$. 
The top 6 panels of Figure \ref{fig:NGC6522pm} show the cluster proper motion obtained in the present work.
Our analysis produced the values
 $(\mu_\alpha^*,\;\mu_\delta)=(3.35\pm0.60,\;-1.19\pm0.34) \;\mbox{mas}\;\mbox{yr}^{-1}$, 
corresponding to a proper motion in Galactic coordinates of
$(\mu_l,\;\mu_b)=(0.4,-3.1) \;\mbox{mas}\;\mbox{yr}^{-1}$. 
Our result is about a factor  of 2 smaller than the one obtained by Terndrup et al. (1998).  
We also tried to derive its proper motion independently, by using archive HST images. 
The results are shown in the bottom six panels of Figure \ref{fig:NGC6522pm}. 
This new estimate of the proper motion of the cluster produced
 the value $(\mu_\alpha,\;\mu_\delta)=(3.6\pm0.6,\;-1.2\pm0.3) \;\mbox{mas}\;\mbox{yr}^{-1}$, 
which is consistent with our previous analysis. 
Note that the calibration star from Terndrup is outside our field.
We finally also note that in Barbuy et al. (2014) a revised radial velocity of 
v$_{\rm r}$ = -14.3$\pm$0.5 km s$^{-1}$ was obtained,
 which is basically compatible with 
the adopted value given in Table \ref{tab:all_clusters}.

\subsection*{NGC 6558}

NGC 6558 has a well-determined low proper motion, combined with a high radial velocity
of -197.2 km s$^{-1}$ (Barbuy et al. 2007).
First and second epoch observations were obtained respectively, at the
NTT+EMMI in June 1993, and NTT+EFOSC2 in May 2012. The second epoch was obtained
under exceptionally good seeing conditions.
 The present decontaminated CMD is very populous
with well-defined sequences. The cleaning was effective in this case, superseding
previous CMDs for this cluster. Vazques et al. (2013) found a value of 
the proper motion
($\mu_{l}$ cos(b),$\mu_{b}$) = (0.30$\pm$0.14, -0.43$\pm$0.13) mas/yr. 
The present proper motion
in Galactic coordinates is ($\mu_{l}$ cos(b),$\mu_{b}$) = (0.37$\pm$0.43, 0.29$\pm$0.62)
mas/yr. 
Considering  the large errors relative to the small motion of the
cluster, and that the telescopes and the techniques are different,
 the results are in a very good agreement.
 An orbit derivation is needed in order to establish its dynamics.

\subsection*{NGC 6540}

NGC 6540 has a detectable small proper motion, and the cleaning is effective, showing that we
correctly identified the CMD features. A very low radial velocity of 17.2 km s$^{-1}$ 
is given in Harris (1996, 2010 version).
First and second epoch observations were obtained respectively, with the direct
CCD camera at the Danish telescope
 in June 1990, and NTT+EFOSC2 in May 2012. The observing runs were the same as for Terzan 1.
Given its low proper motion and radial velocity, the object could be near its apogalacticon.
Bica et al. (1994) estimated a metallicity of [Fe/H]$\sim-1.0$.
This could be another moderate metallicity cluster with a BHB.
The proper-motion-cleaning was not very effective, as shown by about
the same number of BHB stars in the cluster and in the field,
 due to a very small  difference between the main motion of the cluster
and field stars. Nevertheless, the proper-motion dispersion of field 
stars is sizably larger than that of cluster members. We can clearly
see that the CMD of proper-motion-selected stars (middle panel on the bottom
of Figure 23) is fairly free of field stars. Most of the field stars are
indeed included in the bottom-right panel, where they should belong.

\subsection*{AL 3}
First and second epoch observations were obtained respectively, with DFOSC@Danish telescope
 in March 2000, and NTT+EFOSC2 in May 2012.
AL~3 shows a rather high right ascension proper motion as compared with the other central clusters,
but it is comparable to that of the bulge edge cluster NGC 6652.
The proper motion in declination on the other hand is clearly higher in NGC 6652 than in AL~3.
A radial velocity measurement for AL~3 is needed for a kinematical diagnostic.
As compared with Ortolani et al. (2006),
a few stars in the RGB are missing in the present CMD, owing to saturation effects in the second epoch image.
 The proper-motion-cleaned
 CMD confirms that AL~3 is a BHB cluster, and the RGB indicates a
moderate metallicity.

\subsection*{ESO456-SC38}
First and second epoch observations were obtained respectively, with NTT+SUSI
 in May 1994, and NTT+EFOSC2 in May 2012.
This cluster has a relevant proper motion. Despite this, the proper motion cleaning did not produce a
well-defined CMD, because of the loose structure of the cluster and a high differential reddening.
Preliminary radial velocity values of $\sim$-160 km s$^{-1}$ were obtained 
by S. V\'asquez (in preparation).

\subsection*{Palomar 6}

The proper motion of Palomar 6 was based on the longest time base of our sample,
 spanning 20 years.
First and second epoch observations were obtained respectively, with NTT+EMMI
 in June 1993, and NTT+EFOSC2 in May 2013.
 The second epoch was obtained with the highest available spatial resolution.
It has a proper motion of around 3 mas/year. Harris (1996) reports a radial velocity
of 190 km s$^{-1}$, and Lee et al. (2004) measured 180 km s$^{-1}$. 
The cluster is very reddened with E(B-V)=1.33, and its distance
is compatible with a location near the Galactic center. For this reason, high velocity vectors
are expected. The proper motion cleaning was very effective (see lower panels in Fig. \ref{fig:Palomar6pm}),
disentangling the cluster from the bulge population.
The  proper-motion-cleaned
 CMD confirms that it is a 47 Tuc-like cluster, with a red HB (Lee et al. 2004).

\subsection*{NGC 6652}

We added NGC 6652 to the sample for comparison purposes.
This borderline bulge cluster is located at l=1\fdg53, b=-11\fdg38. The edge of the bulge
was defined to be at around 10$^{\circ}$ by Soto et al. (2014) and Rojas-Arriagada et al. (2014). 
Proper motions were derived from HST observations as reported by Sohn et al. (2014).
We find that its kinematics is very different from the sample bulge clusters,
because its orbit is much more extended in the halo, reaching
apocentric distances up to 10 kpc from the Galactic center.

\section{DISCUSSION AND CONCLUSIONS}

We derived absolute proper motion values for ten inner bulge globular clusters,
thanks to first epoch images taken by our group in the 90s,
 combined with recent 
dedicated observations at the ESO NTT telescope. 
The proper motions were computed from second epoch distortion-corrected
positions using field stars as absolute reference frame.

The derived velocity components are as a rule very low suggesting
 that most of these clusters
are trapped in the bulge and/or the bar. These clusters could have been formed
earlier than the bar, and were later trapped inside it (Babusiaux et al. 2014).
The low velocities are made evident, by a comparison with NGC 6652 (Sohn et al. 2014).
This cluster is located near b$\sim$10$^{\circ}$, a region
considered by Soto et al. (2014) and Rojas-Arriagada et al. (2014) as the
 outer bulge. The bulge-halo transition is defined as the
region where some bulge rotation is still detected for the field stars, 
but the ratio of metal-rich/metal-poor 
 field stars components is inverted, with metal-poor stars (halo) prevailing outside
this region.
 The comparison of the kinematics between our inner bulge sample clusters and NGC 6652
shows a clear difference in space velocities, indicating the central clusters to be a distinct 
population relative to the outer bulge/inner halo clusters. It would be important to acquire
proper motions of other clusters in the external regions of the bulge.

Dinescu et al. (1999) found lower velocities for  inner galaxy clusters as compared with
outer halo ones. We found even lower velocities for inner bulge clusters. 

In conclusion, the proper motion analysis of the present sample clusters indicates that 
the moderately metal-poor and
 metal-rich clusters appear to have preferentially low velocity components.
The halo contamination inside the bulge appears to be limited to very few objects, if any.
 The proper-motion-cleaned CMDs confirm 
a typical pattern for the Blue Horizontal Branch moderate metallicity clusters.

The orbits and their evolution will be presented in a forthcoming paper.

\section*{Acknowledgments}
We are thankful to the referee for a careful reading and useful suggestions.
We are grateful to S. Casotto for important clarifications on
dynamical aspects. We are thankful to Y. Momany for helping with
the observations. We thank V. Mezzalira for helpful technical
support. LR acknowledges a CRS scholarship from Swinburne University of Technology. 
BB  and  EB acknowledge  grants from  the
Brazilian  agencies  CNPq and  Fapesp.   SO  acknowledges the  Italian
Ministero dell'Universit\`a e della Ricerca Scientifica e Tecnologica, and the financial 
support from the Universit\`a di Padova.

\begin{table*}

\begin{tabular}{ c  c  c  c  c  c   c   c   c  }

\hline

\\

CLUSTER & $\alpha$ ($^\circ$) & $\delta$ ($^\circ$) & $l$ ($^\circ$)& $b$
($^\circ$)& $\mu_{\alpha}^{*}$ (mas/yr)& $\mu_{\delta}$ (mas/yr)   &
$v_r$ (km/s) & $d_{\odot}$ (kpc) \\

\\

\hline

\\

\\

Terzan 1 & 263.95 & -30.47 & 357.57 & 1.00 & $0.51\pm0.31$ &
$-0.93\pm0.29$ & $114.0\pm 14.0$ & $6.2\pm 0.6$ \\

\\

Terzan 2  & 261.89 & -30.80 & 356.32 & 2.30 & $-0.94\pm0.30$ &
$0.15\pm0.42$ & $109.0\pm15.0$ & $8.7\pm 0.8$ \\

\\

Terzan 4 & 262.66 & -31.59 & 356.02 & 1.31 & $3.50\pm0.69$ & $0.35\pm0.58$
& $-50.0\pm 2.9$ & $9.1\pm 0.9$ \\

\\

Terzan 9 & 270.41 & -26.84 & 3.61 & -1.99 & $0.00\pm0.38$ & $-3.07\pm0.49$
& $59.0\pm 10.0$ & $7.7\pm 0.7$ \\

\\

NGC 6522 & 270.89 & -30.03 & 1.02 & -3.93 & $3.35\pm0.60$ & $-1.19\pm0.34$
& $-21.1\pm 3.4$ & $7.8\pm 0.7$ \\

\\

NGC 6558 & 272.57 & -31.76 & 0.20 & -6.02 & $-0.12\pm0.55$ & $0.47\pm0.60$
& $-197.2\pm 1.5$ & $7.4\pm 0.7$ \\

\\

NGC 6540 & 271.53 & -27.76 & 3.29 & -3.31 & $0.07\pm0.40$ & $1.90\pm0.57$
& $-17.72\pm 1.4$ & $3.7\pm 0.3$ \\

\\

AL 3 & 273.53 & -28.63 & 3.36 & -5.27 & $4.77\pm0.46$ & $0.55\pm0.44$ &
$-$ & $6.5\pm 0.6$ \\

\\

ESO 456--SC38 & 270.45 & -27.83 & 2.77 & -2.50 & $3.08\pm0.29$ &
$2.00\pm0.34$ & $-$ & $6.7\pm 0.6$ \\

\\

Palomar 6 & 265.93 & -26.22 & 2.10 & 1.78 & $2.95\pm0.41$ & $1.24\pm0.19$
& $181.0\pm 2.8$ & $7.3\pm 0.7$ \\

\\

NGC 6652 & 278.94 & -32.99 & 1.53 & -11.38 & $4.75\pm0.07$ & $-4.45\pm0.10$ & $-111.7\pm 5.8$ & $9.6\pm 0.9$ \\

\\
\hline

\\

\end{tabular}

\begin{tabular}{ c  c  c  c  c c  c  c }

CLUSTER & $X$ (kpc) & $Y$ (kpc)& $Z$ (kpc) & $\dot{X}$ (km/s)& $\dot{Y}$
(km/s) & $\dot{Z}$ (km/s) & [Fe/H]\\

\\

\hline
\\

Terzan 1  & 1.3 & 0.3 & 0.1 & $-125\pm14$ & $-25\pm18$ & $-24\pm9$ & -1.3 \\

          & (2.1) & (0.3) & (0.1) & ($-125$) & ($-45$) & ($-24$) \\

\\

Terzan 2  & -1.2 & 0.6 & 0.4 & $-122\pm15$ & $18\pm22$ & $-26\pm14$ & -0.5 \\

          & (-0.4) & (0.6) & (0.4) & ($-122$) & ($-10$) & ($-25$) \\

\\

Terzan 4  & -1.6 & 0.6 & 0.2 & $46\pm3$ & $121\pm32$ & $132\pm31$ & -1.6 \\

          & (-0.8) & (0.6) & (0.2) & ($46$) & ($92$) & ($133$) \\

\\

Terzan 9  & -0.2 & -0.5 & -0.3 & $-76\pm10$ & $99\pm25$ & $-58\pm16$ &
-1.0 \\

          & (0.6) & (-0.5) & (-0.3) & ($-76$) & ($54$) & ($-57$) \\

\\

NGC 6522  & -0.3 & -0.1 & -0.5 & $20\pm4$ & $-11\pm22$ & $-129\pm22$ &
-0.86 \\

          & (0.5) & (-0.1) & (-0.5) & ($23$) & ($-36$) & ($-127$) \\

\\

NGC 6558  & 0.1 & 0.0 & -0.8 & $184\pm3$ & $-12\pm26$ & $32\pm20$ & -0.97 \\

          & (0.9) & (0.0) & (-0.8) & ($184$) & ($-36$) & ($32$) \\

\\

NGC 6540  & 3.8 & -0.2 & -0.2 & $8\pm2$ & $-155\pm19$ & $20\pm8$ & -1.0 \\

          & (4.6) & (-0.2) & (-0.2) & ($8$) & ($-167$) & ($20$) \\

\\

AL 3      & 1.0 & 0.4 & -0.6 & - & - & - & -1.3 \\

          & (1.8) & (0.4) & (-0.6) & (-) & (-) & (-)  \\

\\

ESO 456--SC38     & 0.8 & 0.3 & -0.3 & - & - & - & -0.5 \\

                  & (1.6) & (0.3) & (-0.3) & (-) & (-) & (-)  \\

\\

Palomar 6  & 0.2 & -0.3 & 0.2 & $-195\pm3$ & $85\pm23$ & $76\pm18$ & -1.0 \\

          & (1.0) & (-0.3) & (0.2) & ($-195$) & ($62$) & ($77$) \\

          \\

NGC 6652  & -1.9 & -0.3 & -1.9 & $150\pm8$ & $169\pm19$ & $-256\pm28$ & -0.81 \\

          & (-1.1) & (-0.3) & (-1.9) & ($150$) & ($138$) & ($-255$) \\

\\

\hline

\end{tabular}

\caption{Kinematical properties. Upper panel: input data; lower panel: derived Galactocentric vectors and
velocities, and literature metallicities; first and second lines correspond to distance of the Sun to
the Galactic center of 7.5 kpc and 8.3 kpc.}

\label{tab:all_clusters}

\end{table*}

\begin{figure*}
\centering
\begin{tabular}{c c c}
\includegraphics[scale=0.35]{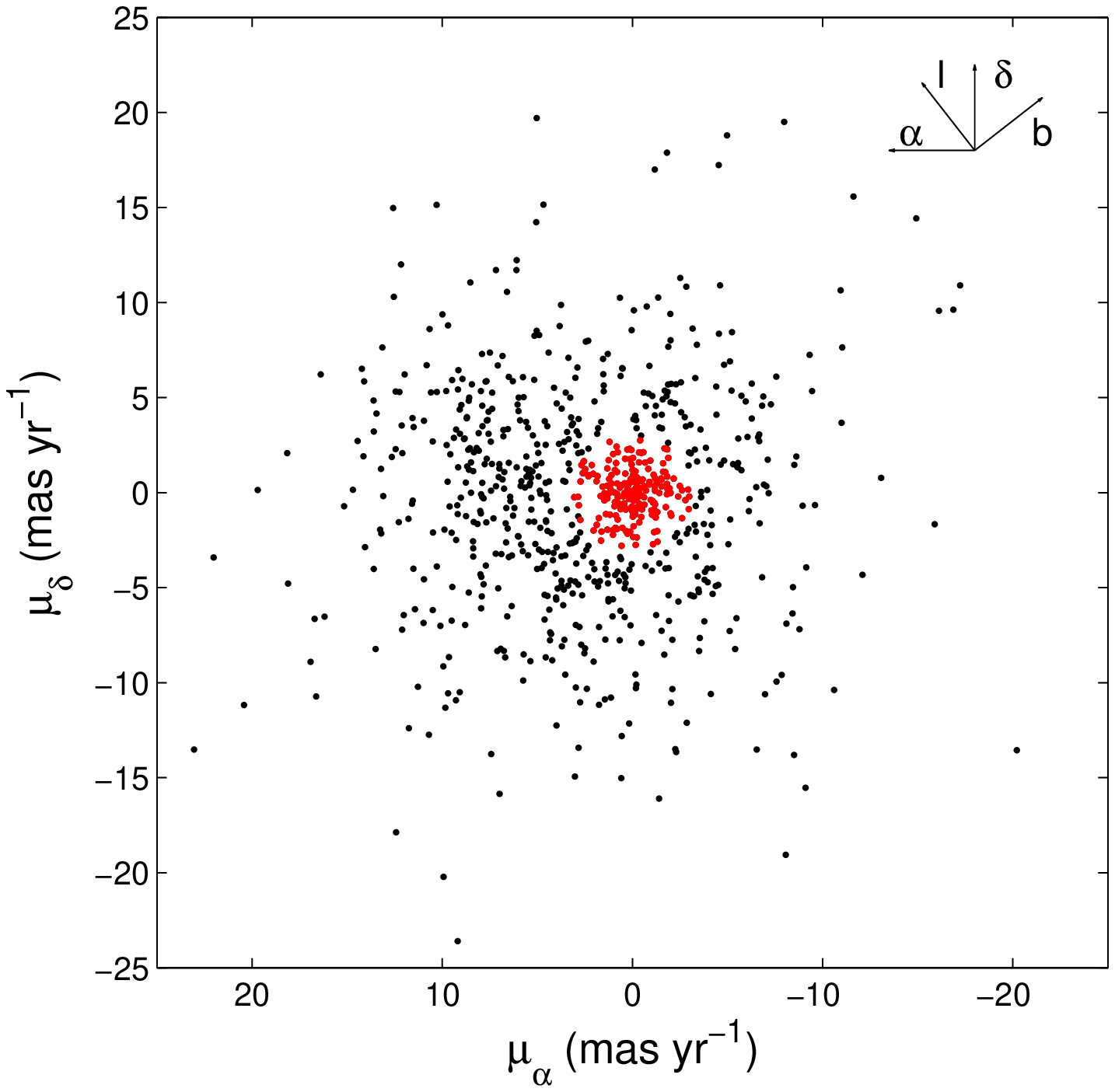} 
\includegraphics[scale=0.35]{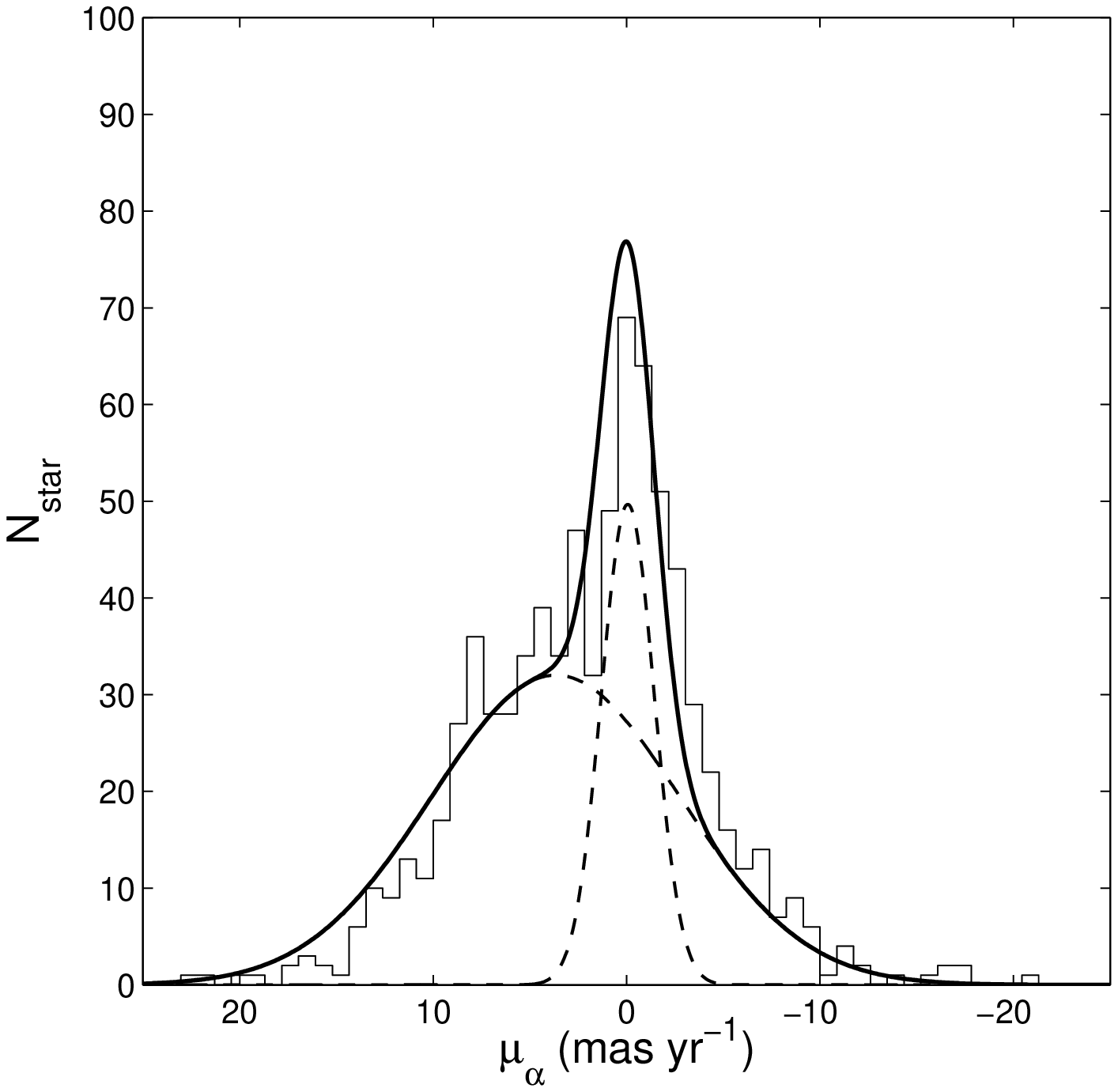} 
\includegraphics[scale=0.35]{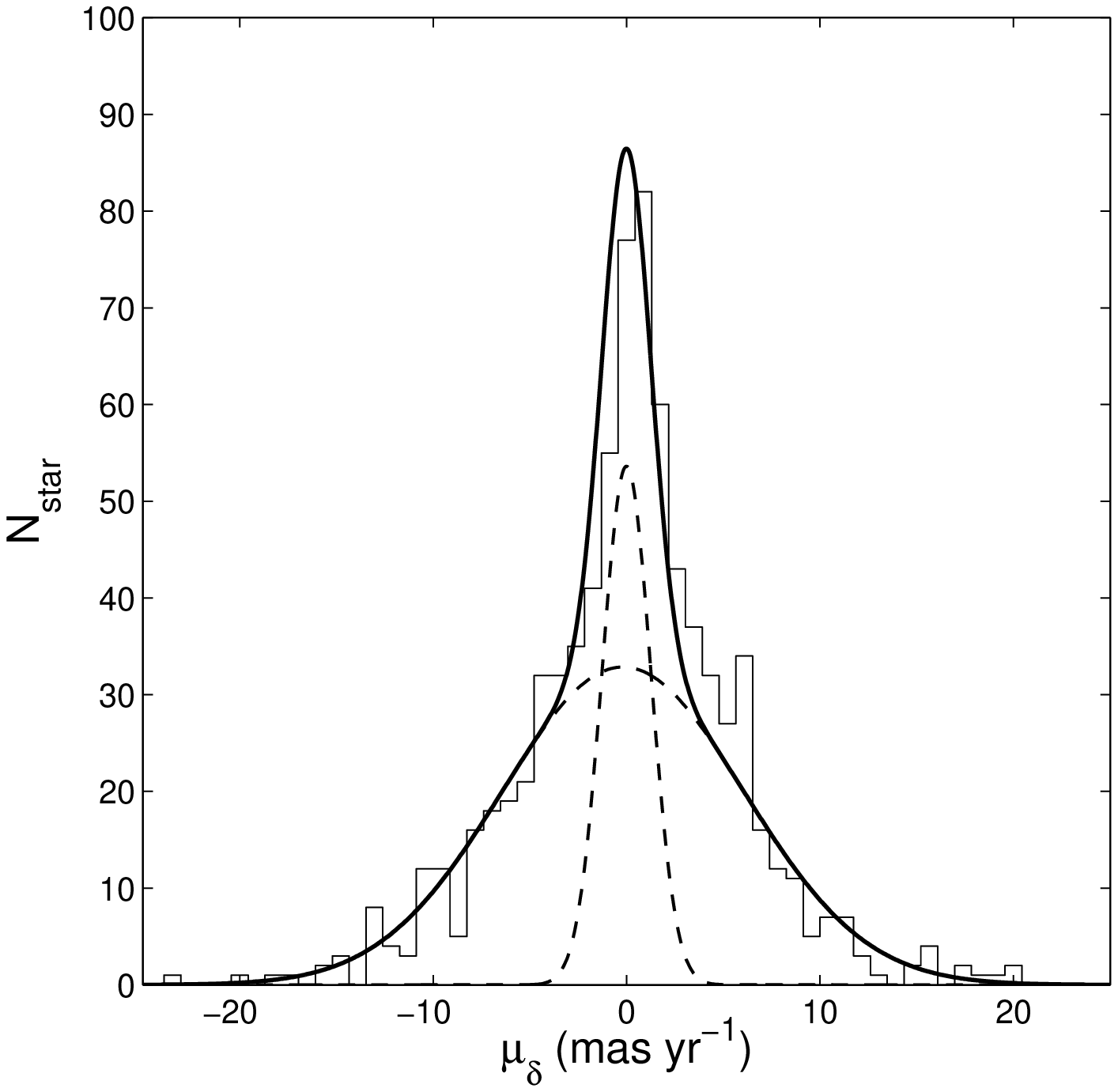}\\
\includegraphics[scale=0.35]{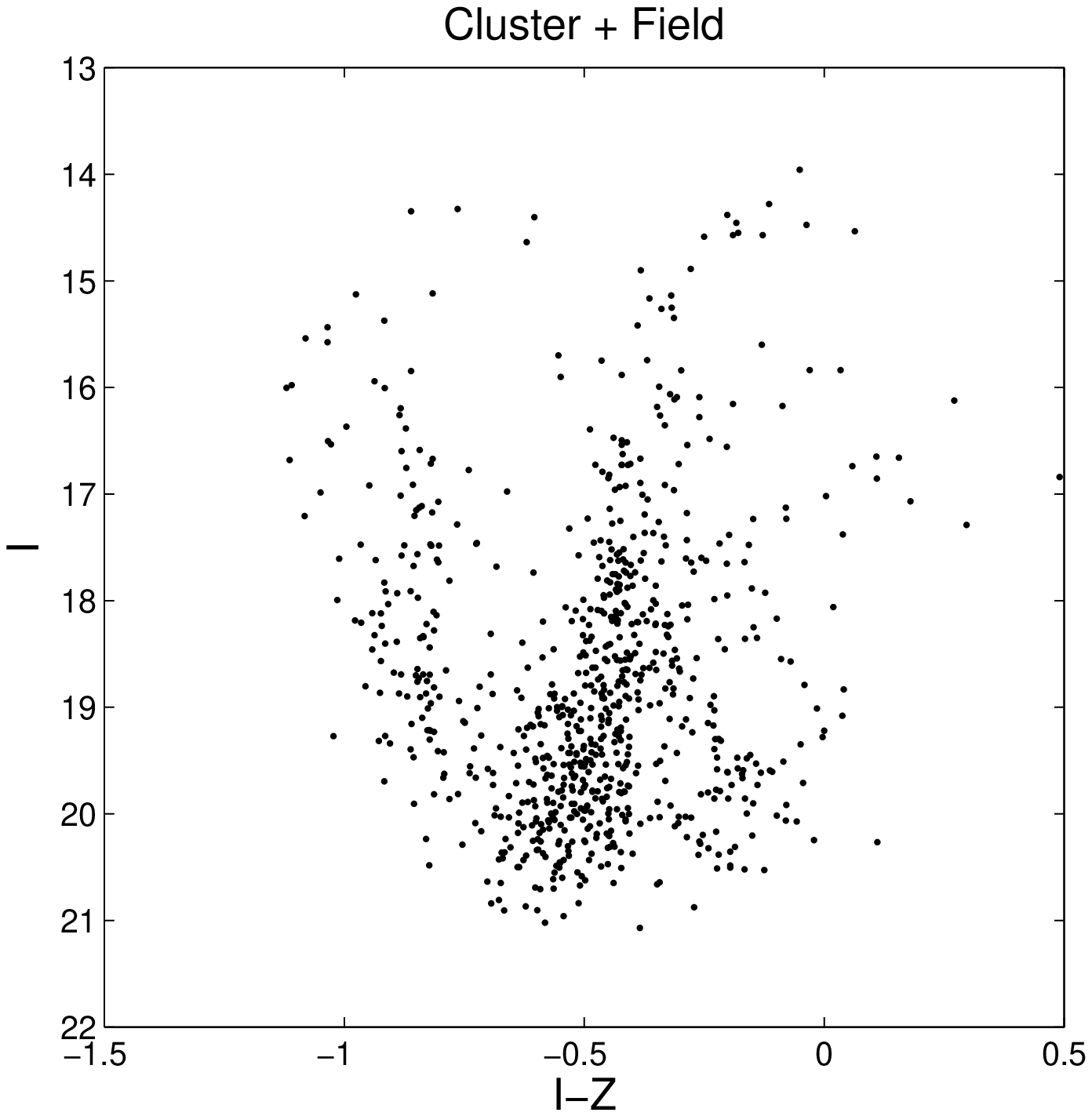} 
\includegraphics[scale=0.35]{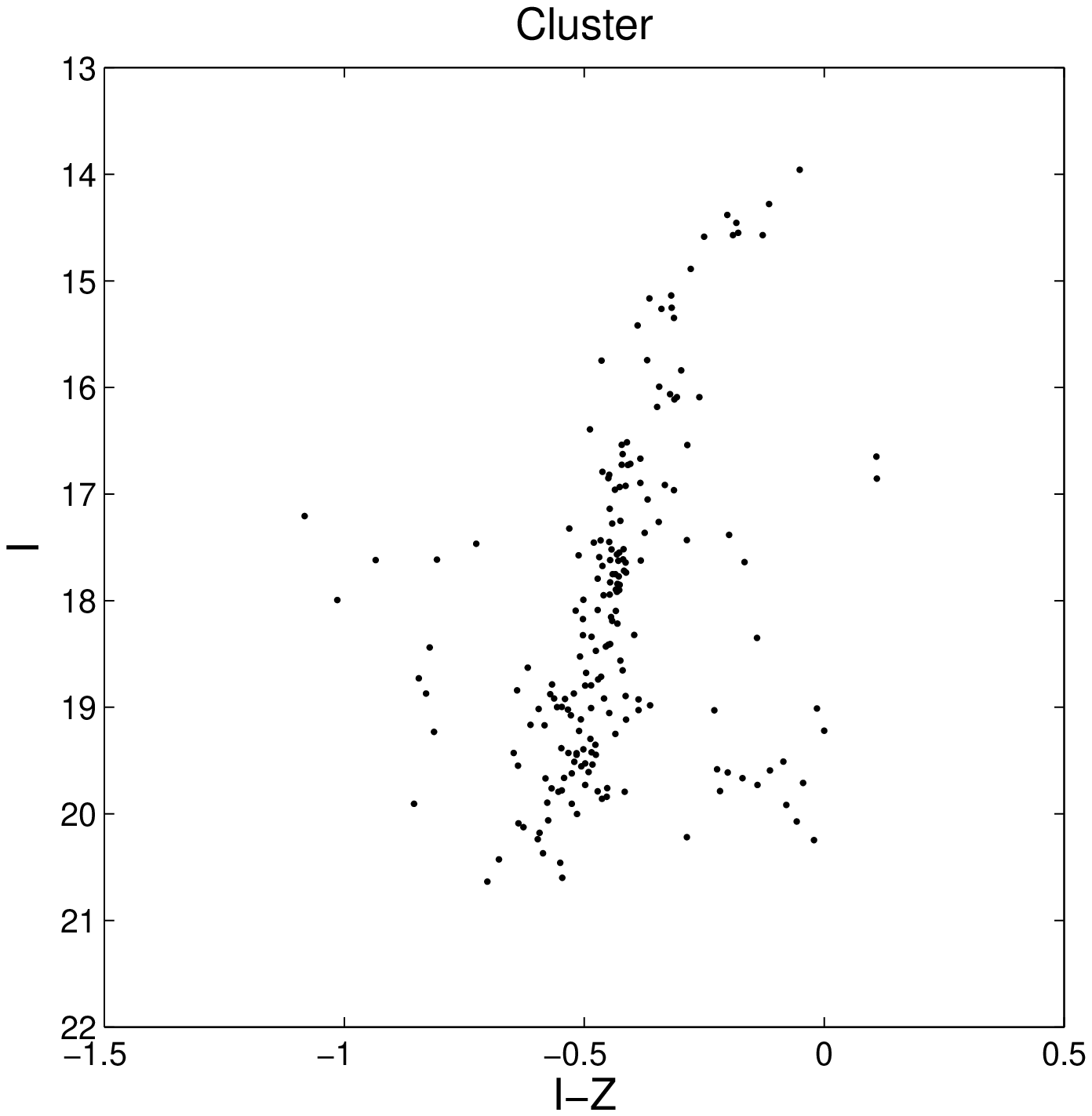} 
\includegraphics[scale=0.35]{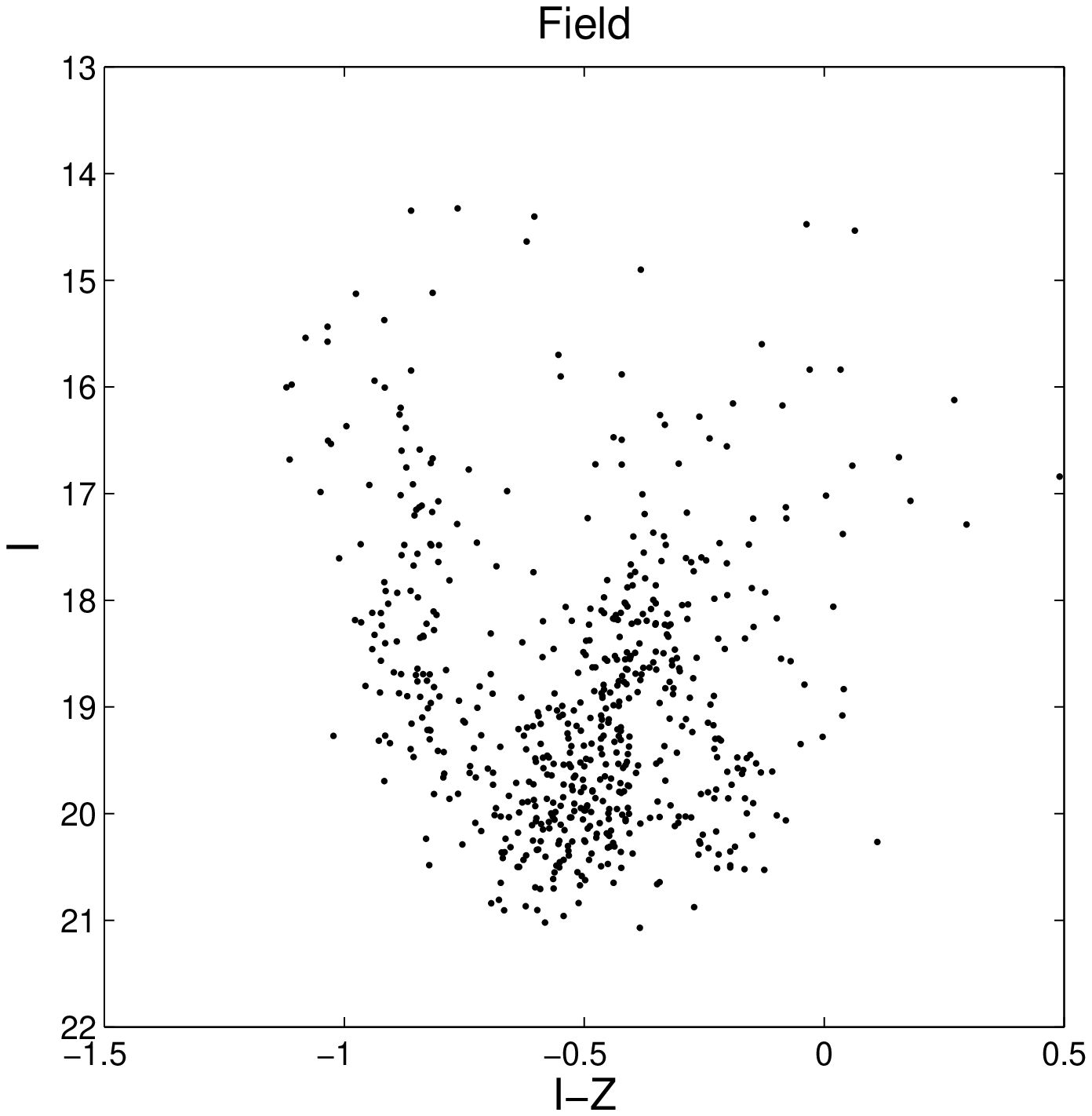}
\end{tabular}
\caption{Proper motion and CMD decontamination of Terzan 4.}
\label{fig:Terzan4pm}
\end{figure*}

\begin{figure*}
\centering
\begin{tabular}{c c c}
\includegraphics[scale=0.35]{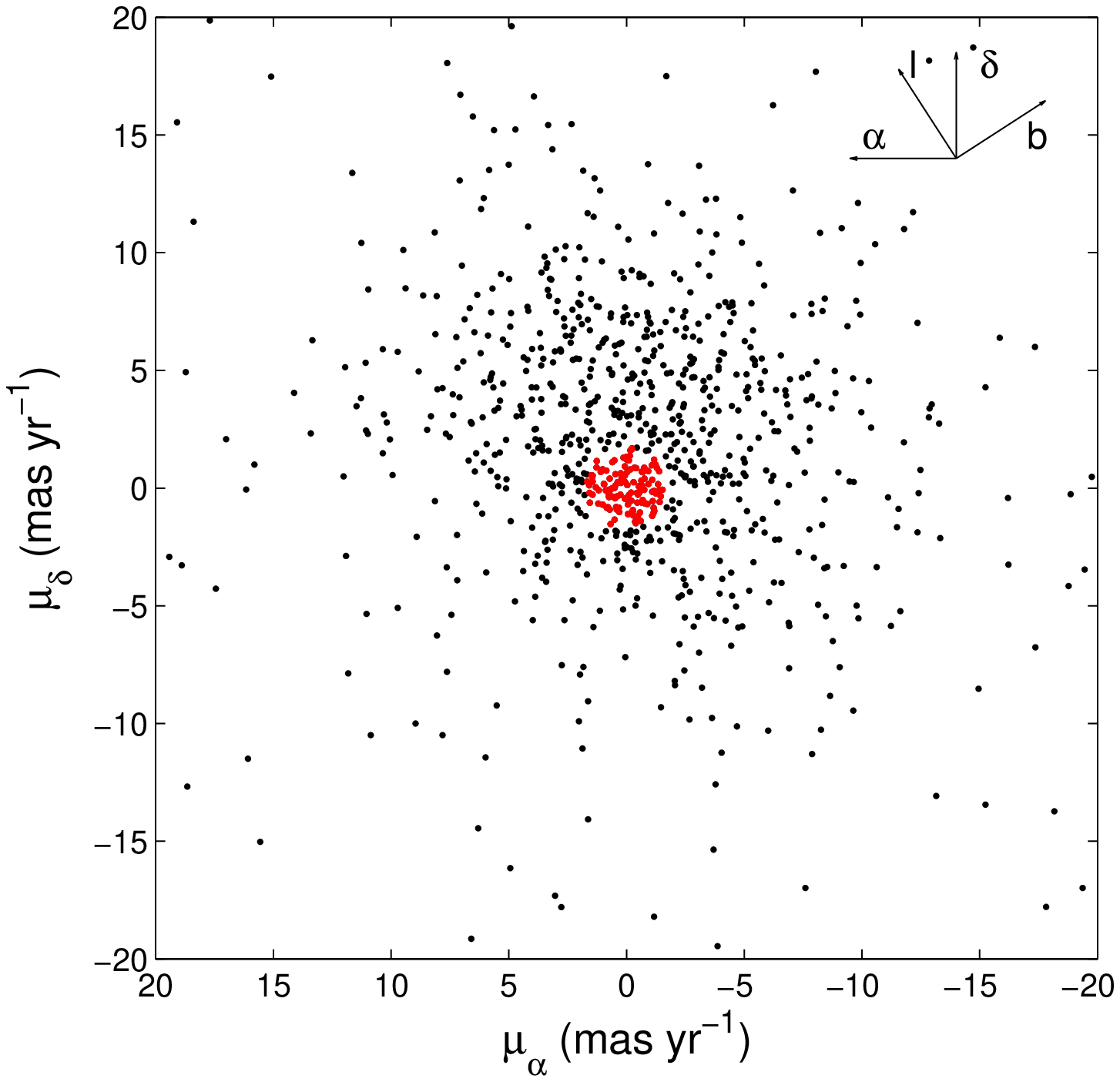} 
\includegraphics[scale=0.35]{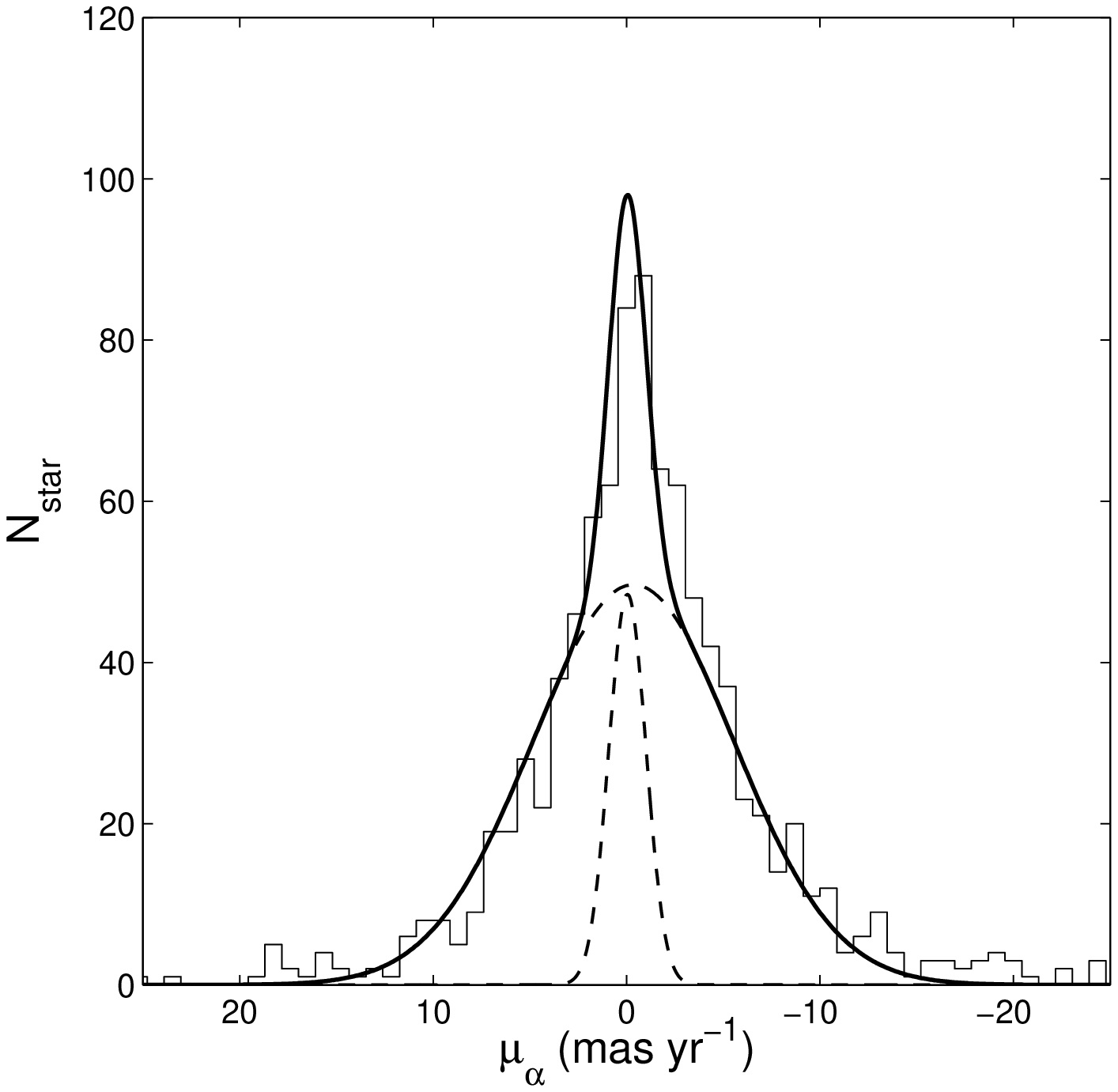} 
\includegraphics[scale=0.35]{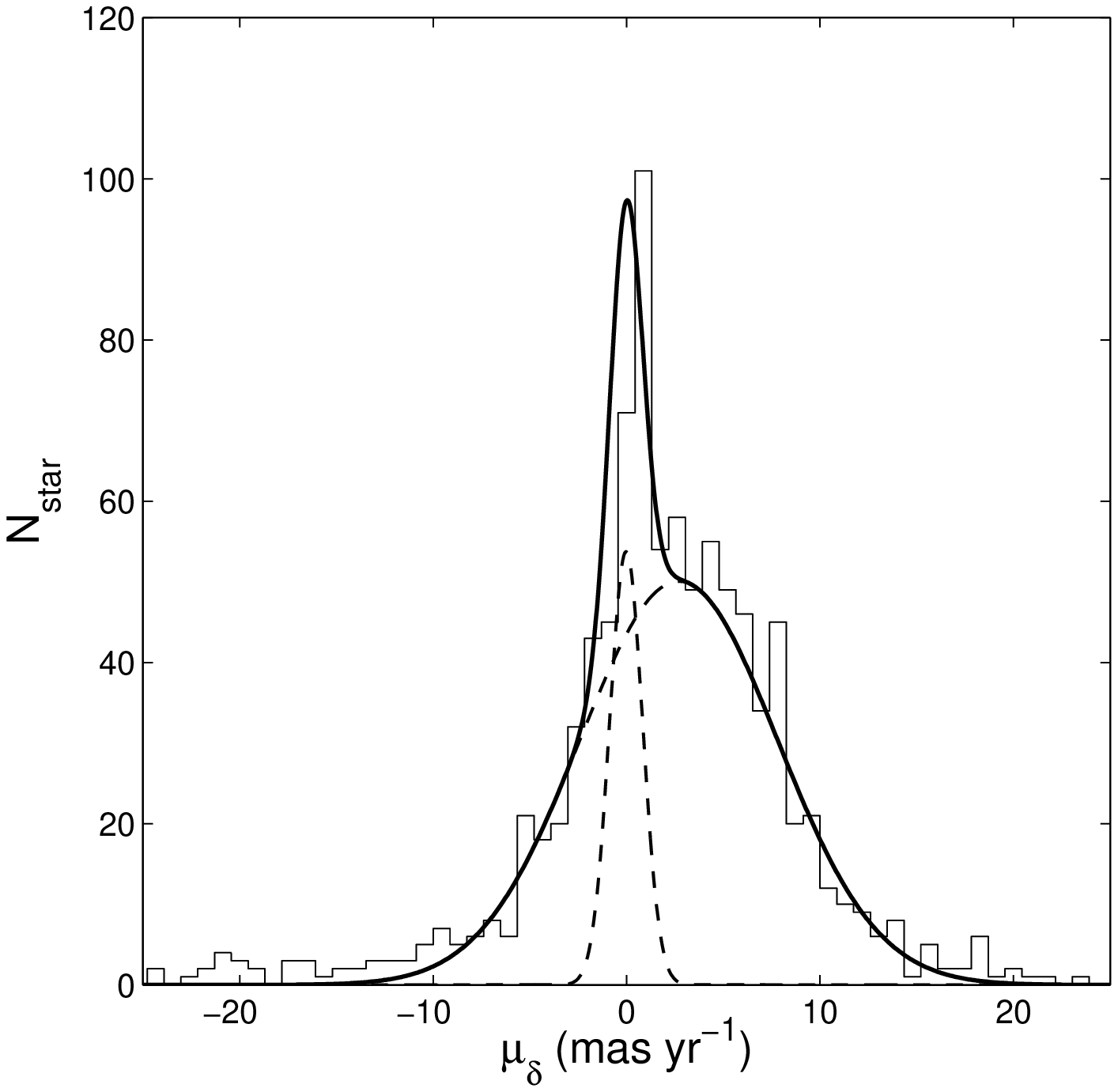}\\
\includegraphics[scale=0.35]{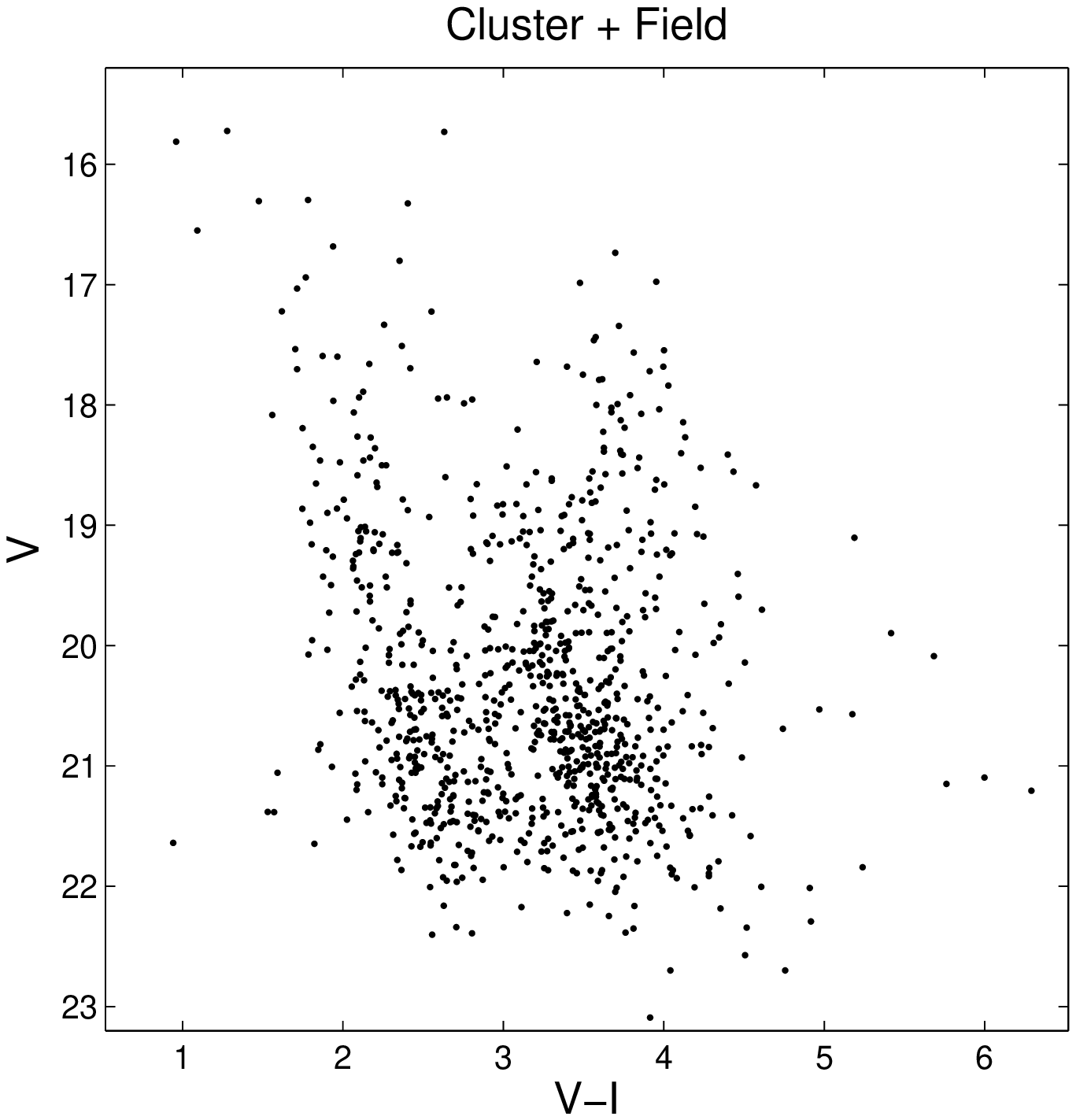} 
\includegraphics[scale=0.35]{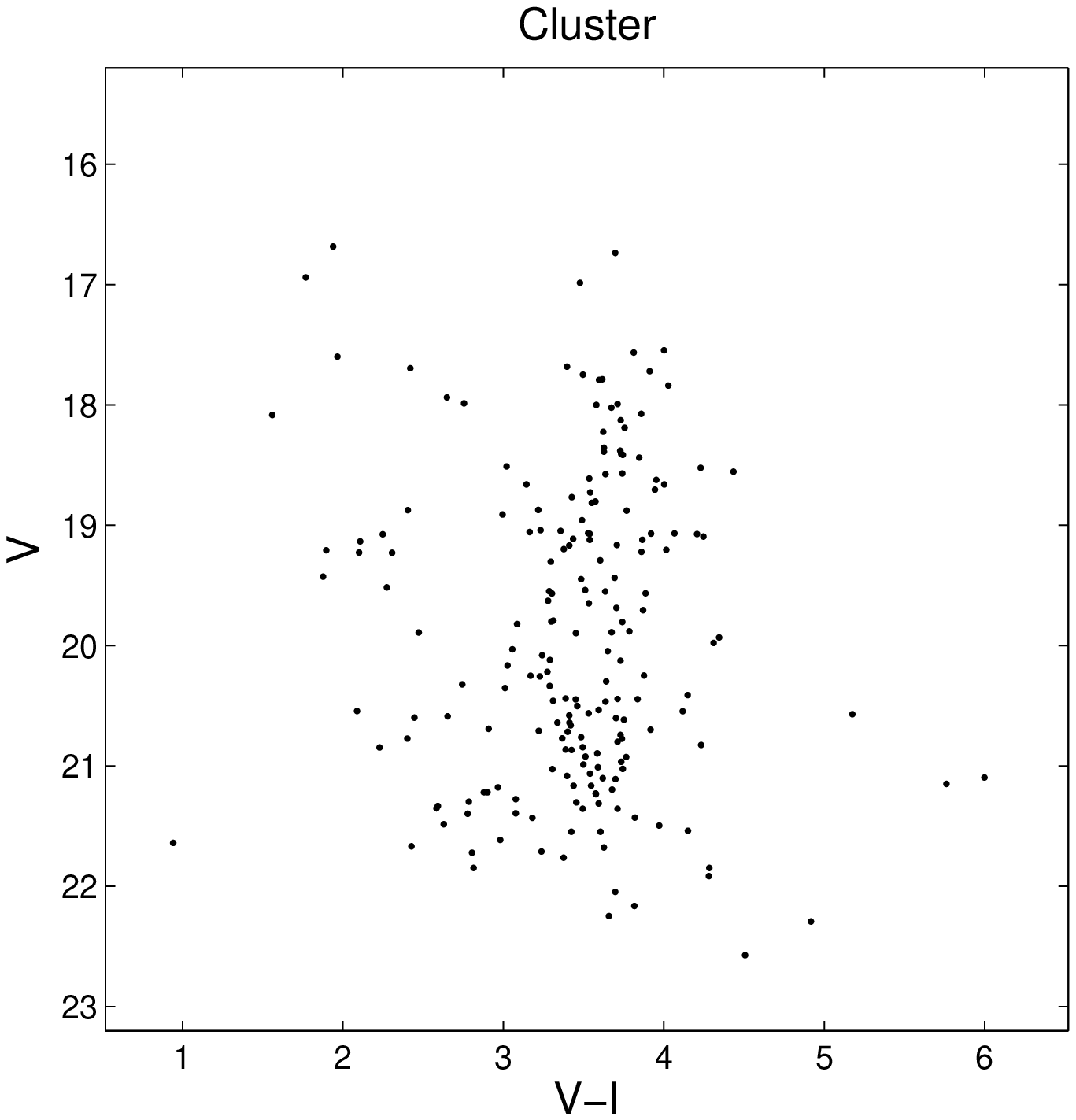} 
\includegraphics[scale=0.35]{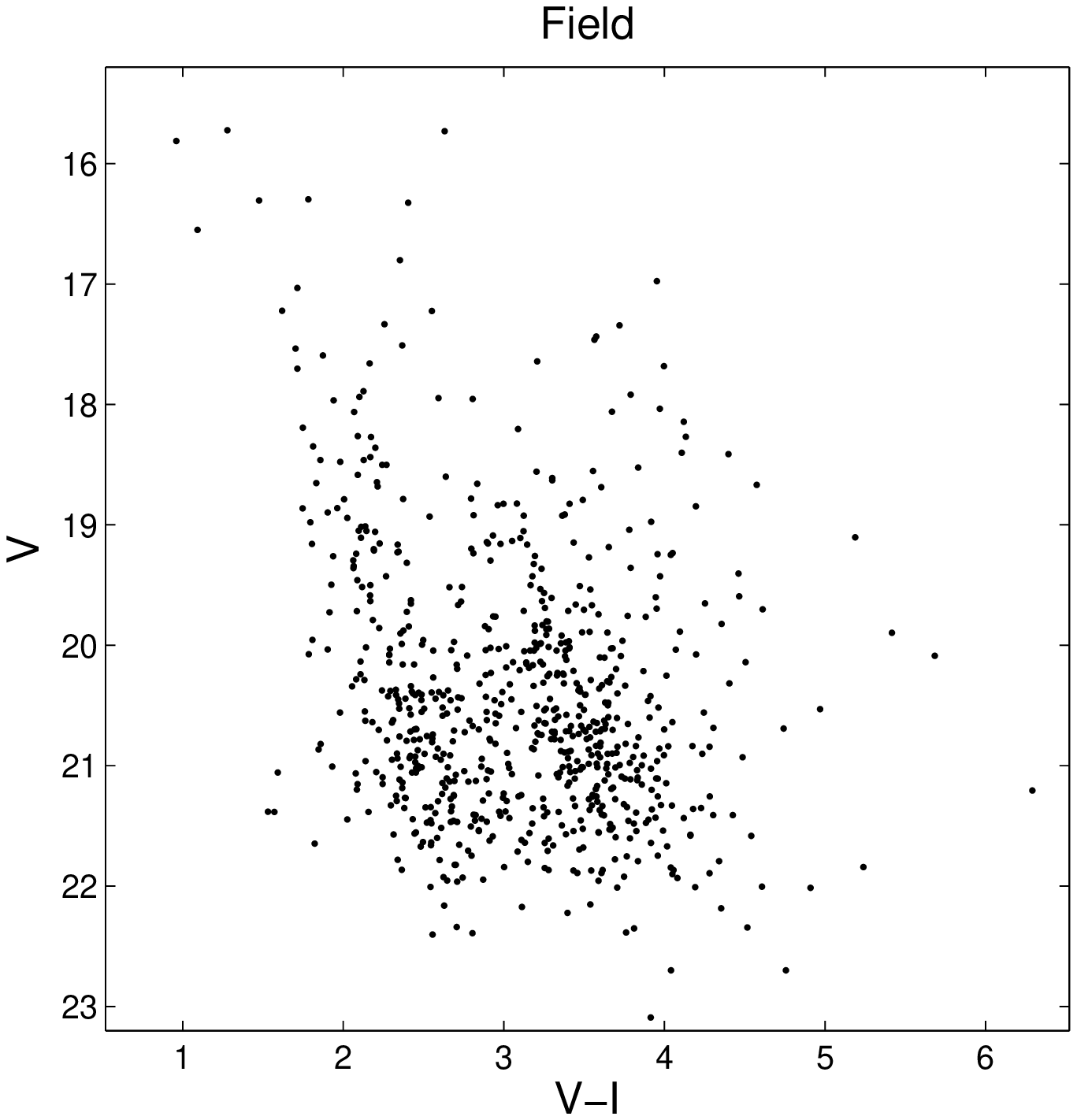}
\end{tabular}
\caption{Proper motion and CMD decontamination of Terzan 9.}
\label{fig:Terzan9pm}
\end{figure*}

\begin{figure*}
\centering
\begin{tabular}{c c c}
\includegraphics[scale=0.35]{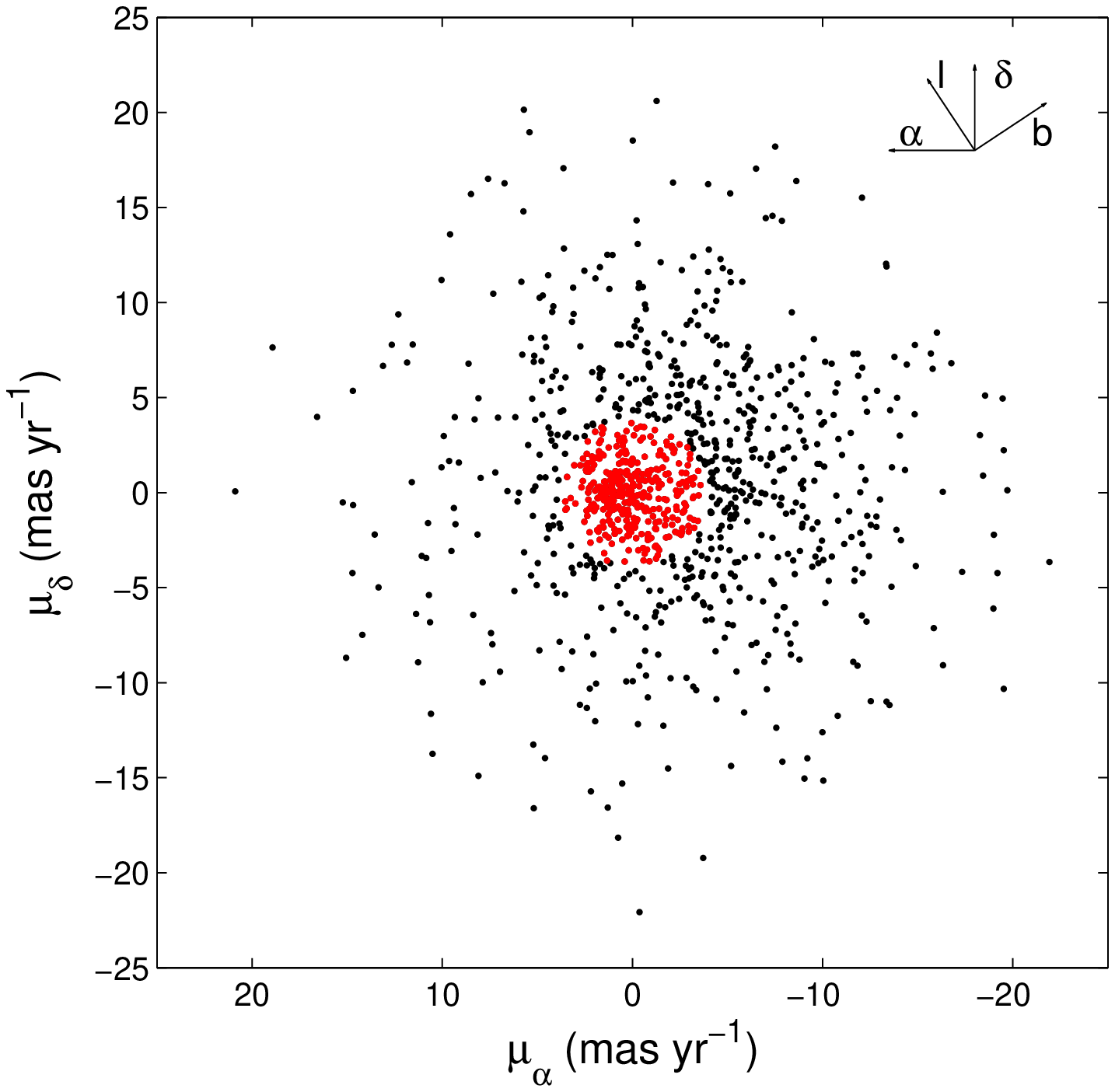} 
\includegraphics[scale=0.35]{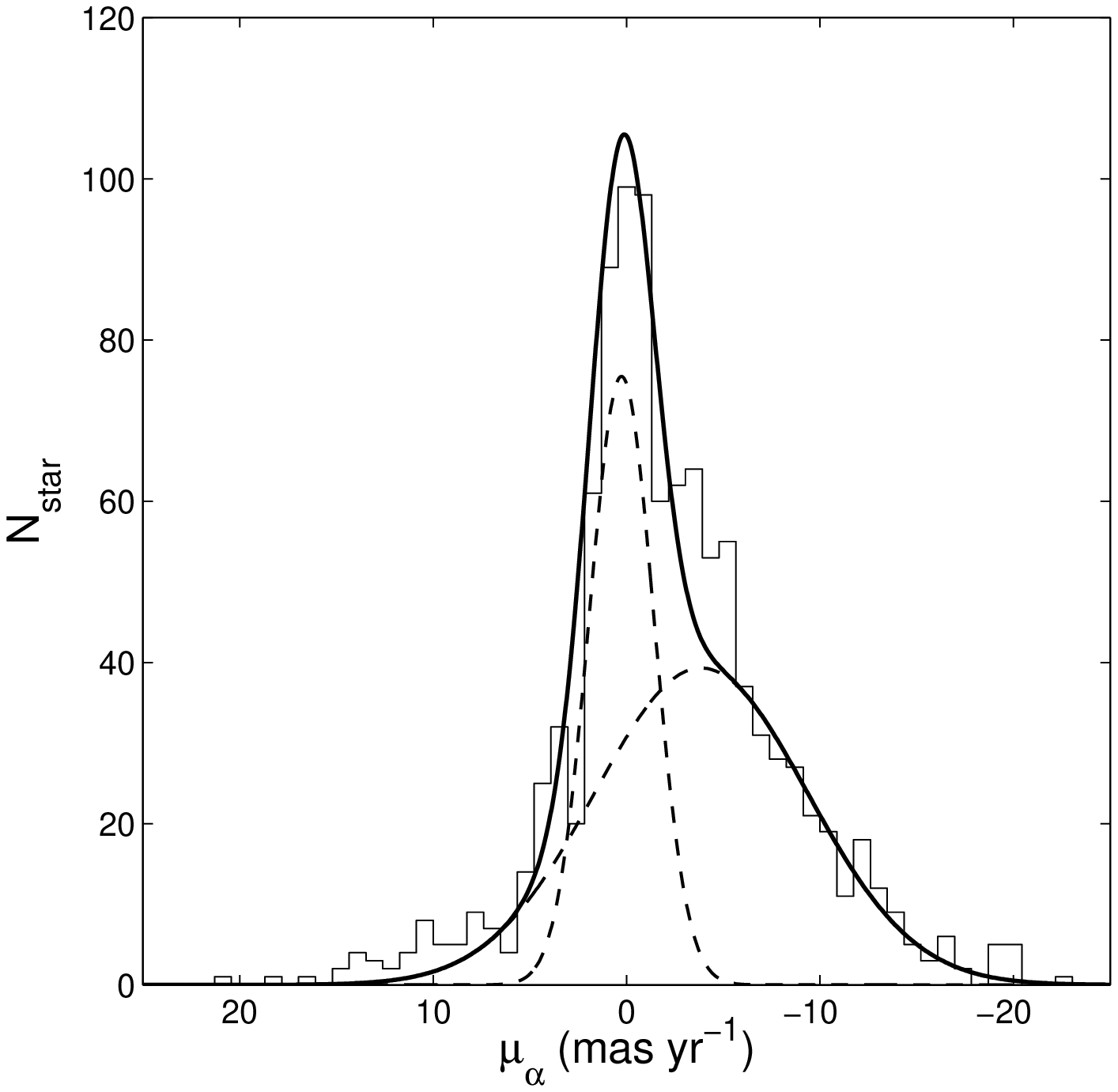} 
\includegraphics[scale=0.35]{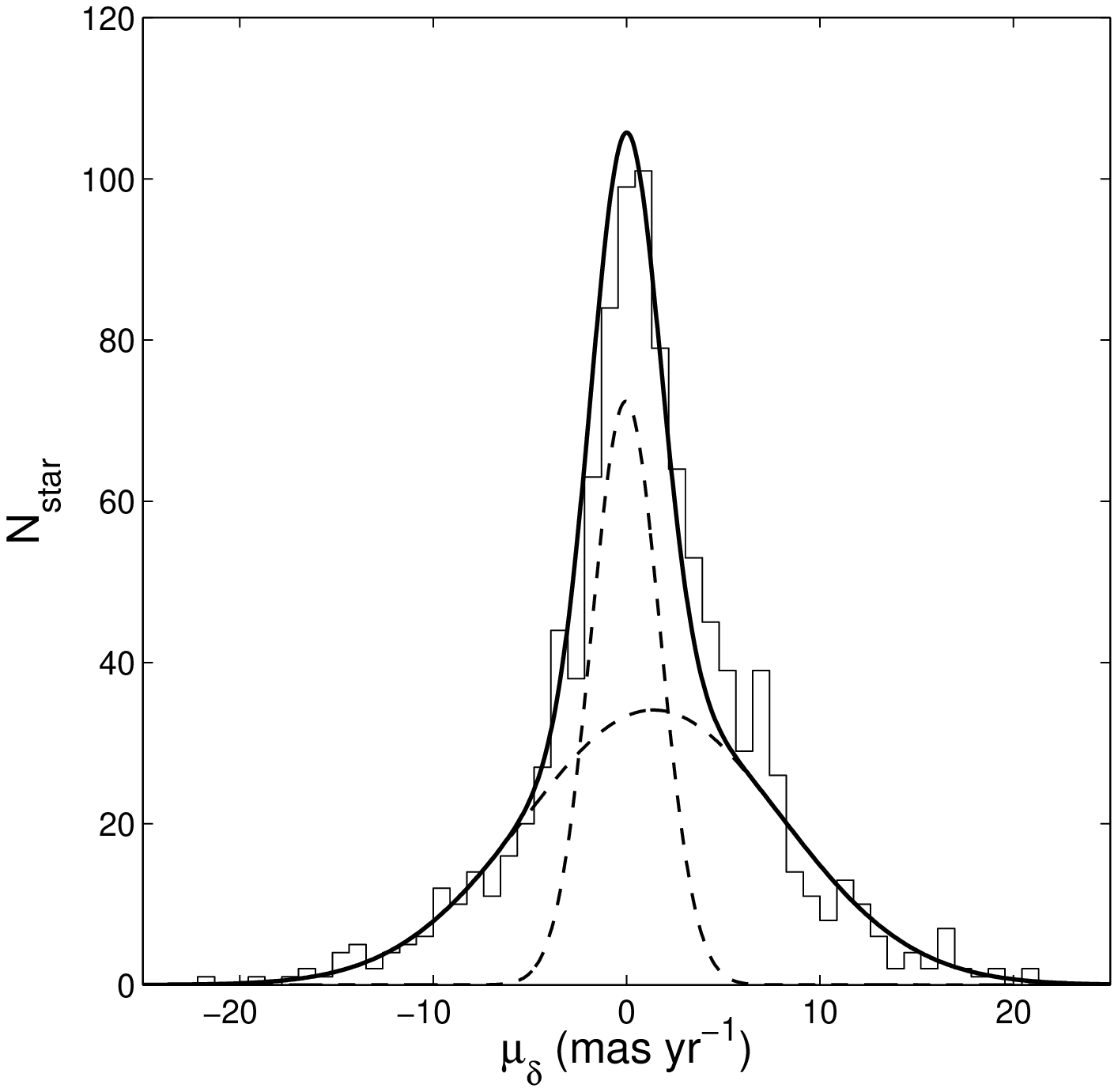}\\
\includegraphics[scale=0.35]{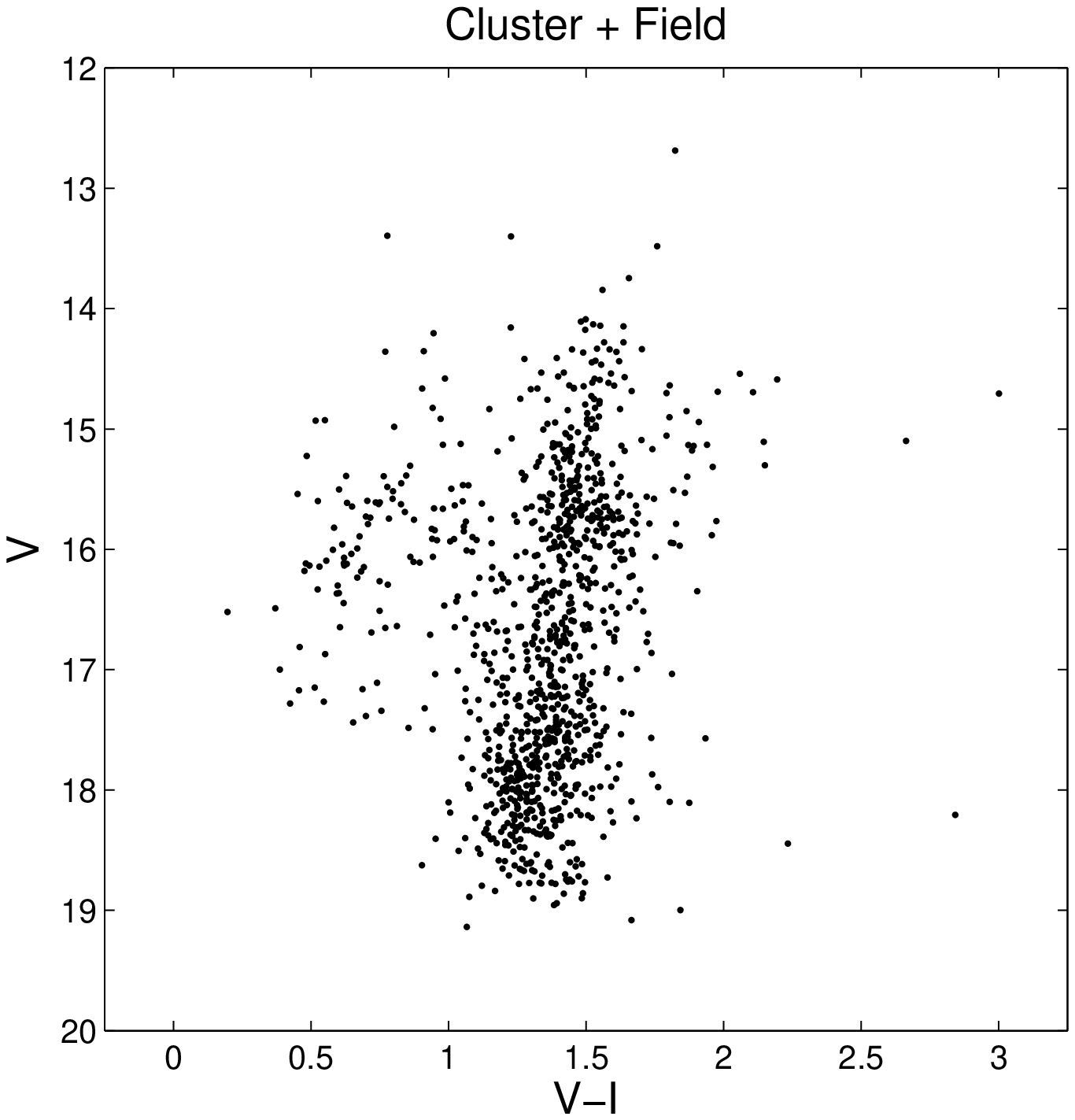} 
\includegraphics[scale=0.35]{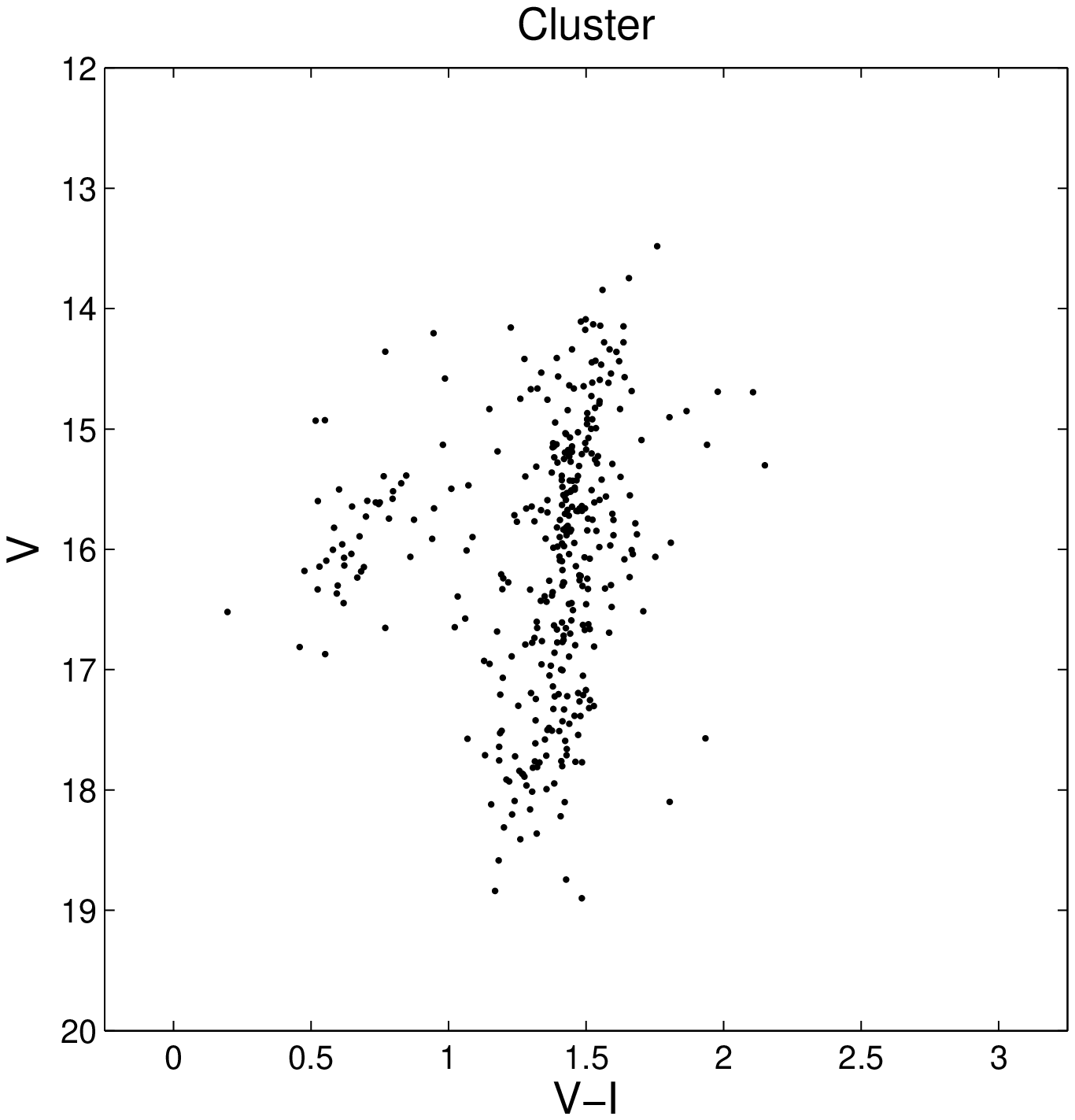} 
\includegraphics[scale=0.35]{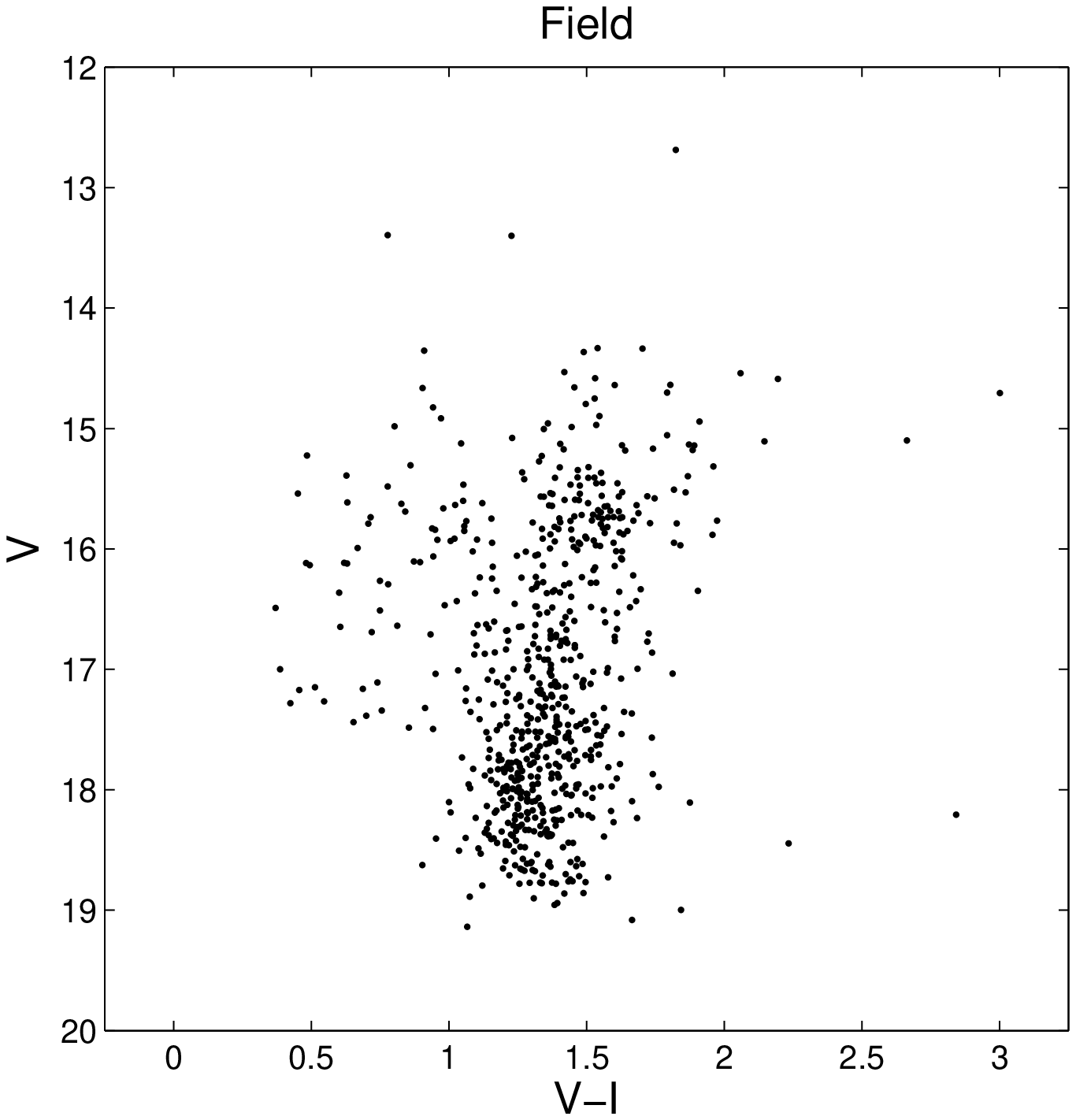}
\end{tabular}
\vskip 10mm
\begin{tabular}{c c c}
\includegraphics[scale=0.35]{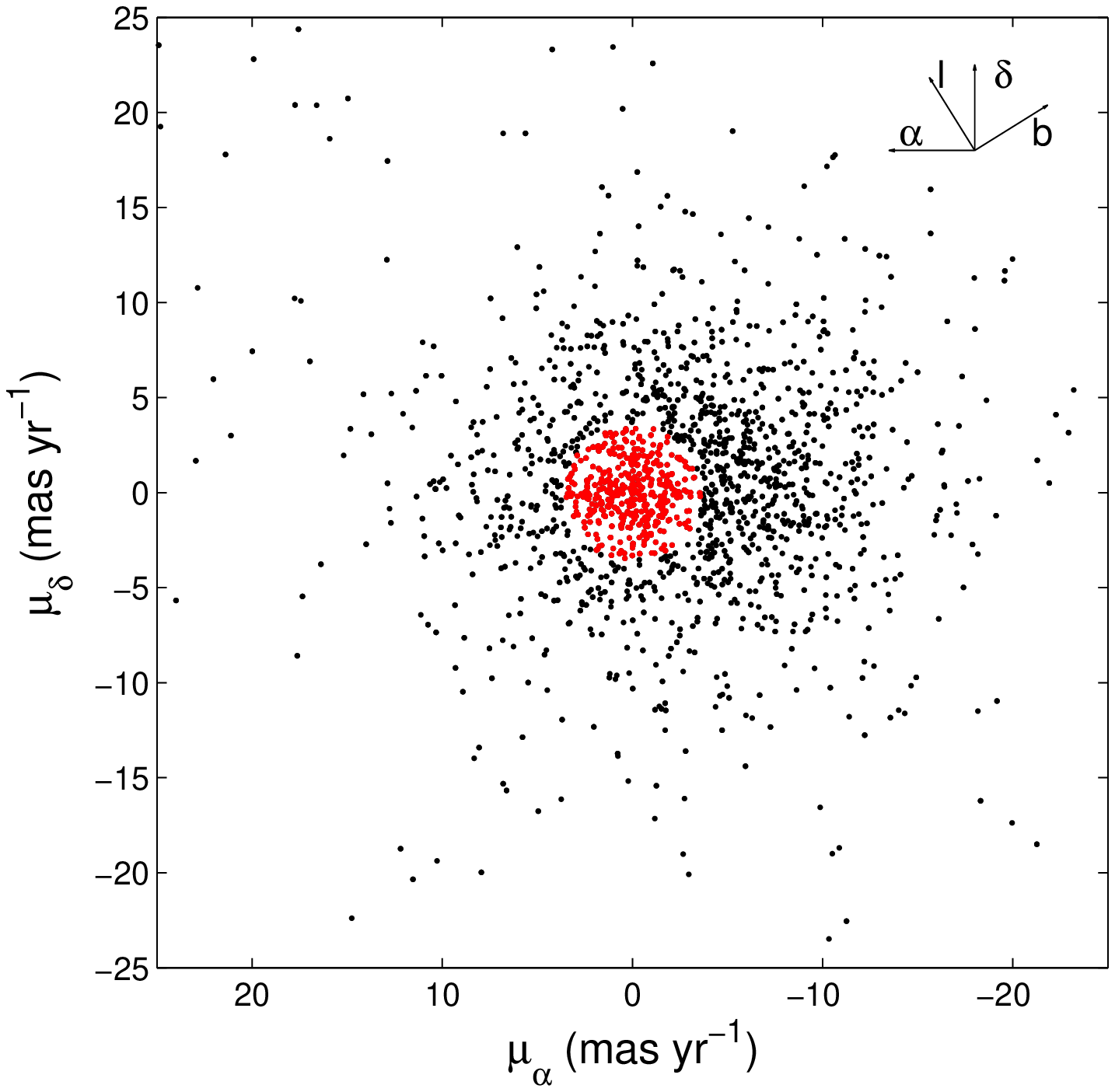} 
\includegraphics[scale=0.35]{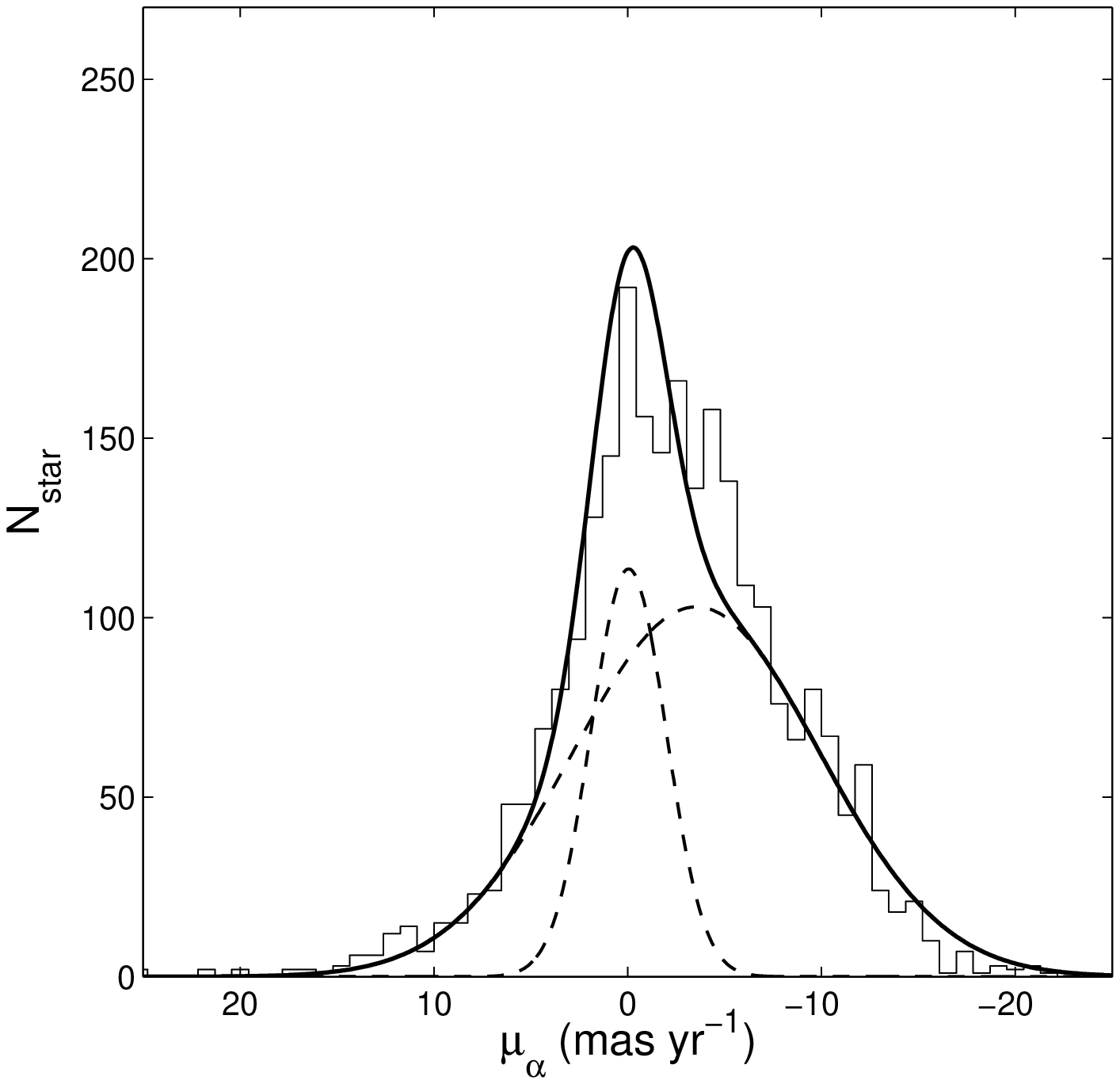} 
\includegraphics[scale=0.35]{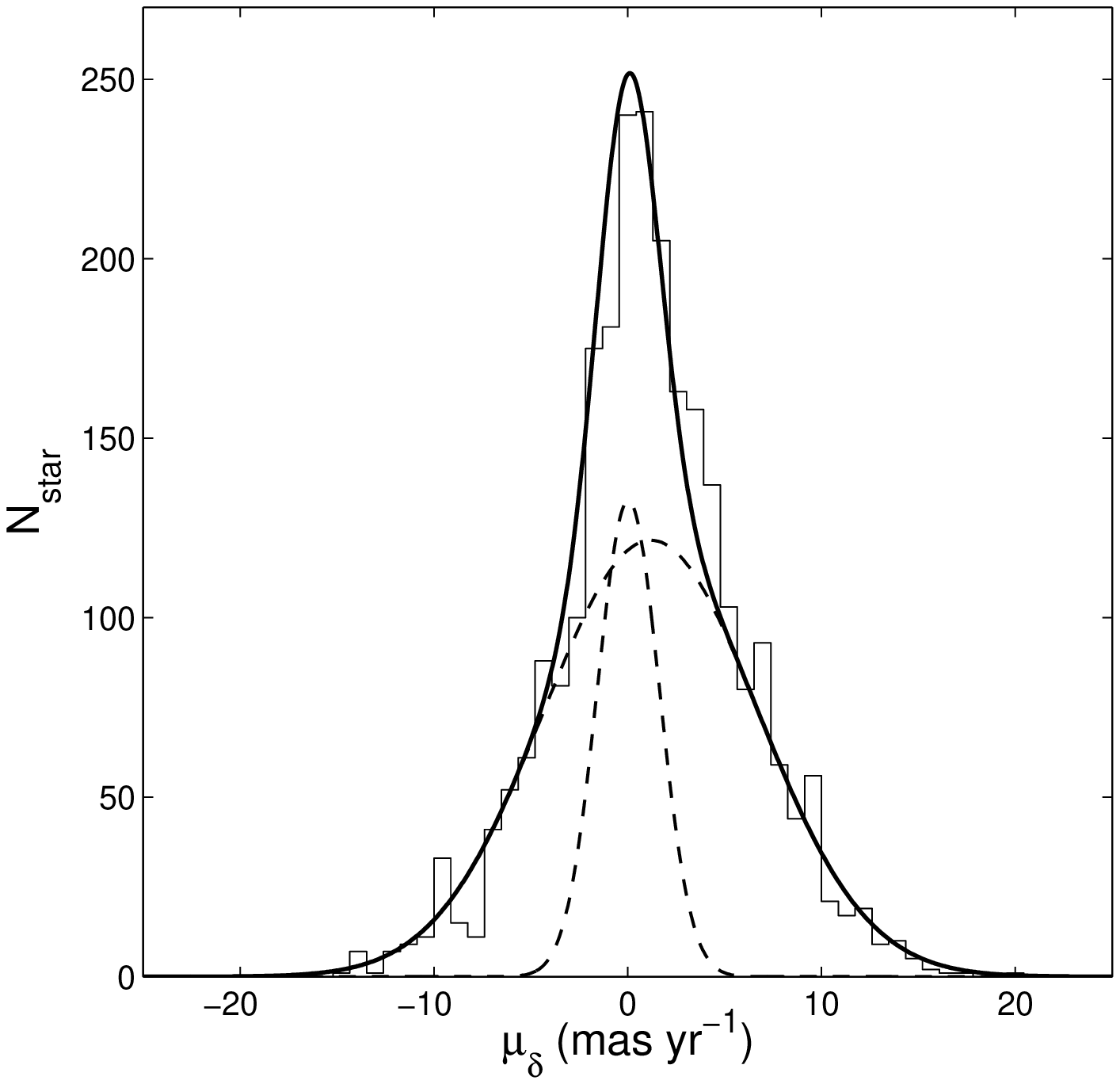}\\
\includegraphics[scale=0.35]{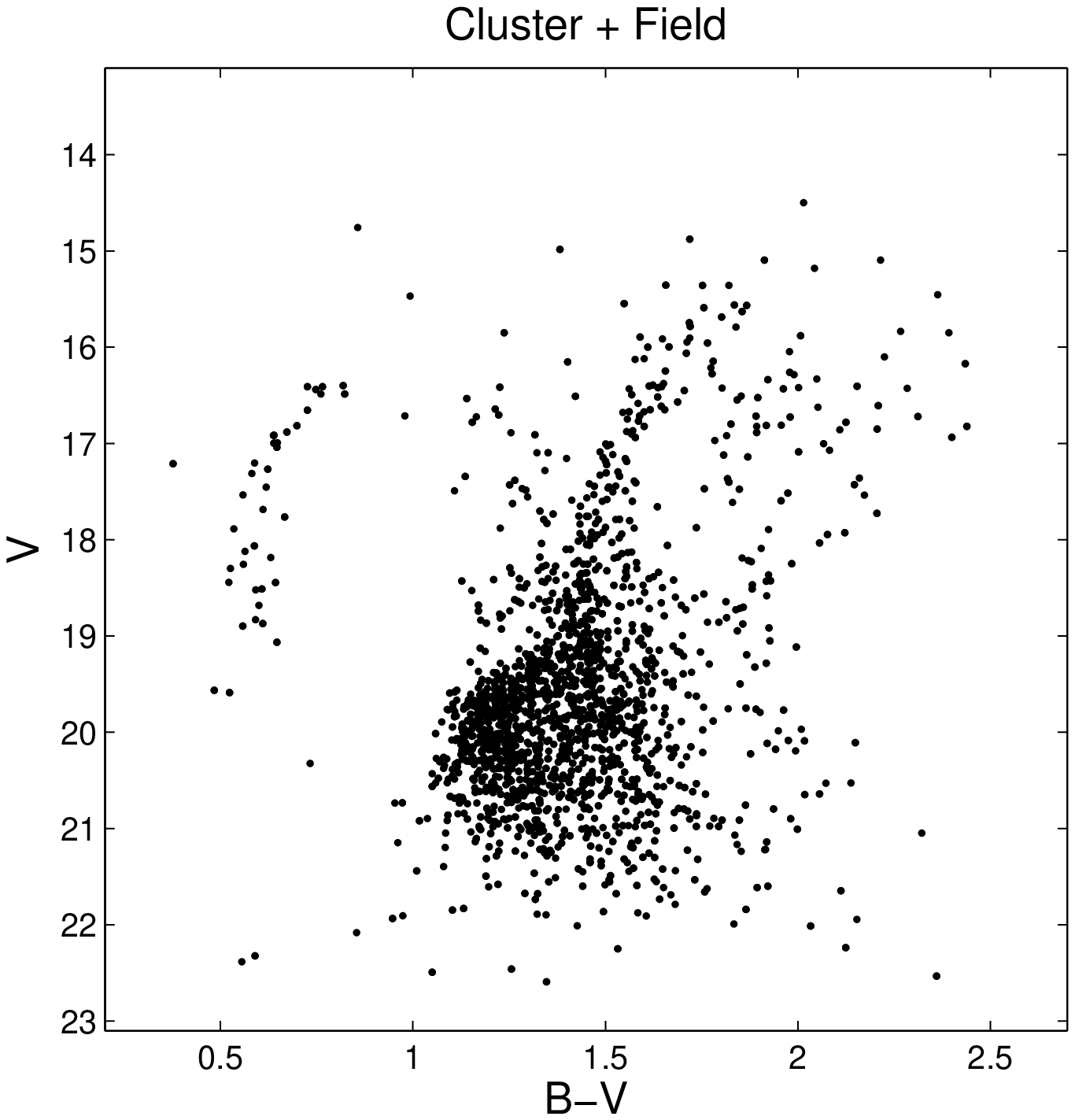} 
\includegraphics[scale=0.35]{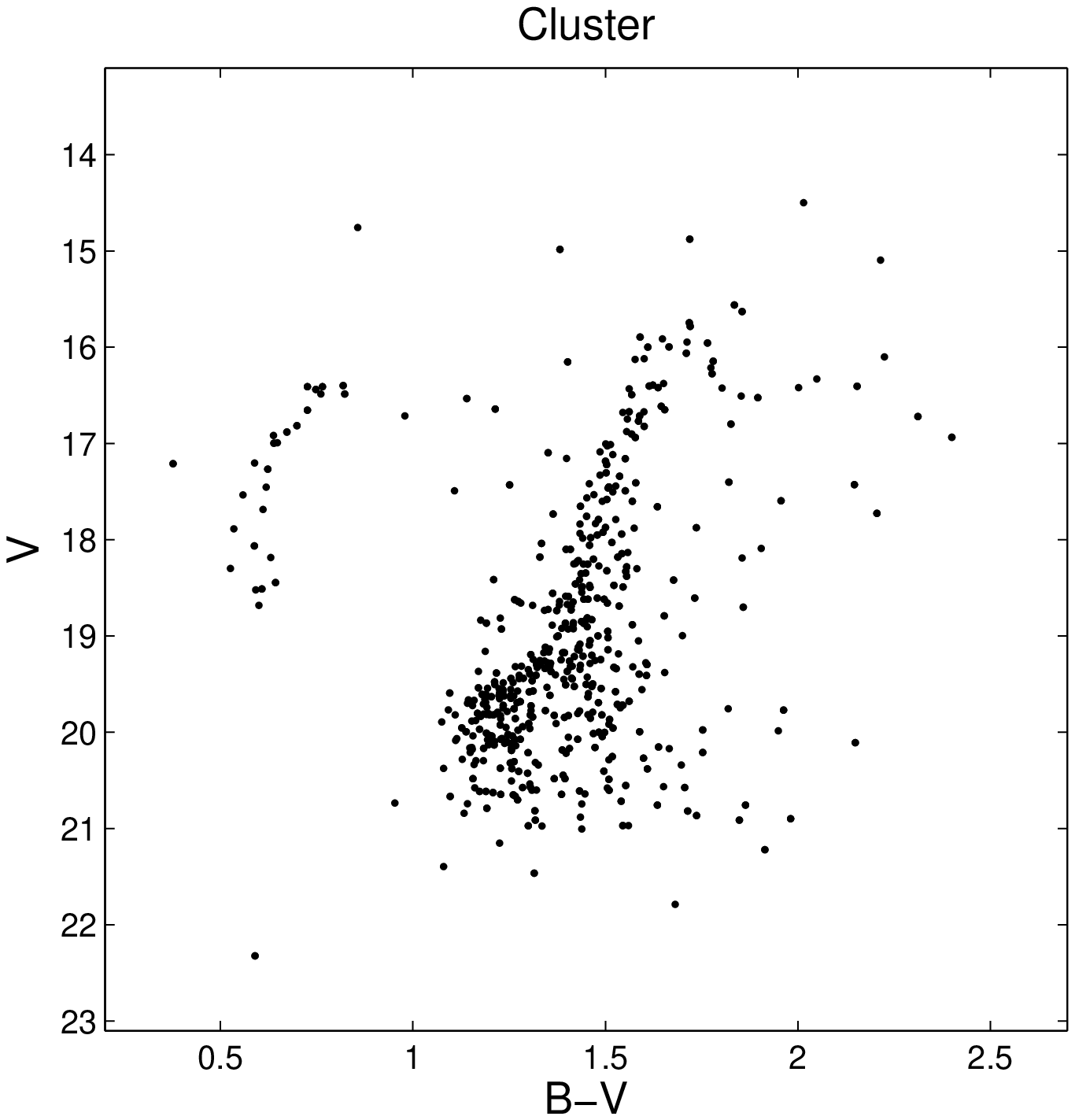} 
\includegraphics[scale=0.35]{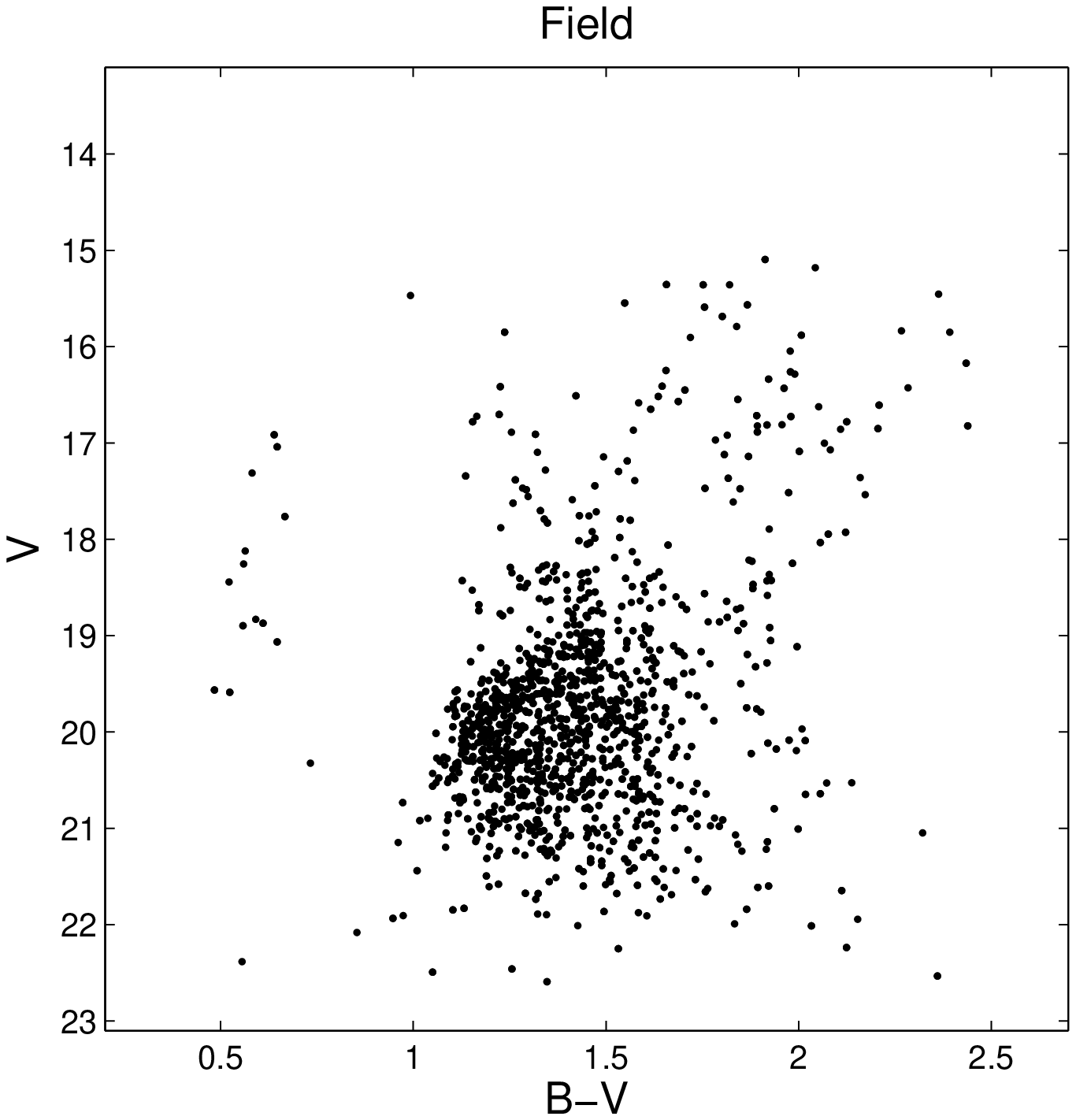}
\end{tabular}
\caption{Proper motion and CMD decontamination of NGC 6522 from Danish Telescope--NTT and NTT--HST data.}
\label{fig:NGC6522pm}
\end{figure*}

\begin{figure*}
\centering
\begin{tabular}{c c c}
\includegraphics[scale=0.35]{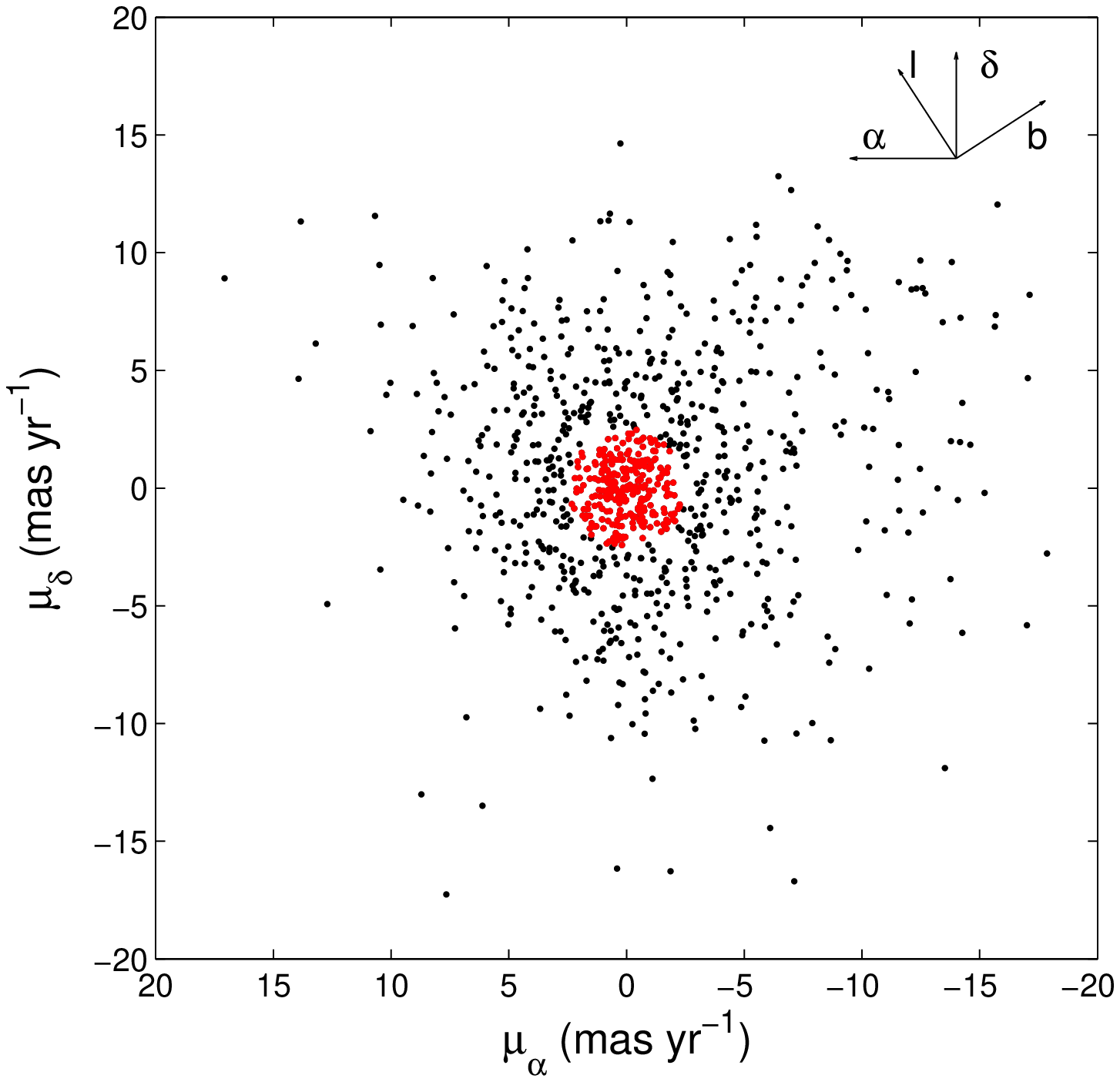} 
\includegraphics[scale=0.35]{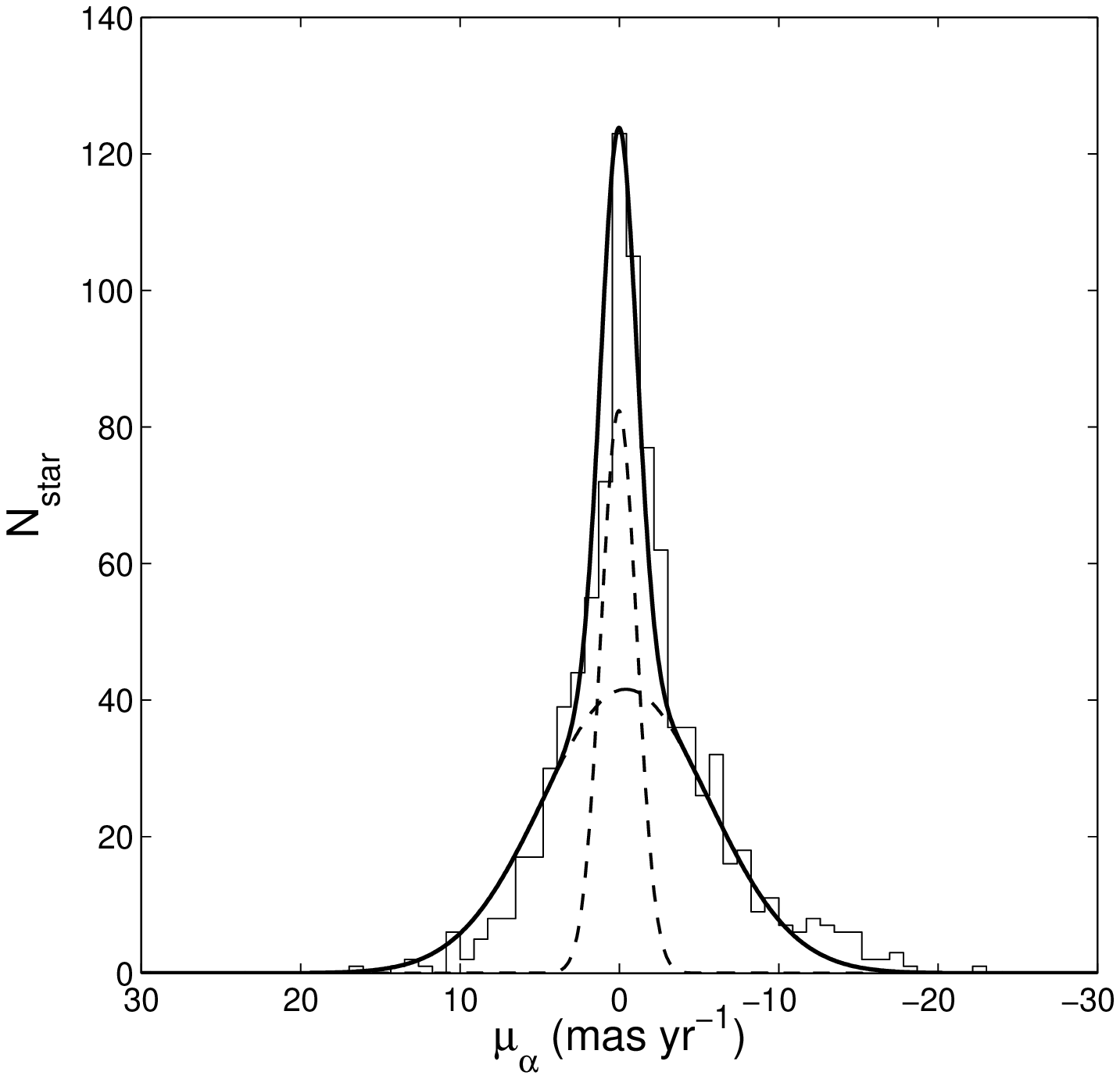} 
\includegraphics[scale=0.35]{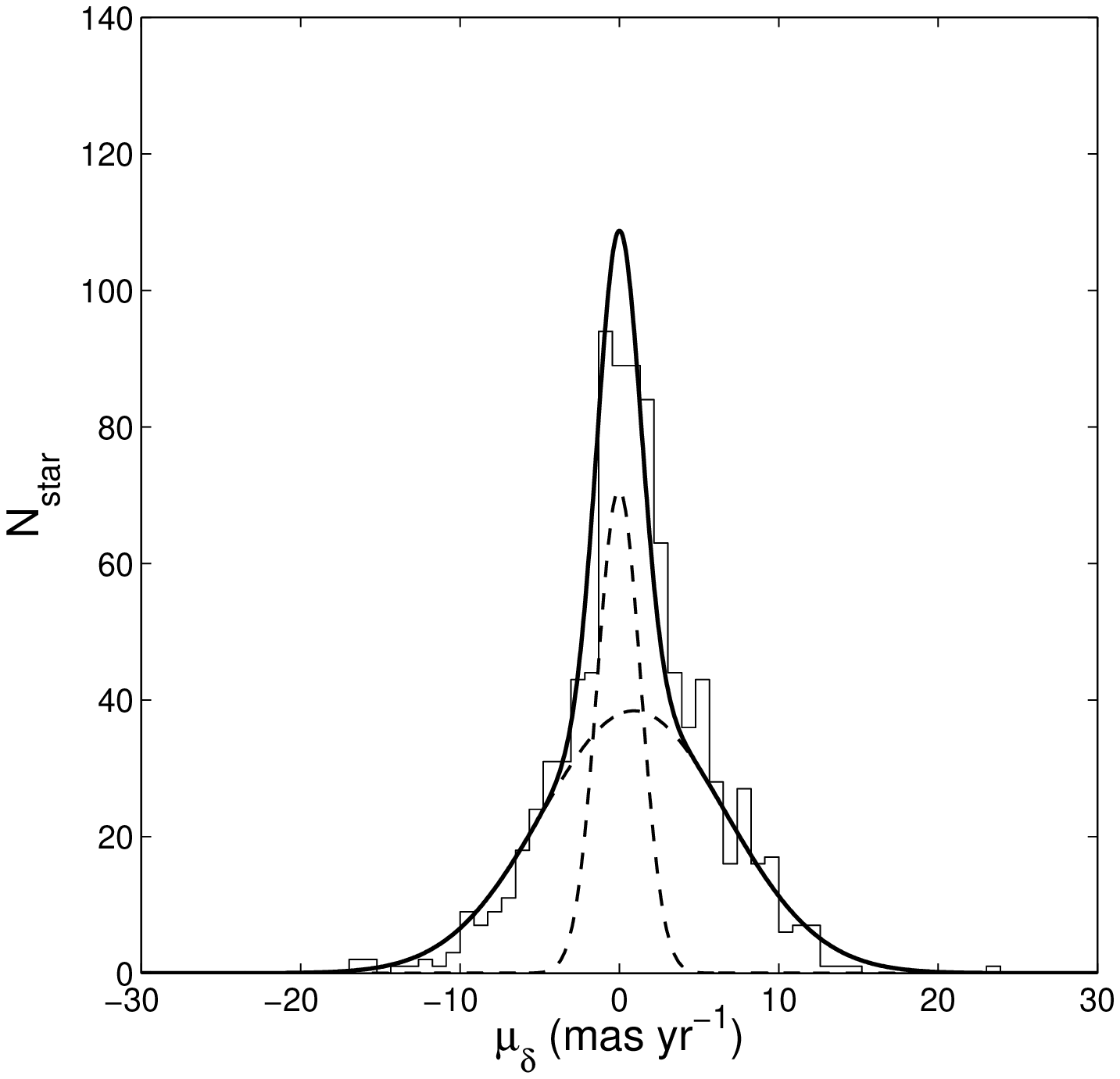}\\
\includegraphics[scale=0.35]{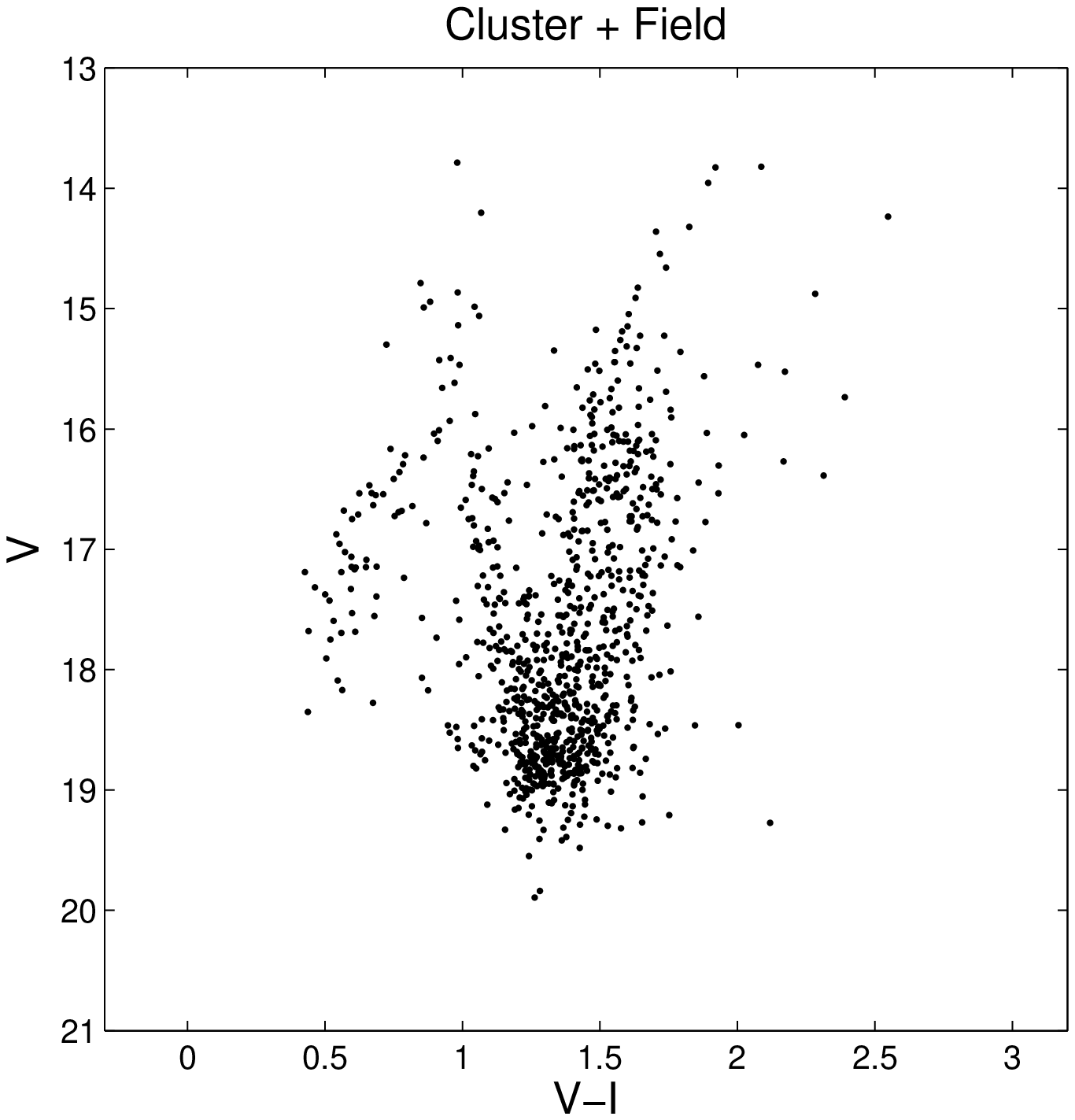} 
\includegraphics[scale=0.35]{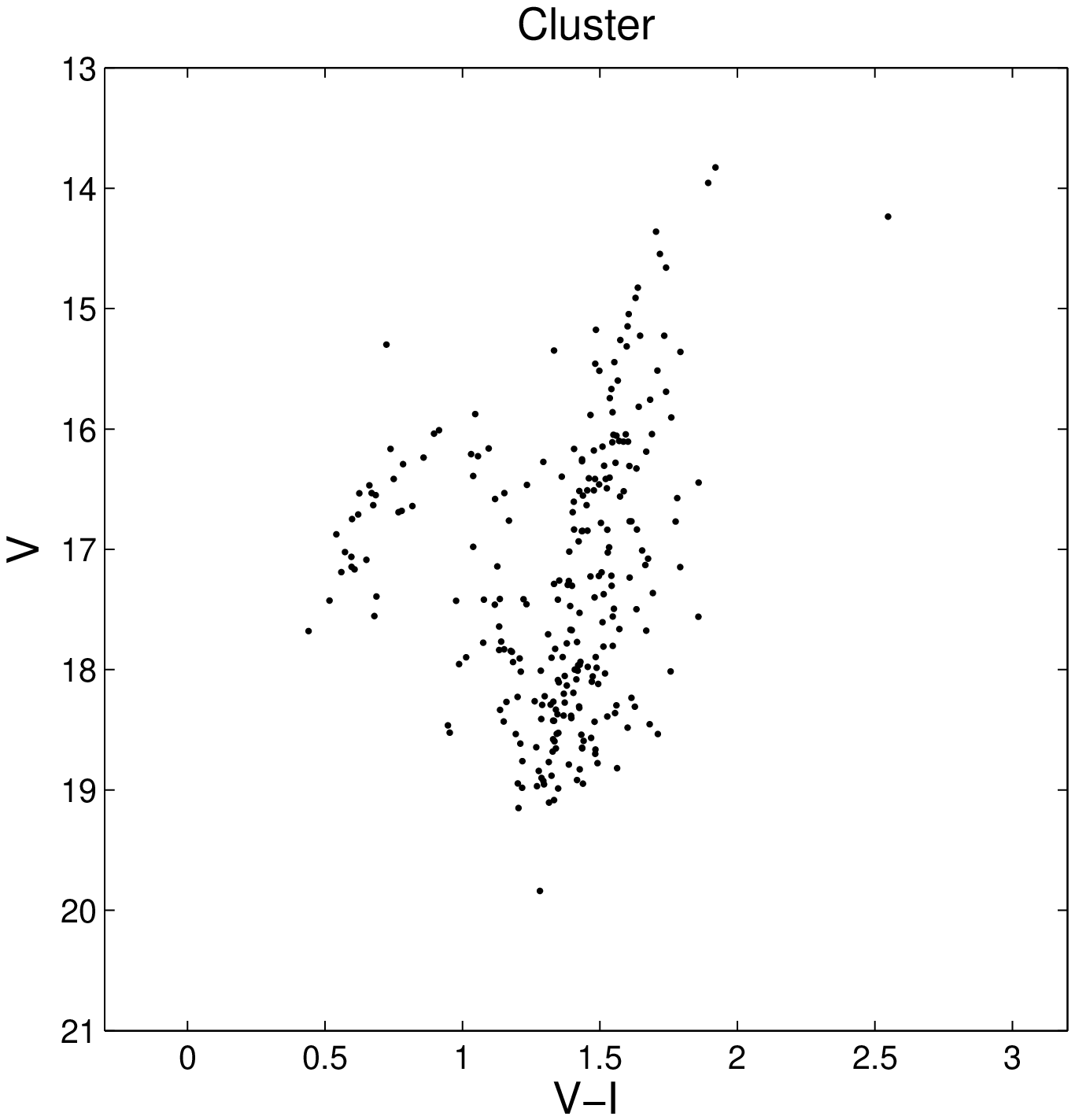} 
\includegraphics[scale=0.35]{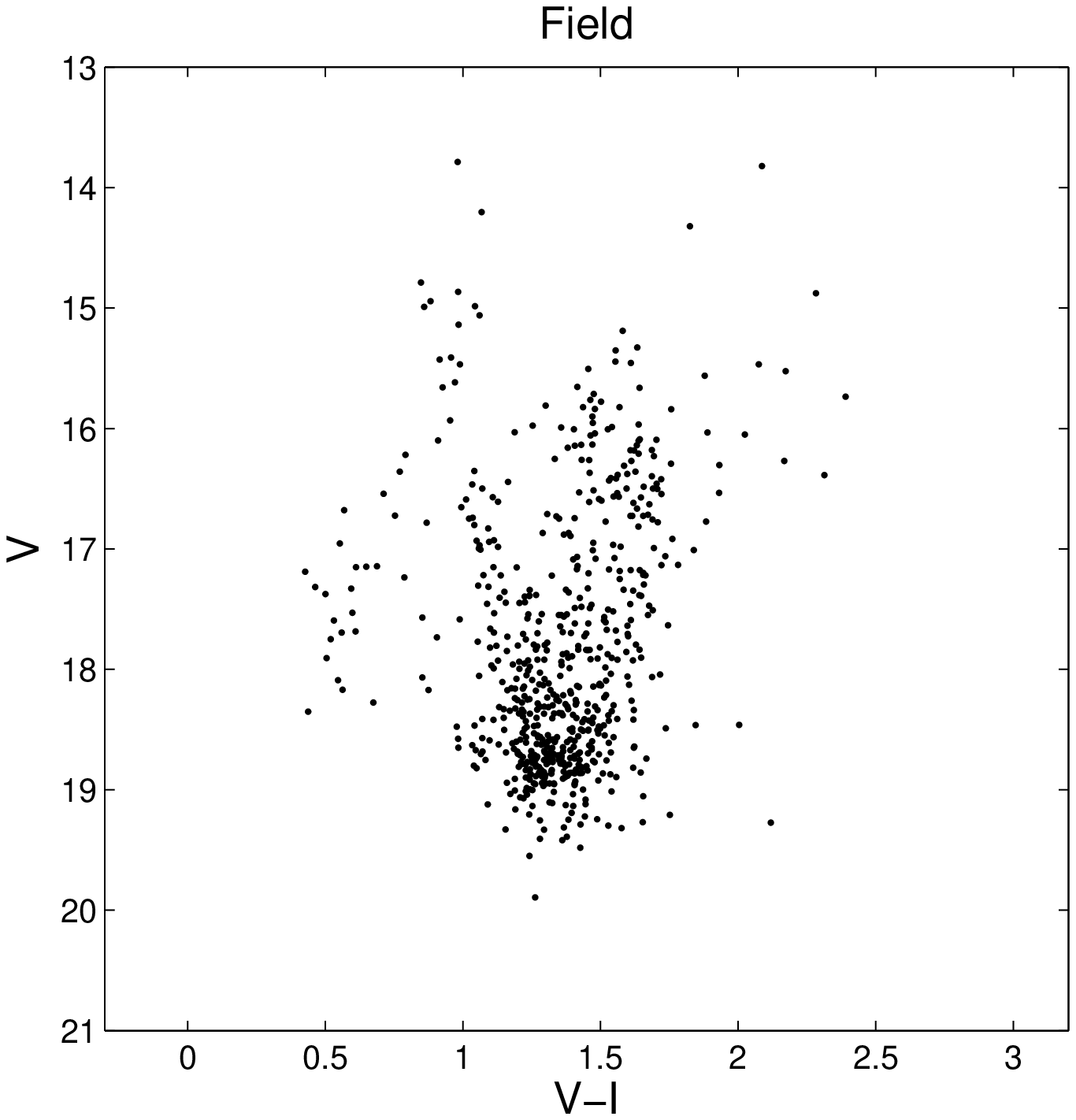}
\end{tabular}
\caption{Proper motion and CMD decontamination of NGC 6558.}
\label{fig:NGC6558pm}
\end{figure*}

\begin{figure*}
\centering
\begin{tabular}{c c c}
\includegraphics[scale=0.35]{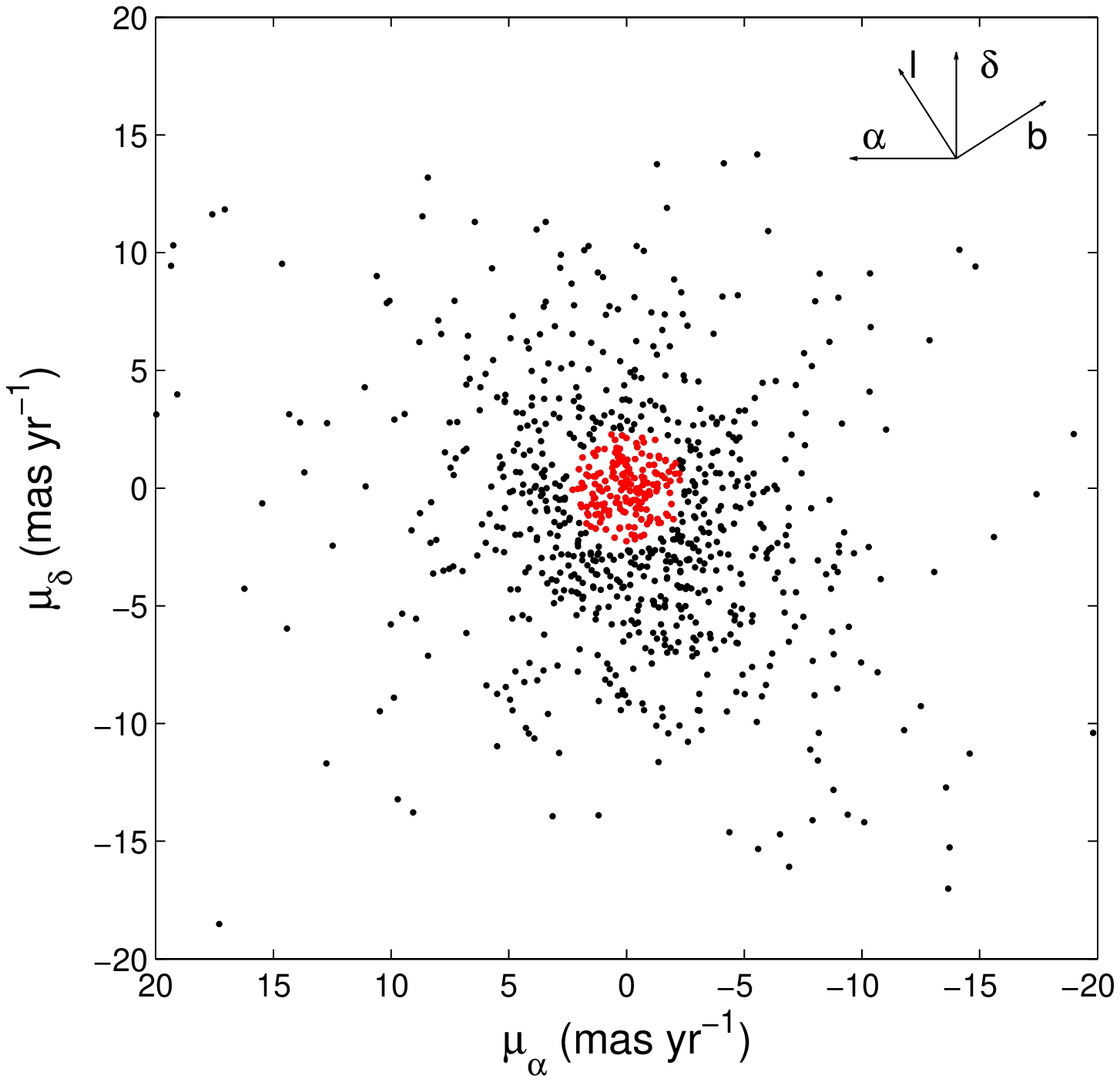} 
\includegraphics[scale=0.35]{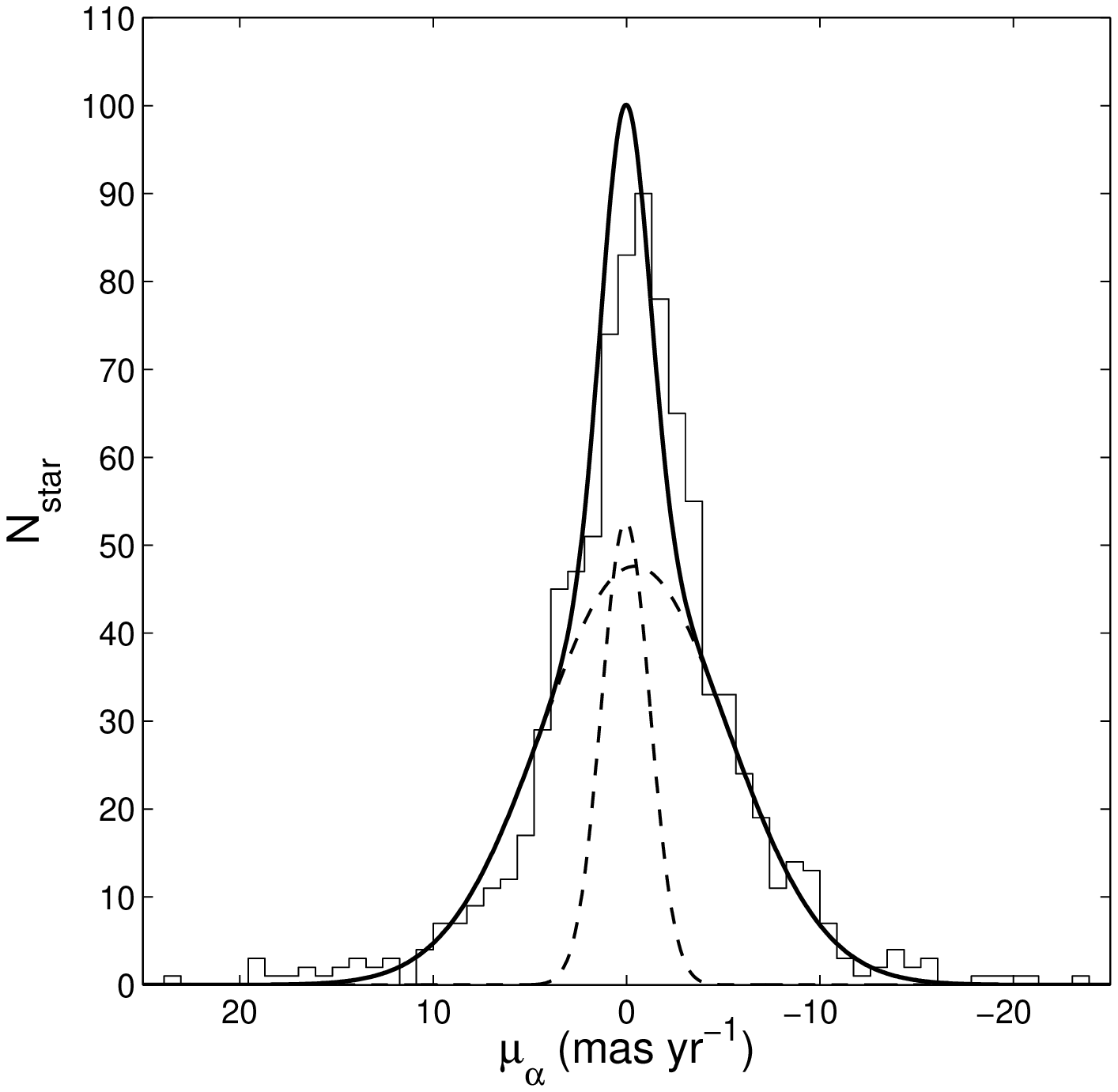} 
\includegraphics[scale=0.35]{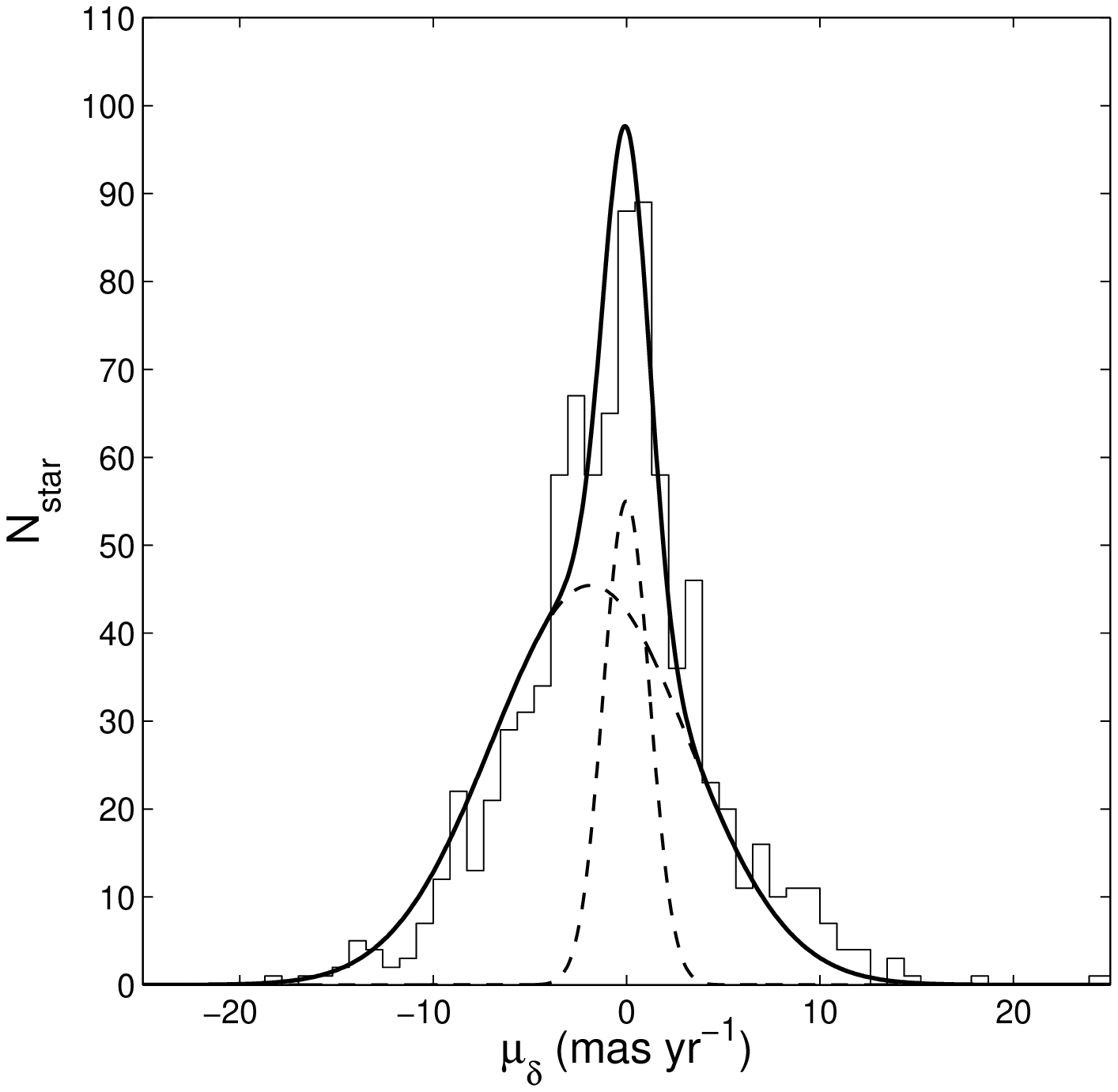}\\
\includegraphics[scale=0.35]{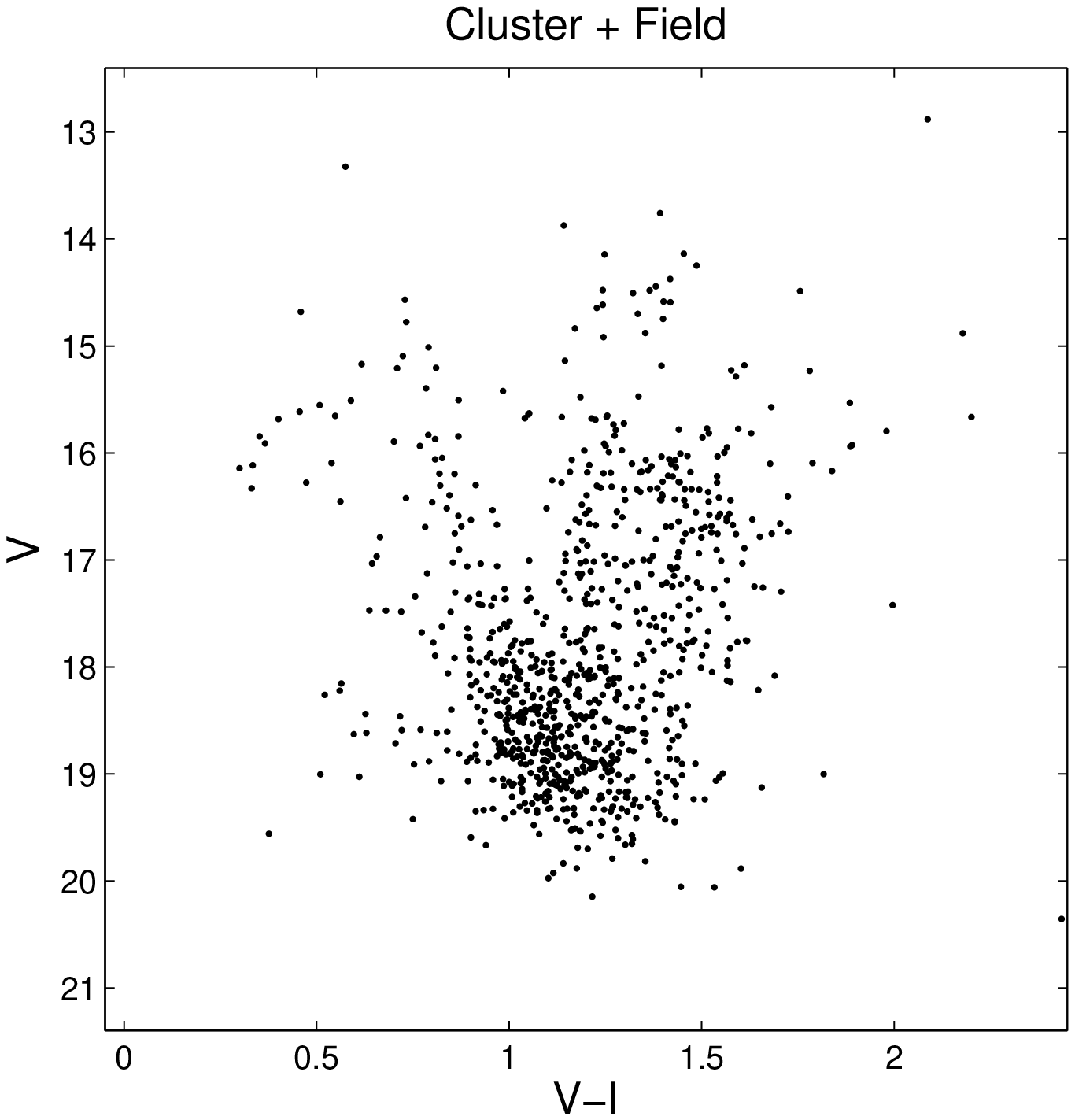} 
\includegraphics[scale=0.35]{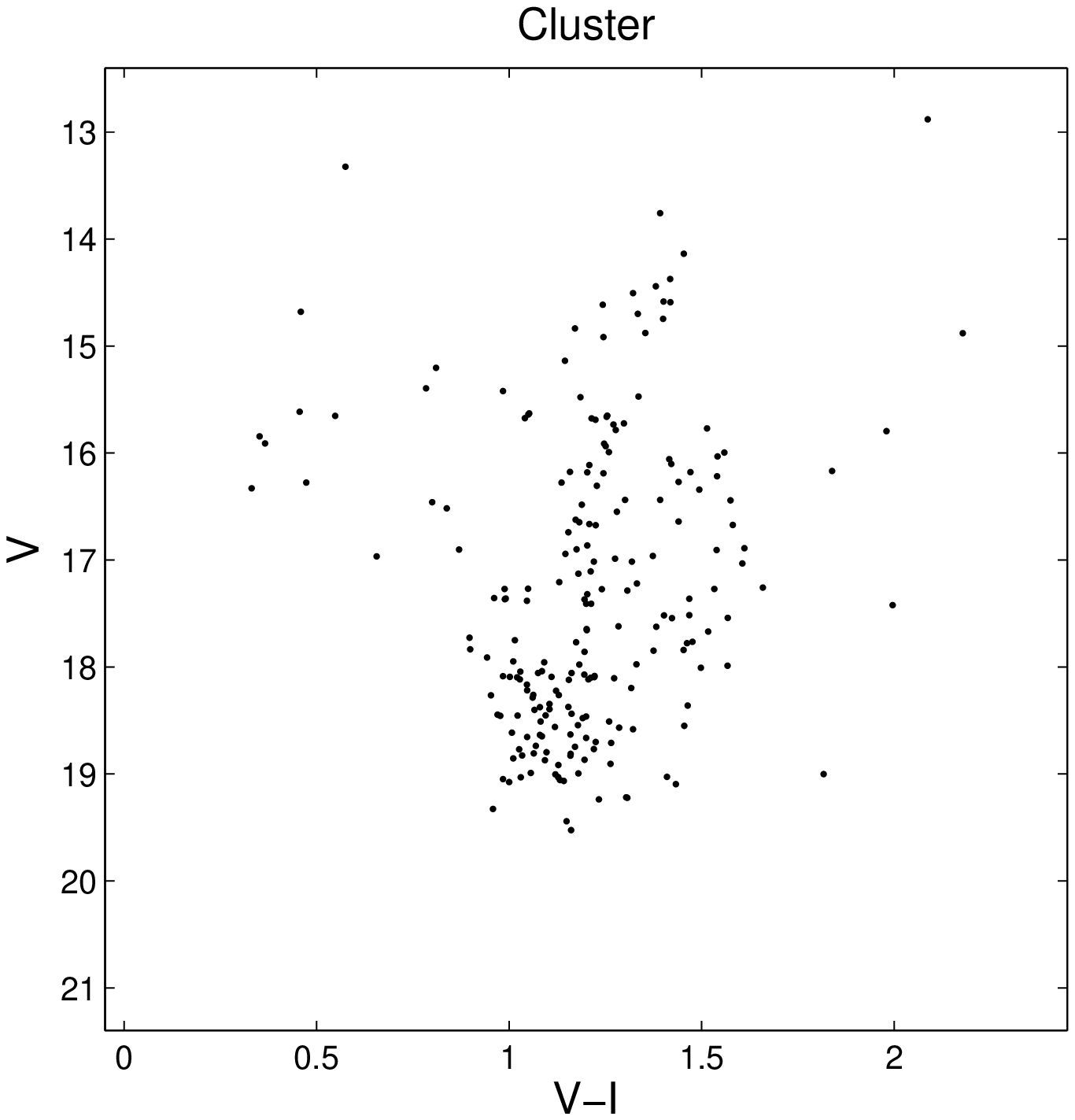} 
\includegraphics[scale=0.35]{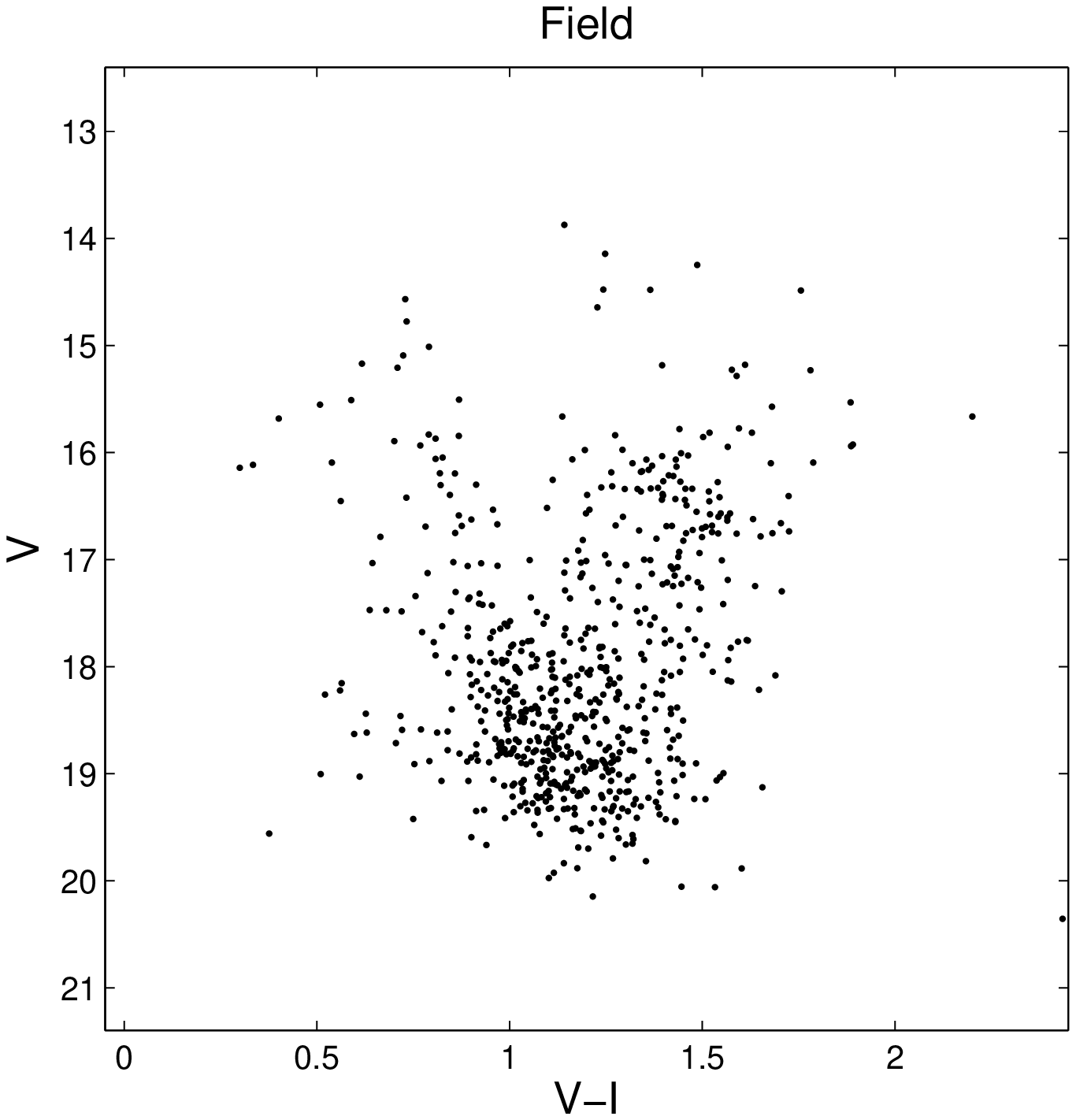}
\end{tabular}
\caption{Proper motion and CMD decontamination of NGC 6540.}
\label{fig:NGC6540pm}
\end{figure*}

\begin{figure*}
\centering
\begin{tabular}{c c c}
\includegraphics[scale=0.35]{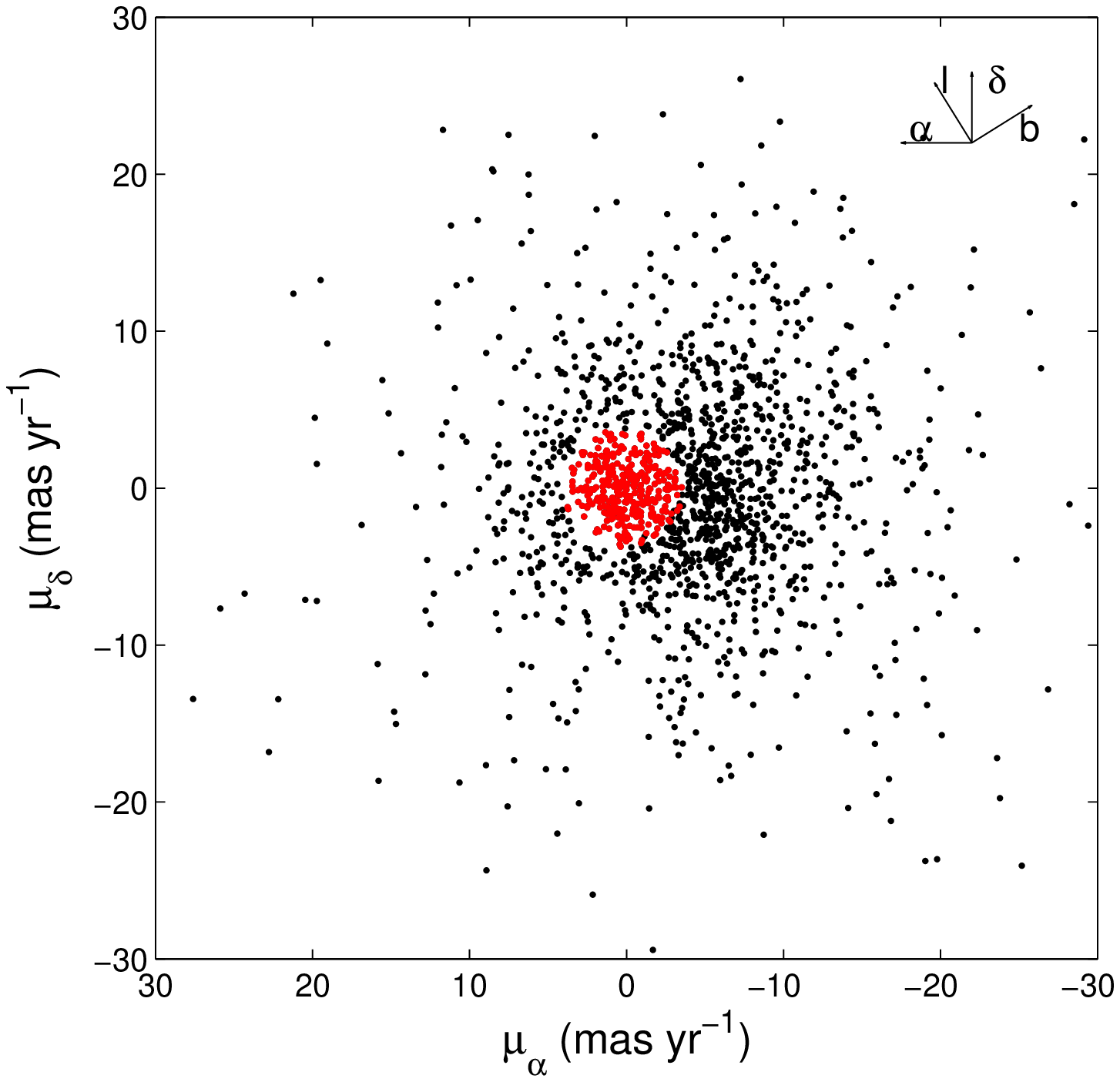} 
\includegraphics[scale=0.35]{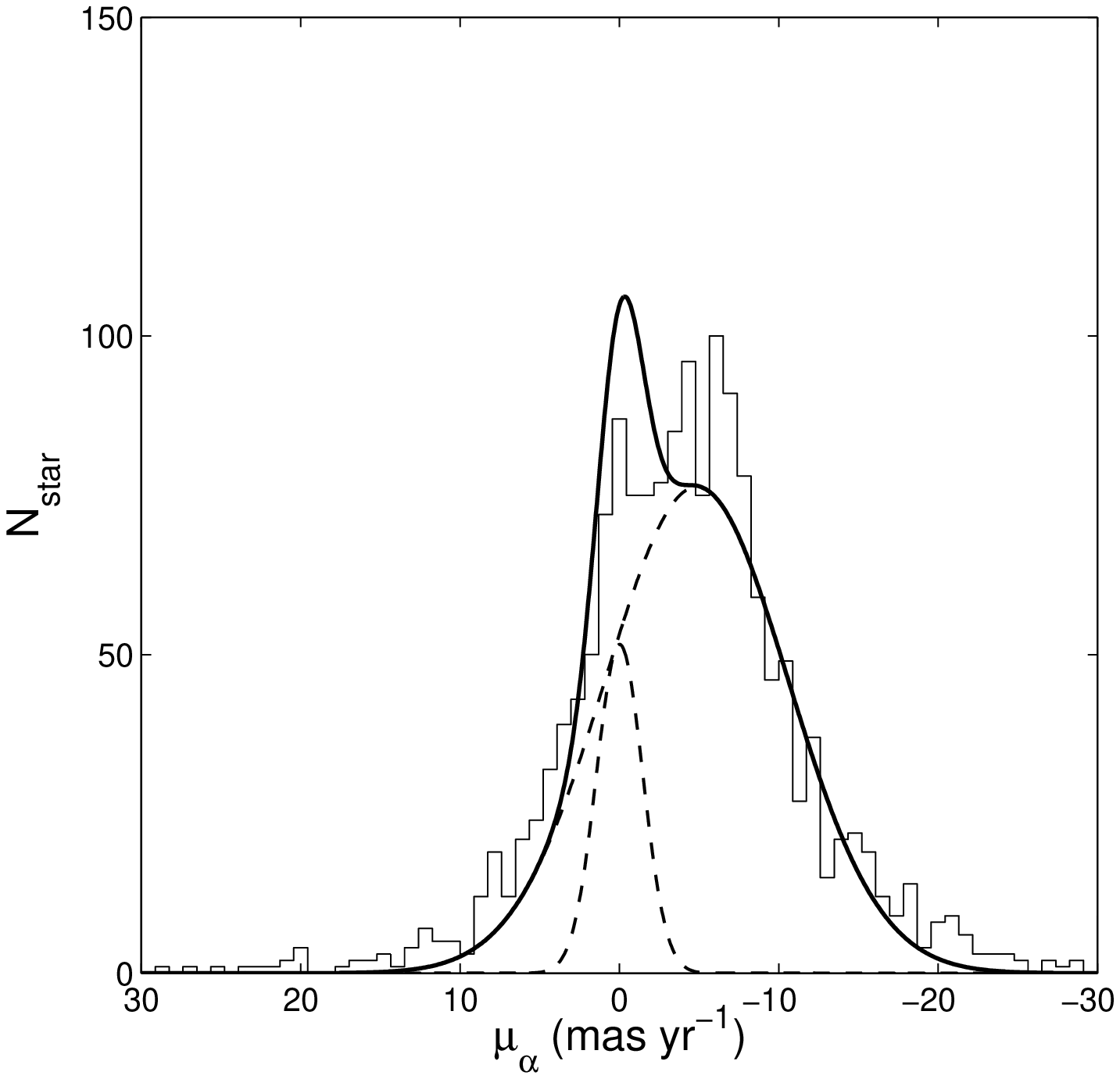} 
\includegraphics[scale=0.35]{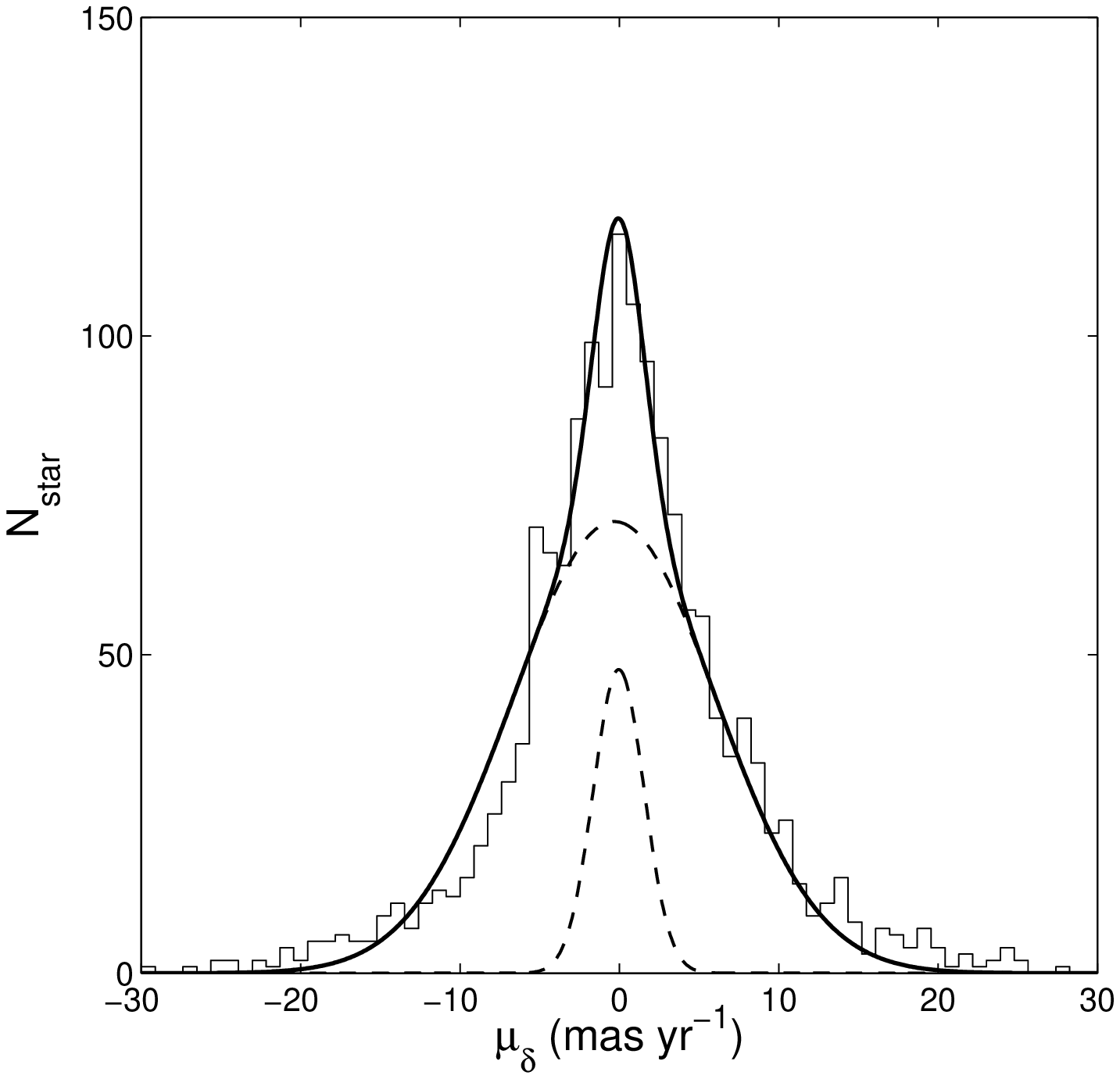}\\
\includegraphics[scale=0.35]{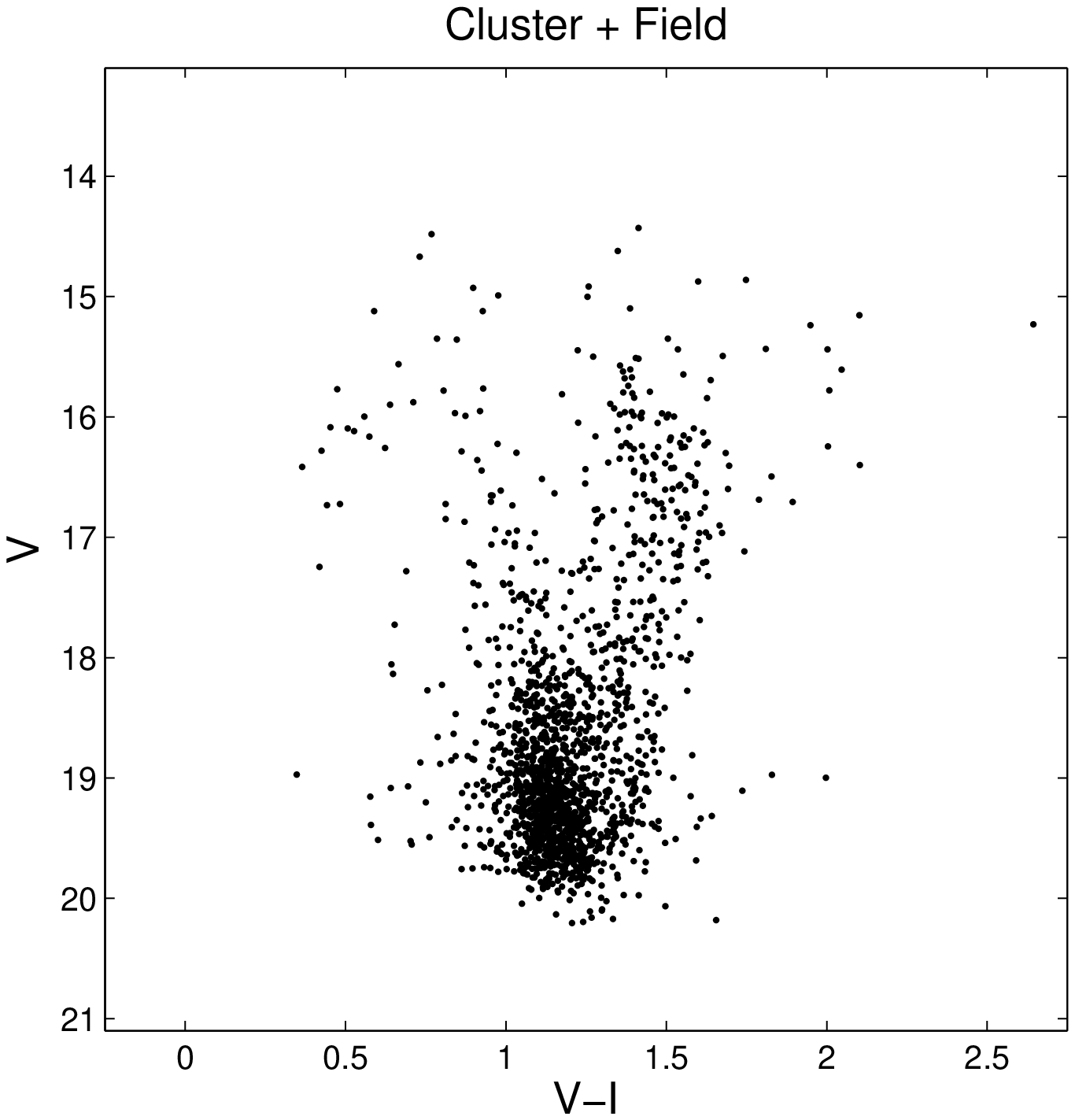} 
\includegraphics[scale=0.35]{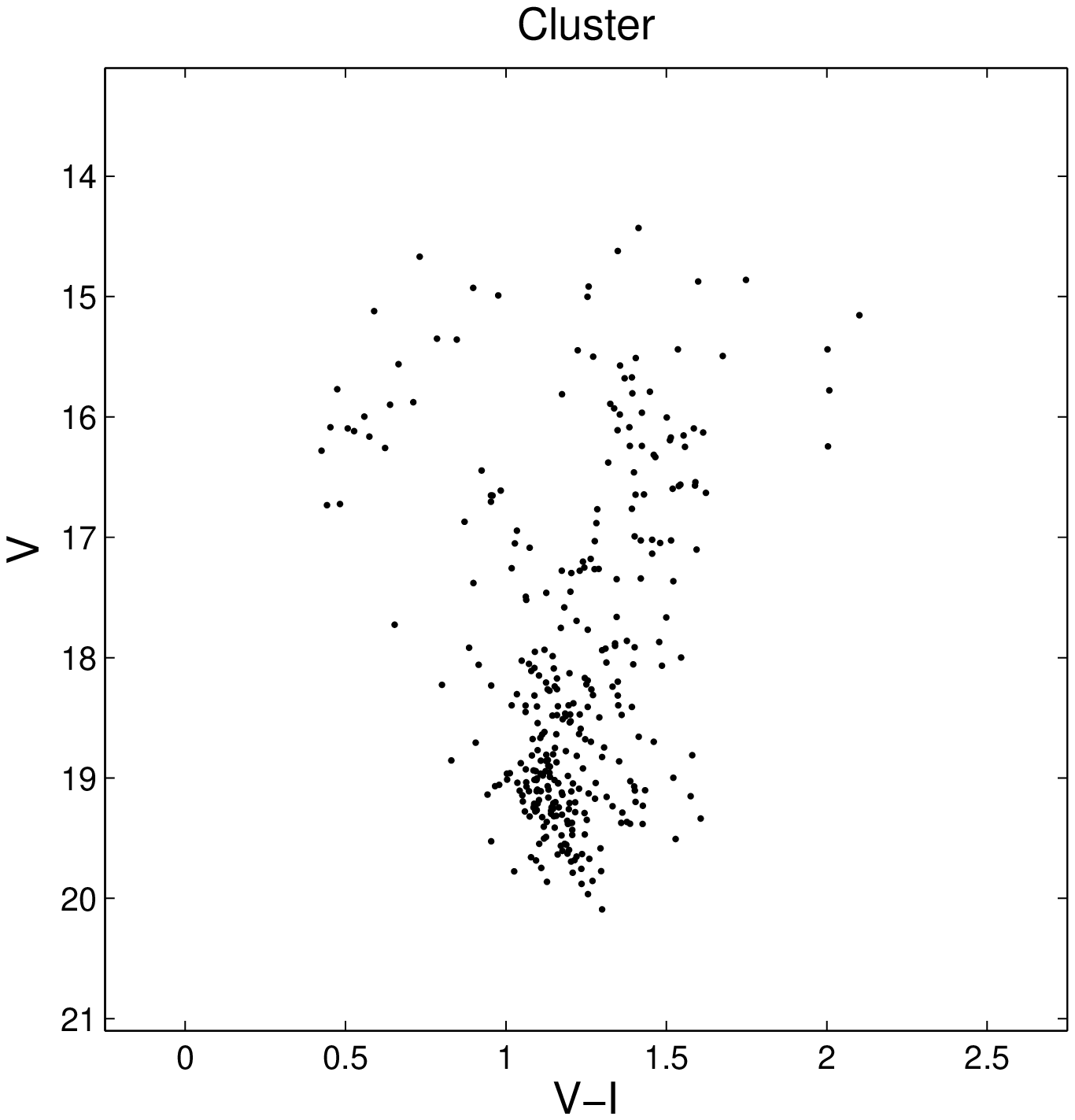} 
\includegraphics[scale=0.35]{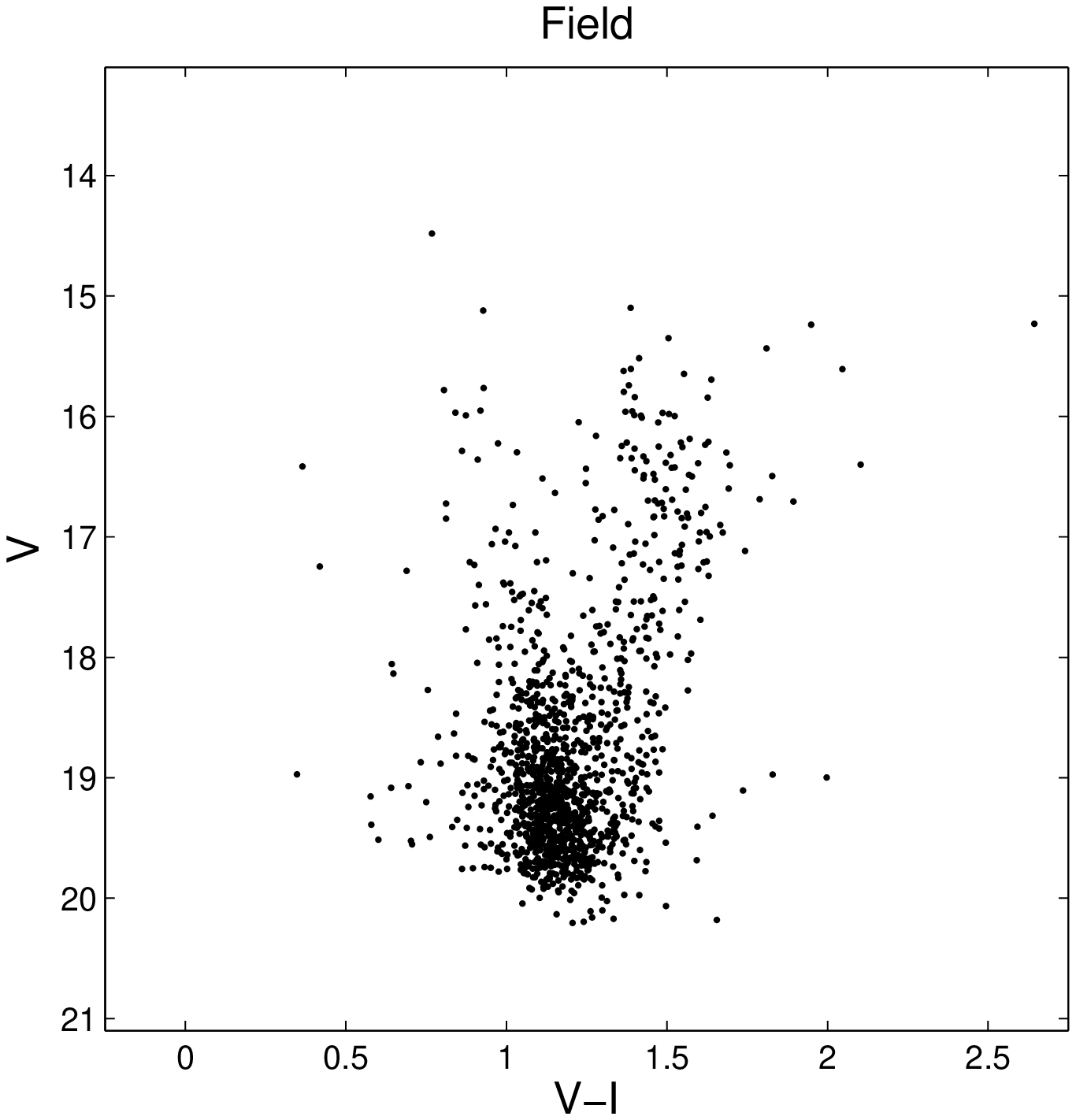}
\end{tabular}
\caption{Proper motion and CMD decontamination of AL 3.}
\label{fig:AL3pm}
\end{figure*}

\begin{figure*}
\centering
\begin{tabular}{c c c}
\includegraphics[scale=0.35]{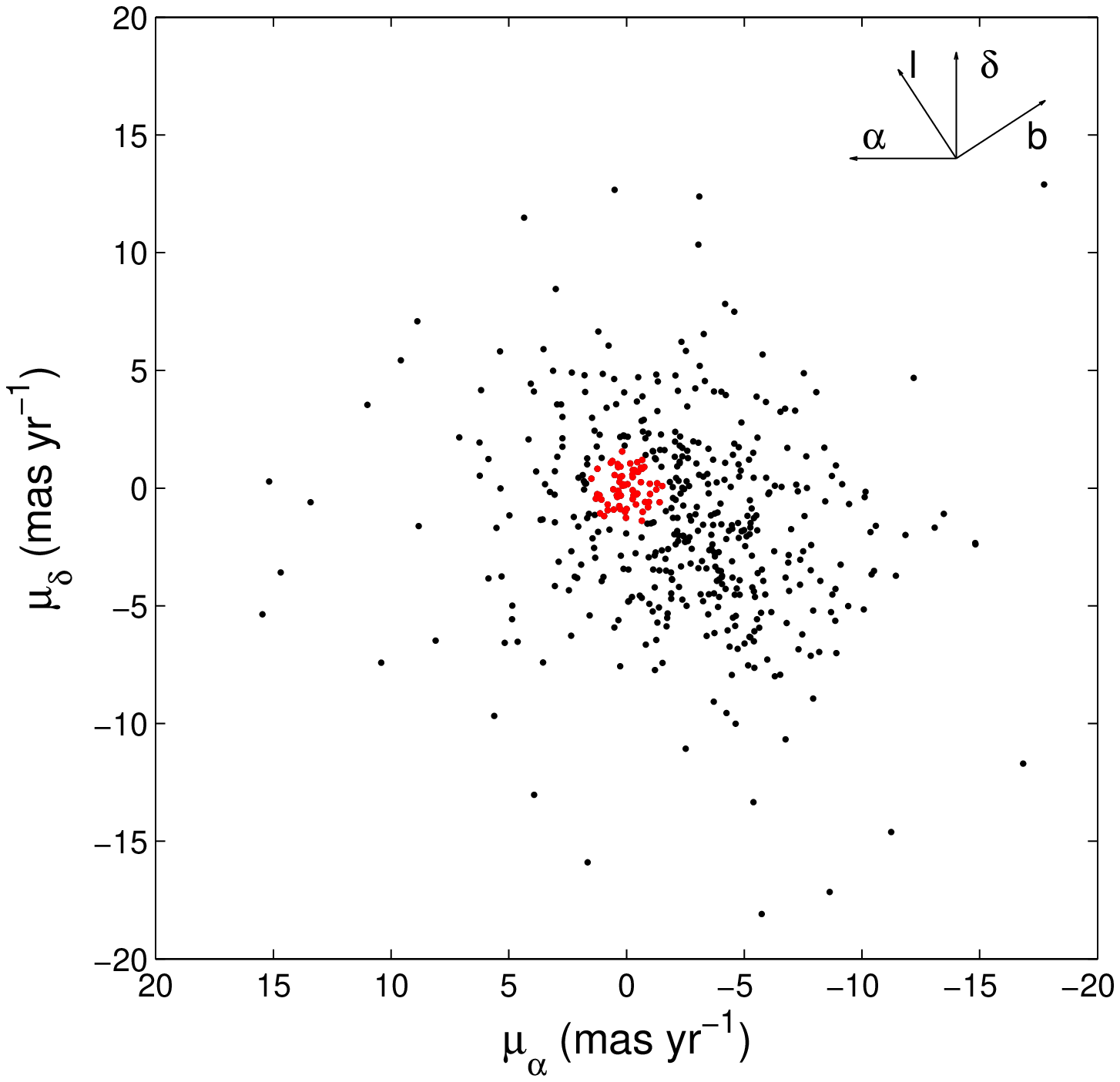} 
\includegraphics[scale=0.35]{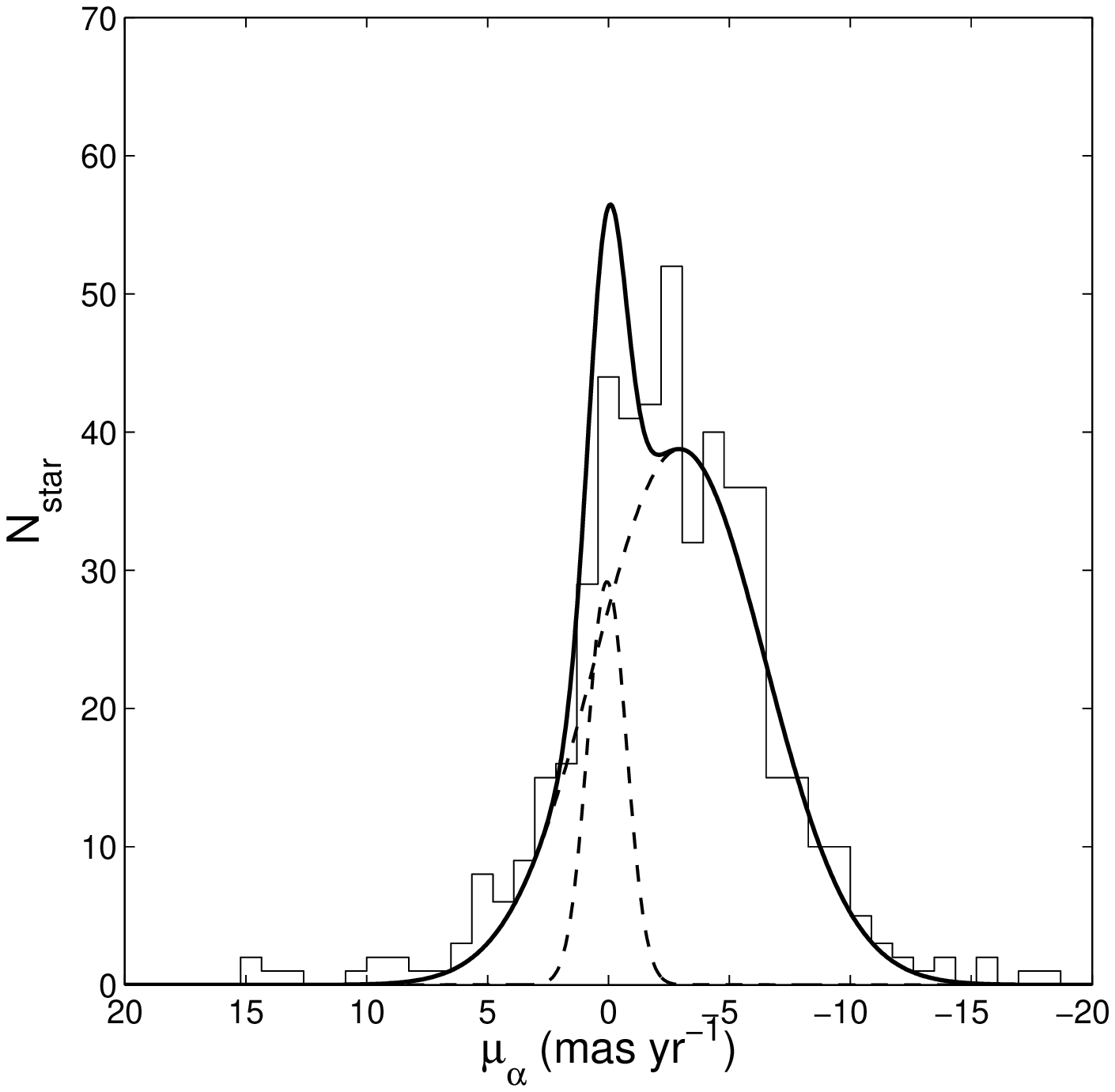} 
\includegraphics[scale=0.35]{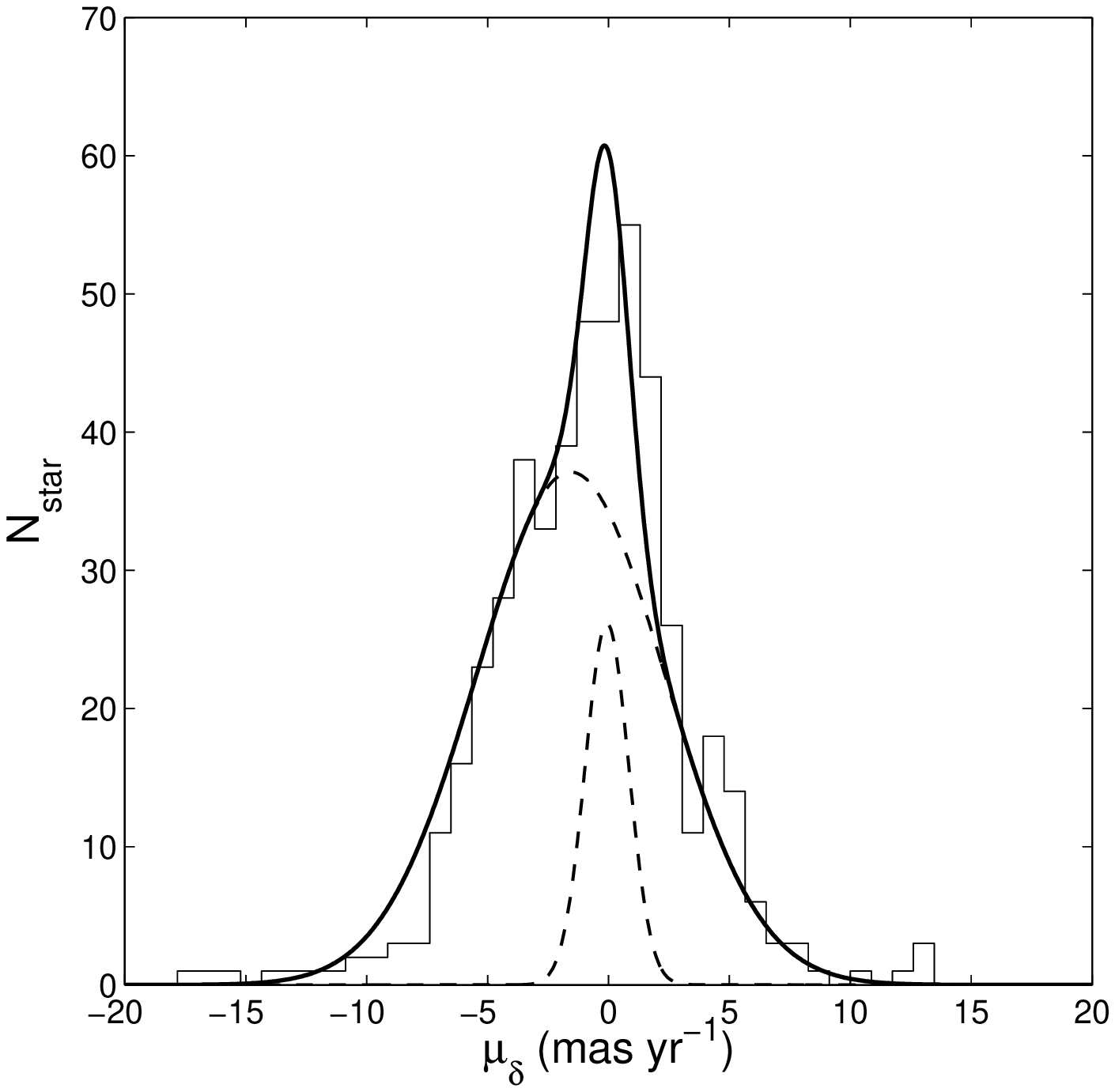}\\
\includegraphics[scale=0.35]{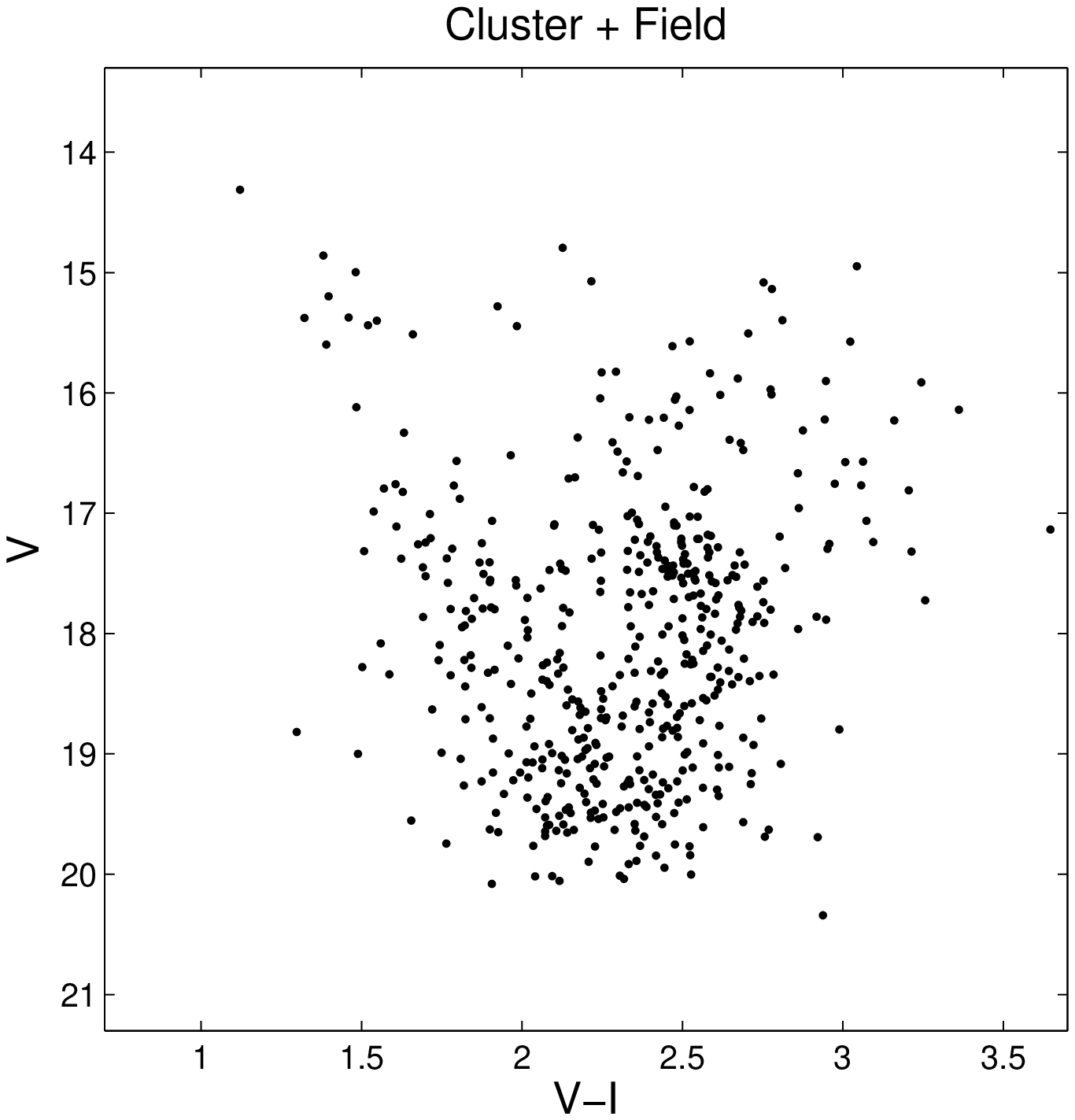} 
\includegraphics[scale=0.35]{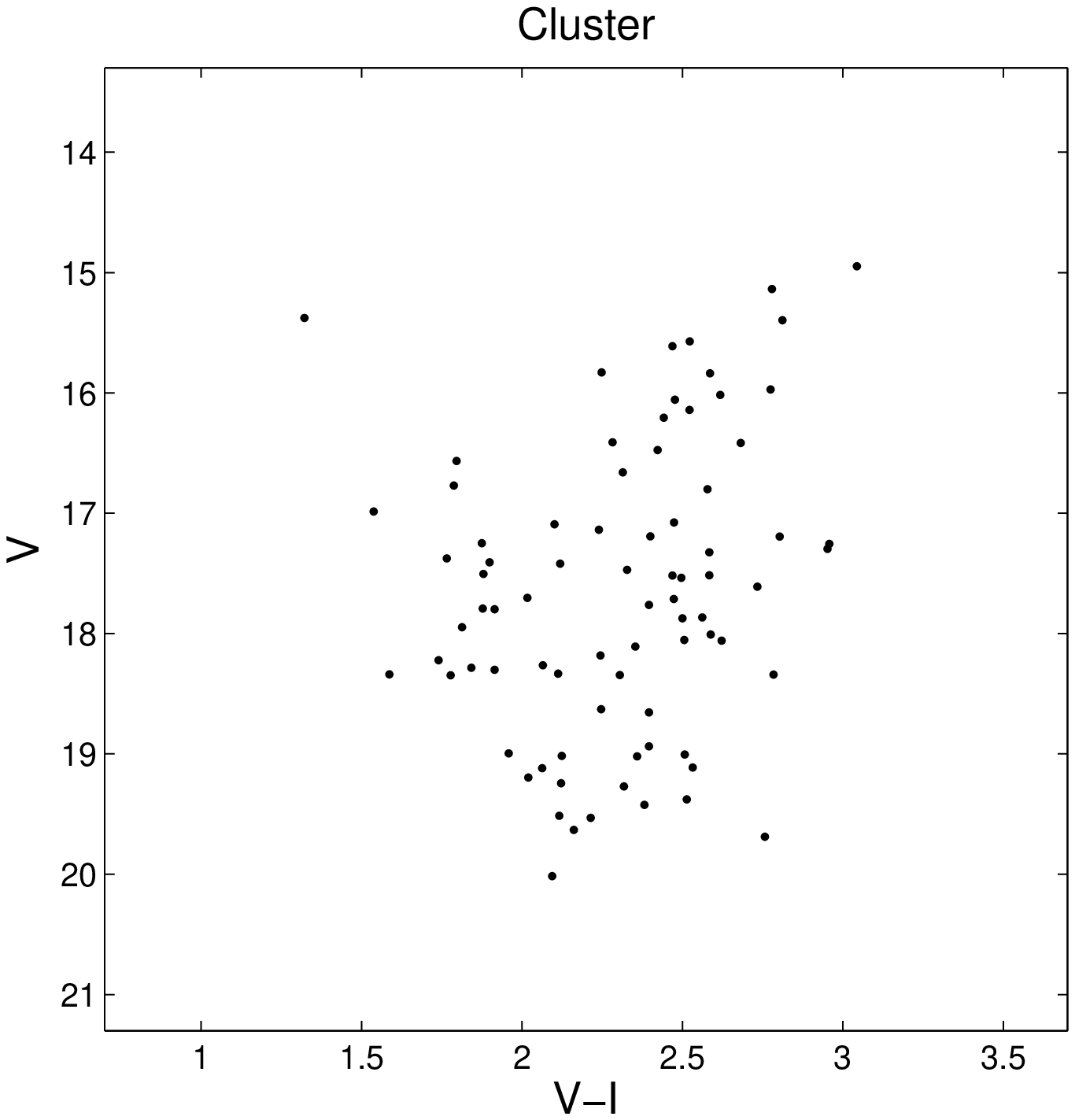} 
\includegraphics[scale=0.35]{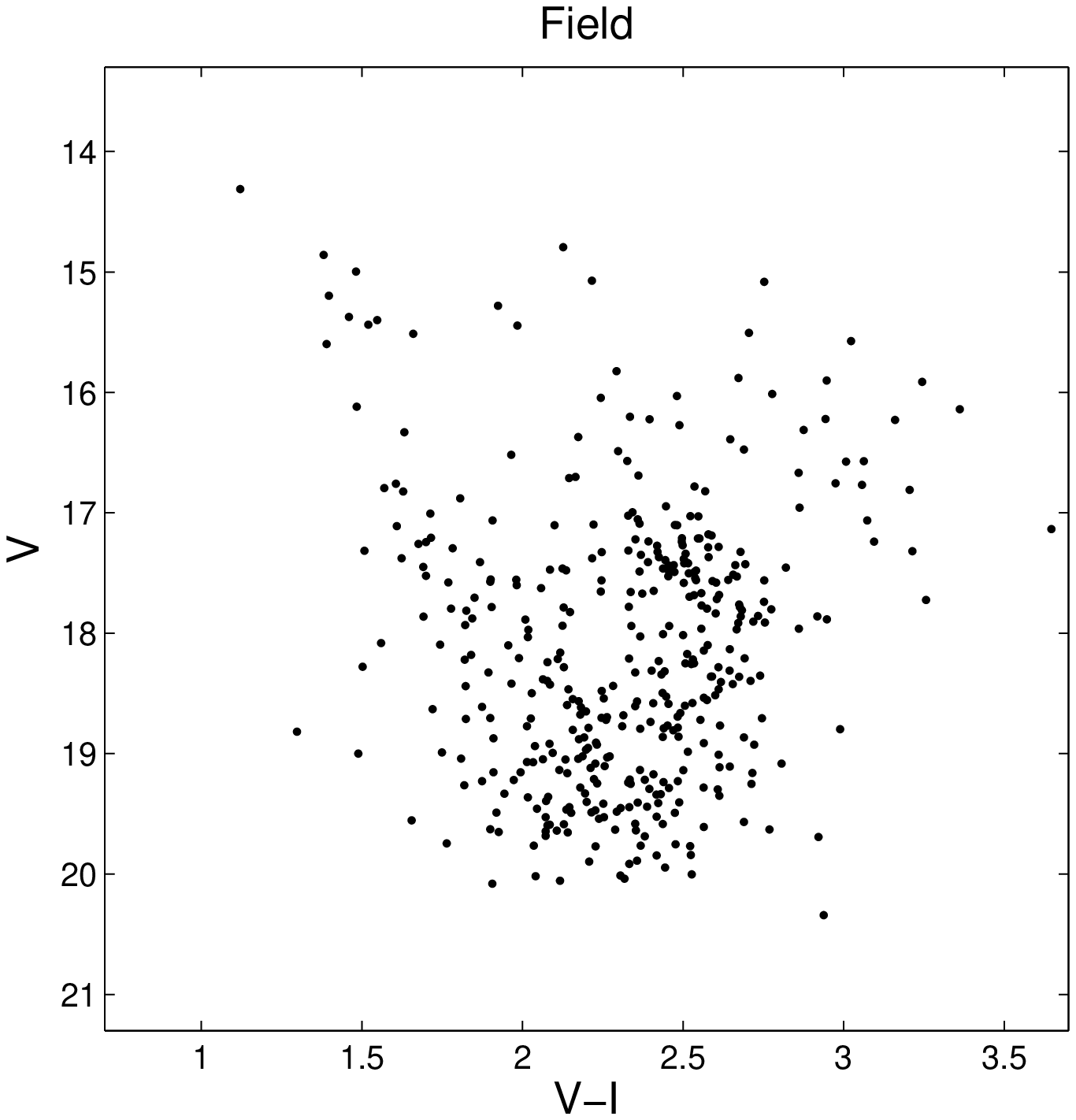}
\end{tabular}
\caption{Proper motion and CMD decontamination of ESO456-SC38.}
\label{fig:Djorg2pm}
\end{figure*}

\begin{figure*}
\centering
\begin{tabular}{c c c}
\includegraphics[scale=0.35]{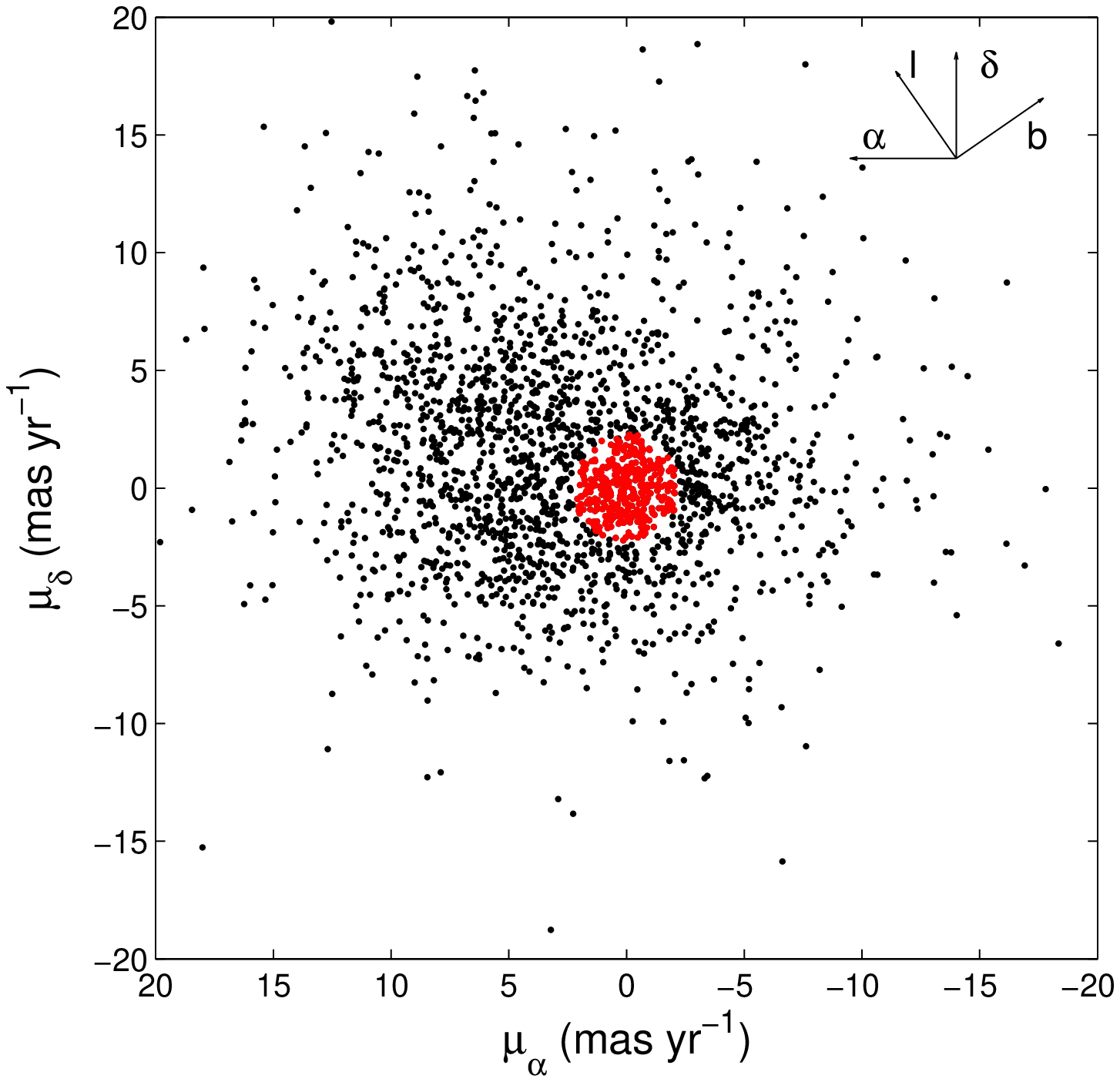} 
\includegraphics[scale=0.35]{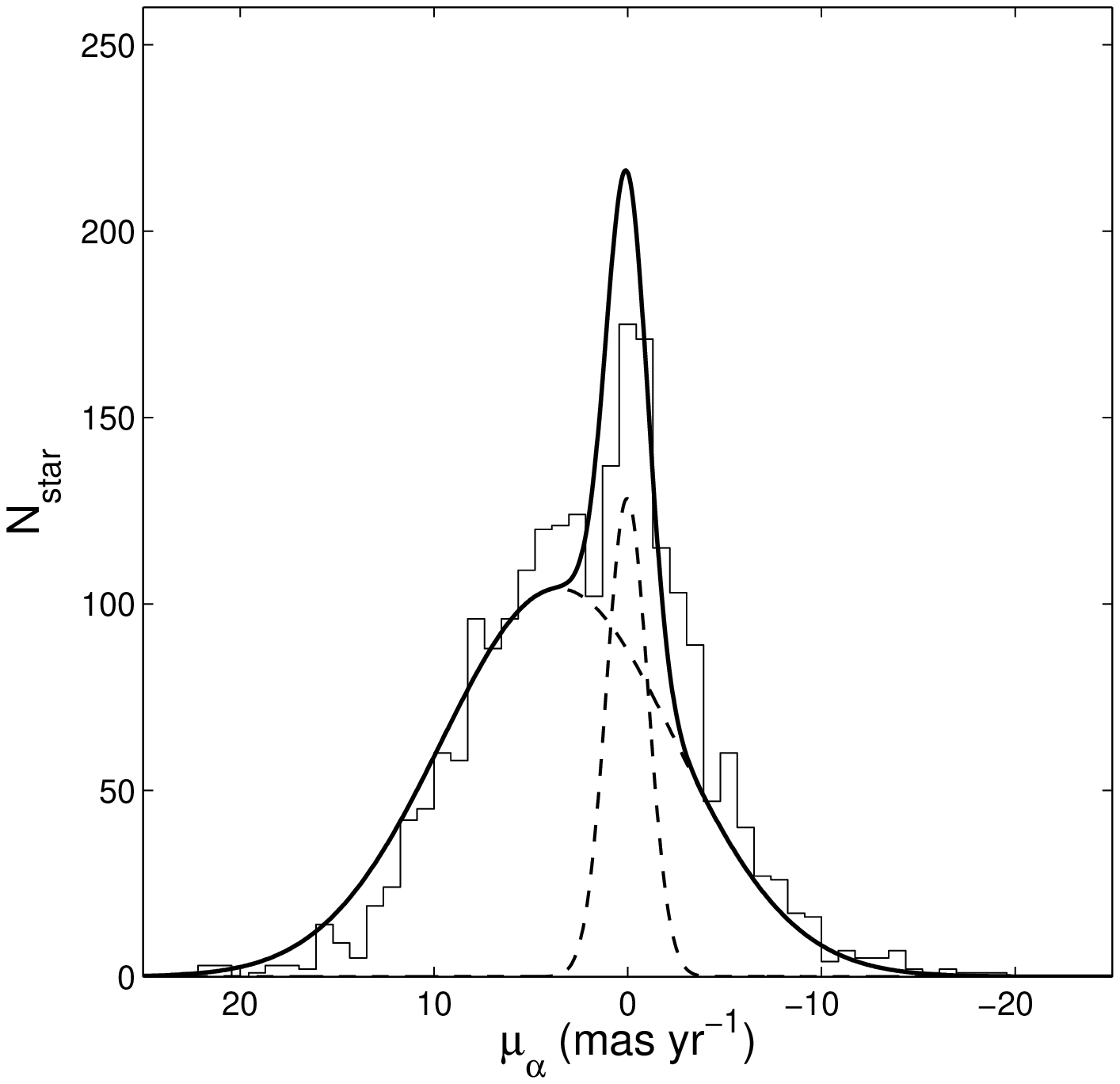} 
\includegraphics[scale=0.35]{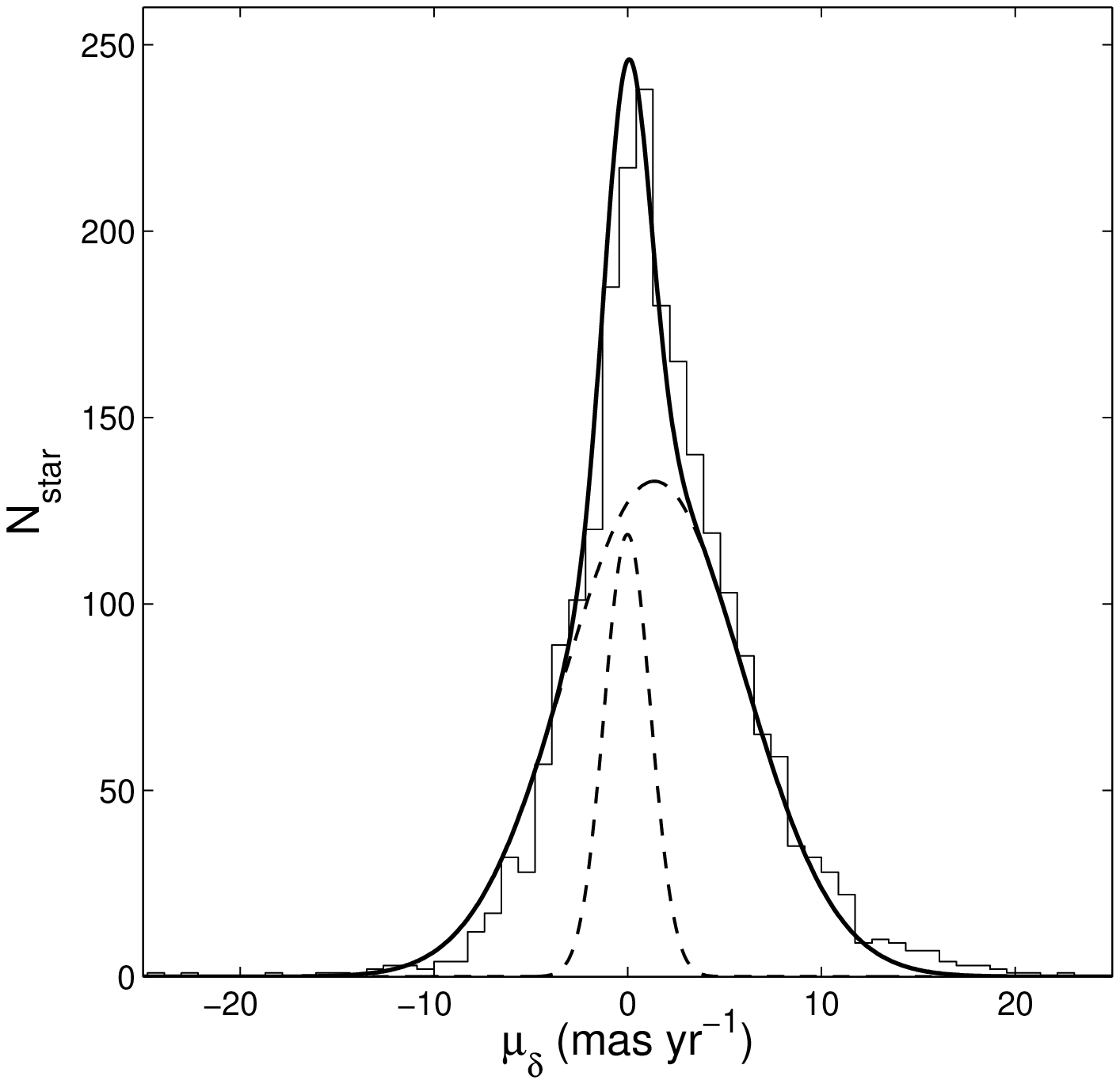}\\
\includegraphics[scale=0.35]{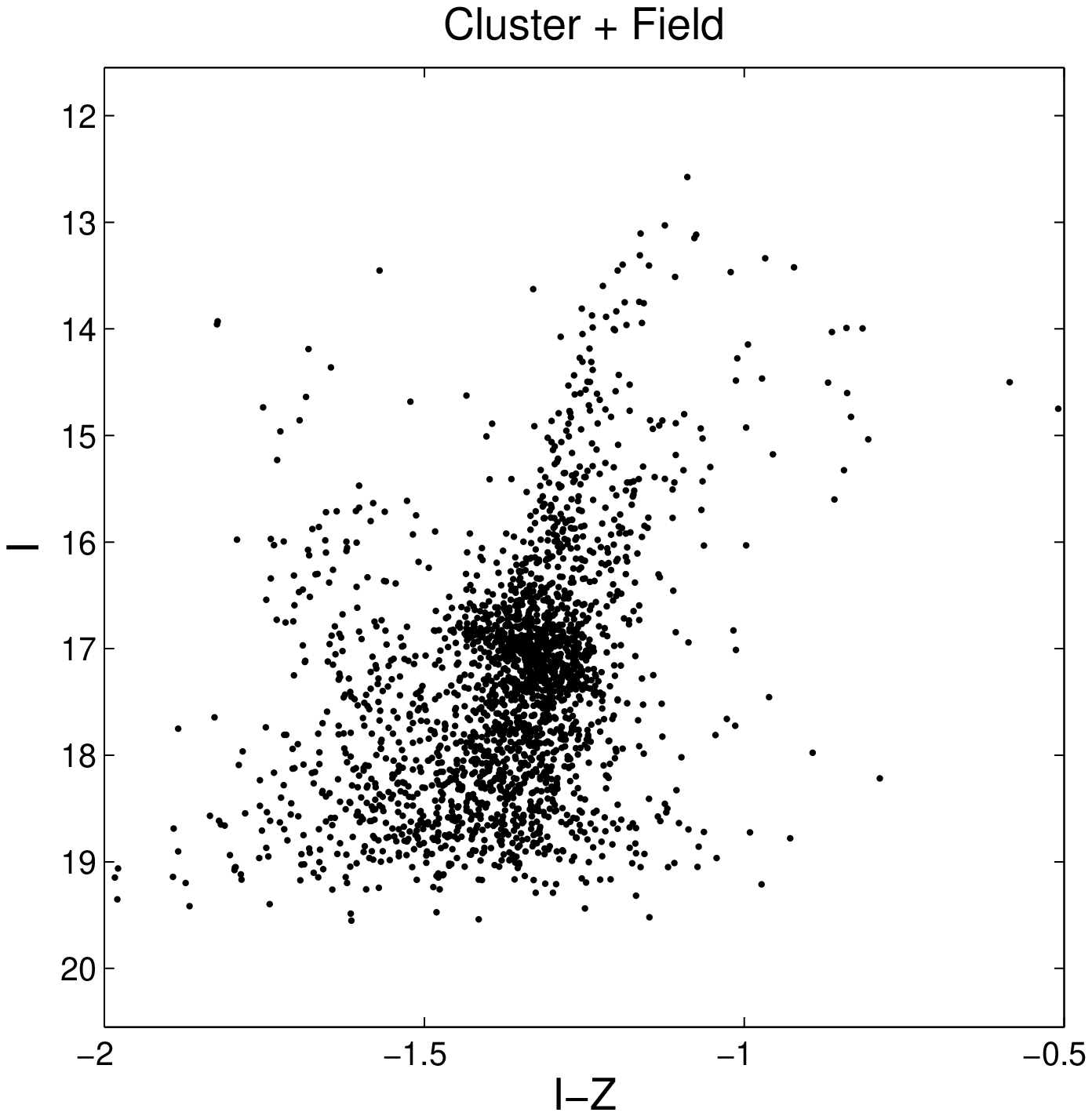} 
\includegraphics[scale=0.35]{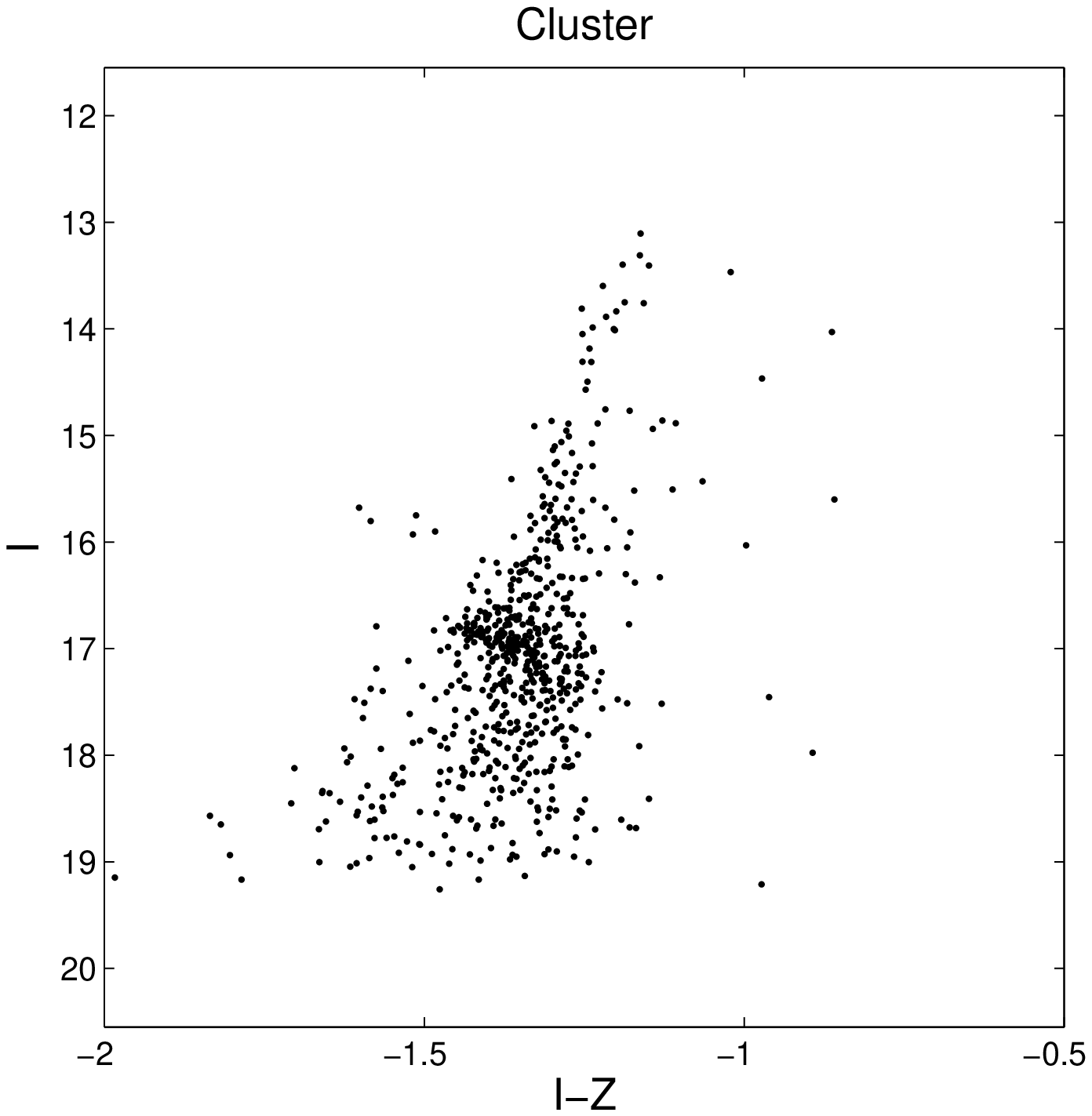} 
\includegraphics[scale=0.35]{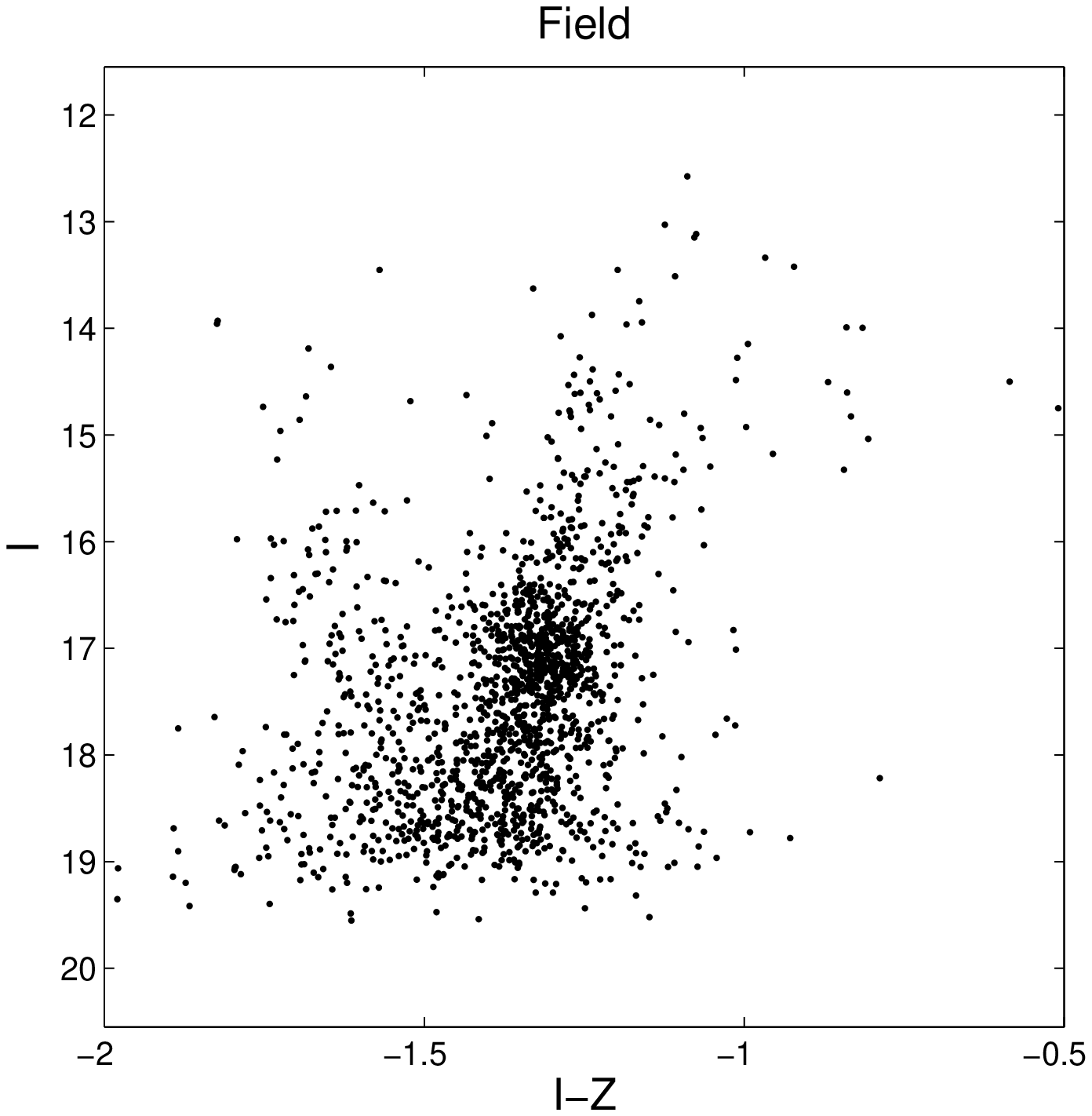}
\end{tabular}
\caption{Proper motion and CMD decontamination of Palomar 6.}
\label{fig:Palomar6pm}
\end{figure*}




\clearpage

\label{lastpage}
\end{document}